\documentclass[%
reprint,
 amsmath,amssymb,
 aps,
]{revtex4-2}
\usepackage{booktabs}
\usepackage{aas_macros}
\usepackage{graphicx}
\usepackage{dcolumn}
\usepackage{bm}
\usepackage{hyperref}
\usepackage{color}
\usepackage{natbib}
\usepackage{CJK}
\usepackage[normalem]{ulem}
\usepackage[usenames,dvipsnames,svgnames,table]{xcolor} 

\usepackage[english]{babel}

\newcommand{\jcapp}{{J.~Cosmol.~Astropart.~Phys.}}

\newcommand{\rvw}[1]{{#1}}

\begin{document}
\begin{CJK*}{UTF8}{gbsn} 
\preprint{APS/123-QED}

\title{A binary black hole metric approximation from inspiral to merger}

\author{Luciano Combi$^{1,2}$, Sean M. Ressler$^3$}

\affiliation{
$^1$Perimeter Institute for Theoretical Physics, Waterloo, Ontario, Canada, N2L 2Y5\\
$^2$Department of Physics, University of Guelph, Guelph, Ontario, Canada, N1G 2W1 \\
$^3$Canadian Institute for Theoretical Astrophysics, University of Toronto, Toronto, On, Canada M5S 3H8} 

\begin{abstract}
We present a {semi-analytical} binary black hole (BBH) metric approximation that models the entire evolution of the system from inspiral to merger. The metric is constructed as a boosted Kerr-Schild superposition following post-Newtonian (PN) trajectories at the fourth PN order in the inspiral phase. During merger, we interpolate the binary metric {in time} to a single black hole remnant with properties obtained from numerical relativity {(NR)} fitting formulas. The new metric can model binary black holes with arbitrary spin direction, mass ratio, and eccentricity at any stage of their evolution in a fast and computationally efficient way.
\rvw{We analyze the properties of our new metric and compare it with a full numerical relativity evolution}. Hamiltonian constraints are well-behaved even at merger, and the mass and spin measured self-consistently on the black hole apparent horizon deviate on average only $\lesssim  10\%$ compared to the full numerical evolution. 
\rvw{We perform General Relativistic Magneto-hydrodynamical (GRMHD) simulations for two cases: merging black holes in a uniform gas, and inspiralling black holes accreting from a magnetized circumbinary disk. We demonstrate that, in both cases, the properties of the gas, such as the accretion rate, are remarkably similar between the two approaches, with small average differences.} We demonstrate that our approximate metric significantly reduces computational cost compared to full numerical relativity, enabling a new class of high-resolution, long-term binary accretion simulations. The numerical implementation of the metric is now open-source and optimized for numerical work.
\end{abstract}

\maketitle
\end{CJK*}

\section{Introduction}

Mergers of compact objects, such as black holes and neutron stars, are the most powerful gravitational wave sources in the Universe. When matter is involved in these mergers, e.g., in merging neutron stars, they are likely to produce copious amounts of electromagnetic radiation. In the last decade, multimessenger detection of gravitational wave sources by LIGO/VIRGO in joint efforts with ground-based and space telescopes has opened a new window to the cosmos \citep{abbott2017gravitational}.

Binary black hole mergers may also have an electromagnetic counterpart if embedded in a gas-rich environment. Supermassive binary black holes, in particular, are promising multi-messenger sources \citep{kelly2021electromagnetic,bogdanovic2022electromagnetic}. After two galaxies merge, the supermassive black holes of each galaxy can dive into the galaxy's gravitational well and form a binary \citep{Begelman1980}.  Because most galaxy mergers drive a lot of gas {inwards}, a circumbinary accretion disk can form around the source \citep{Artymowicz1996}. {If enough {orbital} angular momentum is extracted from the system, either by the gas or stellar components, the separations will be sufficiently small to trigger efficient emission of GW. In this case, the binary can merge in less than a Hubble time  \citep{Peters:1963ux, Peters:1964zz}. Strong evidence for the existence of these close-separation supermassive black hole binaries has been provided by the recent detection of a nanohertz gravitational wave background by several pulsar timing arrays \citep{afzal2023nanograv, agazie2023nanograv}. Future space-borne GW detectors such as LISA are expected to detect the mergers from these massive sources \citep{LISA2017}.

An accreting binary system would likely appear very similar to a typical active galactic nucleus (AGN) with subtle distinguishing spectral and time-domain features \citep{graham2015systematic, Charisi2016}. That said, binary accretion flows differ quite dramatically from single BH accretion disks near the source. For aligned equal-mass binaries, the time-dependent non-axisymmetric potential produces strong torques that carve a low-density cavity around the binary \citep{Roedig2014, MacFadyen2008, Noble12, Shi:2012ApJ}; accretion then proceeds by thin streams that plunge from the circumbinary disk to the BHs with sufficient specific angular momentum to form mini-disks \citep{bowen2017, RyanMacFadyen17, westernacher2021, DOrazio13}. For misaligned, unequal mass binaries, the secondary object can interact with the disk surrounding the primary{,} producing quasi-periodic flares \citep{Komossa2006, Sukova2021,Lehto1996, britzen2023precession}. Other black hole binary systems could also be in gas-rich environments, such as stellar mass binaries embedded in the plane of an AGN disk (see \citep{dittmann2023evolution}) or an extreme mass-ratio inspiral (EMRI), where a much smaller black hole eventually merges with a central supermassive black hole \citep{franchini2023quasi}.}

Modeling binary black hole sources along with the surrounding gas presents several challenges. Given the non-linear interactions inherent in a turbulent plasma, quantitative models require magneto-hydrodynamical (MHD) simulations. With the wide separation of scales between the circumbinary accretion disk and the accretion near each BH, simulations must be run for thousands of orbits to reach a steady state \citep{Munoz2020, Miranda2017}. For this reason, most of the previous literature on binaries has focused on two-dimensional hydrodynamical Newtonian simulations, which are easily evolved for very long timescales \citep{duffell2024santa}. To get reliable electromagnetic predictions from these systems, however, we need to resolve the near-horizon flow, where most of the luminosity comes from, and thus 3D GRMHD simulations are essential \citep{cattorini2023grmhd, Gold2019}. 

{Unlike mergers involving neutron stars, the gas around binary black hole mergers is expected to be much less massive than the black holes themselves. 
As a result, the effect of the gas on the spacetime is adiabatic and only relevant on very long timescales.
Therefore, it is generally a good approximation to solve Einstein's field equations, neglecting the energy-momentum tensor of the gas and evolving the spacetime in a vacuum.}

An appropriate black hole binary metric can be obtained from Einstein's equations either by numerical relativity (i.e., solving the equations numerically) or by constructing an analytical approximation {to the equations}. 
The latter is particularly accurate for distances far from the black holes, or at all distances from the black holes if the binary separation is large. There are various ways of constructing such a binary metric approximation. For evolving the outer parts of the binary, a post-Newtonian metric is sufficient; this has been implemented in \citep{noble2012x, zilhao2014pn, noble2021} for circumbinary disk simulations in 3D GRMHD, which are evolved for $\sim 100$ orbits. For evolving the matter around the horizons together with the outer parts, it is possible to construct a matching metric where different approximations, valid at different zones, are sewn together \citep{Mundim2014, ireland16}. This matching approach was used in GRMHD and GRHD simulations to evolve both the circumbinary disk and mini-disks \citep{bowen2019, bowen2017, avara2023accretion, mignon2023origin}. In the special case of equal-mass binaries in circular orbits, a helically symmetric solution can be constructed with the conformal thin-sandwich (CTC) approach.  This method has been used to compute relaxed accretion disk solutions for numerical relativity GRMHD evolution at small separations \citep{farris2011Binary, Farris2012}. 

A somewhat simpler approach is to use a boosted superposition of black holes \citep{combi2021superposed, armengol2021circumbinary, gutierrez2022electromagnetic}. This approach has been used for simulating accretion flows in the entire domain relevant for spinning BHs \citep{combi2021superposed} and ray-tracing \citep{gutierrez2022electromagnetic, davelaar2022selfa, davelaar2022selfb}. The main advantage of a semi-analytical metric approximation is that the computational time and required numerical complexity are greatly reduced when used in a GRMHD evolution compared with numerical relativity. A semi-analytical metric allows for the design of numerical algorithms focused on resolving exclusively the MHD, e.g., using special grids adapted to the flow. This allows for the exploration of certain regimes such as large separations {and small-mass ratios} that are prohibitively expensive in full numerical relativity. Indeed, most of numerical relativity simulations of binary accretion {have only} explored small separations {mass ratios $\gtrsim 0.1$}, or evolved the system for {only} tens of orbits \citep{giacomazzo2012, kelly2021electromagnetic, Gold2014b, paschalidis2021minidisk}

{At very close separations, numerical relativity is required for an accurate calculation of the gravitational wave signal. It is unclear, however, what impact the detailed non-linear dynamics of the binary spacetime has on matter evolution during merger \citep{Gold_2014a, Kelly2017}. Because the timescale for the merger to occur is shorter or comparable to the dynamic timescale associated with the gas, one might wonder if a valid approximation might be constructed in this strong-field regime as well.}

In this work, we present an approximation for a binary black hole metric that covers the inspiral, merger, and post-merger evolution. Our main goal is to provide an accurate, computationally cheap, and versatile way of performing GRMHD simulations and ray-tracing calculations on a BBH spacetime without the need to evolve Einstein's equations at all separations. The framework we present has several important advantages, as it can be easily ported to most GRMHD codes and yields significantly faster simulations compared to those utilizing numerical relativity, especially at large binary separations and when using large amounts of CPUs. Our approximation is built as a superposition of two Kerr-Schild black holes following the methods in Refs. \citep{combi2021superposed, armengol2021circumbinary} but generalizing this in various ways: a) we extend the approximation for any binary separation, spin, mass-ratio, and eccentricity using trajectories calculated at 4PN order; b) we propose a smooth interpolation from inspiral to merger making the approximation (for all practical purposes) valid through the entire evolution; c) we perform a one-to-one comparison of the metric with numerical relativity evolving GRMHD simulations in the binary spacetime, and d) we make the metric implementation open-source and public. Although we have previously shown that the approach works very well at separations of $r_{12}>20\,M$, here we push the approximation to an extreme case where the initial separation is very small, $r_{12}\sim 10\,M$, and the binary evolves to merger. In Ref.~\cite{ressler2024}, we used this metric approximation to model a {small-mass-ratio} binary black hole {system} evolving on inclined orbits with respect to an accretion disk, including spin precession due to spin-orbit effects.
{We have also applied the metric to the dual jet interactions in an equal-mass, high-spin black hole binary system \citep{Ressler2025a}, as well as a small-mass-ratios binary system an inclined orbit interacting with a thin, radiatively cooled accretion disk \citep{Ressler2025b} (i.e., the canonical model for OJ 287).} 
Here, we perform a series of tests to show the validity of the approximation by performing a direct comparison with numerical relativity, analyzing the spacetime properties of the metric in each approach, as well as the behavior of a fluid evolving on top of this metric.

\rvw{The paper is organized as follows: in Section \ref{sec:build}, we show in detail how we obtain our approximation, how we solve the post-Newtonian trajectories, and how we model the merger in our metric; in Section \ref{sec:numrel} we show the methods we use to compare the approximation with numerical relativity; in Section \ref{sec:results_10} we present the results comparing merging binary black holes surrounded by uniform fluid; in Section \ref{sec:results_20} we present results comparing inspiralling binary black holes at a separation of $20\,M$ surrounded by a magnetized circumbinary disk torus; in Section \ref{sec:results_comp}, we show results on computational performance between the two approaches. In Section \ref{sec:con}, we present our conclusions. }

\textbf{Notation:} we follow the notation and conventions in Ref.~\cite{poisson} and use a $+2$ signature for the metric, $a,b,c,.. = 0,1,2,3$ for spacetime indices, $i,j,k,... = 1,2,3$ for three-dimensional space indices, and $A=1,2$ for distinguishing each black hole. We adopt geometric units where $c=G=1$.

\section{Building a binary black hole metric approximation}
\label{sec:build}

In this section, we construct a semi-analytical approximation for the four-dimensional metric of a binary black hole that can be used throughout the entire evolution, including merger.
We superpose two black holes using Kerr-Schild coordinates and we apply a time-dependent boost to each term using trajectories dictated by the post-Newtonian equations of motion.  We consider general binary configurations, with arbitrary eccentricity and spin direction, including all relevant PN effects up to 4PN order. To model the merger itself, we construct an interpolation of the binary metric between the inspiral phase to the final remnant black hole, whose properties we calculate beforehand using numerical relativity fitting formulae. We describe now how to build this approximation, starting from the Kerr-Schild metric.

\subsection{Kerr-Schild metric with arbitrary spin}

 We start with the Kerr-Schild metric in Cartesian coordinates $\lbrace X^a \rbrace$ with arbitrary spin given by
\begin{equation}
	g_{ab} = \eta_{ab} + 2 H(X^{i}) l_{a} l_{b},
 \label{eq:ksmetric}
\end{equation}
with the null covector given by
\begin{equation}
 l_a dX^a = dt + \frac{1}{r^2 + a^2} \left[ r X^i - \epsilon^{i}_{jk} a^j X^k + \frac{(a^i X^j \delta_{ij})}{r}  a^i \right] \, dX_{i},
\end{equation}
the Boyer-Lindquist radius is defined as
\begin{equation}
	r^2 = \frac{1}{2}(R^2-a^2) + \left[ \frac{1}{4}(R^2-a^2)^2 + (a^i X^j \delta_{ij})^2 \right]^{1/2},
\label{eq:covectorBH}
\end{equation}
the Cartesian radius is $R^2 = X^2 +Y^2 +Z^2$, and the function $H$ is defined as
\begin{equation}
	H(X^i) = \frac{M r^3}{r^4+(a^i X^j \delta_{ij} )^2},
\label{eq:hfun}
\end{equation}
where $\epsilon^{i}_{jk}$ is the Levi-Civita symbol, $\delta_{ij}$ is the Kronecker delta, $a^i$ is the spin vector, and $a= \sqrt{a^i a^j\delta_{ij}}$. 
This is a compact form of writing a rotated Kerr-Schild metric  \cite{ma2021extending}. We will use these coordinates to build a moving superposition of black holes.

\subsection{Boosted superposition of black holes}

In the Kerr-Schild form \eqref{eq:ksmetric}, the metric is represented by a superposition of a background flat metric, $\eta_{ab}$, and a black hole term, $2H(X^i) l^a l^b$. 
This provides a natural way of adding a second black hole to the spacetime.
Because Einstein's equations are highly non-linear, this superposition is, of course, not a vacuum solution; however, it naturally approaches an exact solution as the separation of the black holes increases. We will show in the next section that it is a good approximation for the purposes of investigating the behavior of matter around dynamical spacetime.

In the post-Newtonian sense, black holes in a binary are moving on accelerated orbits.
To include the orbital motion of the BHs in the superposition, we build a time-dependent boost transformation that can be considered as a generalization of the linear Lorentz boost.
A time-dependent boost in general can be defined as a transformation from an inertial frame to an accelerated frame that is moving in a time-like trajectory. We now show how to build this transformation for an arbitrary orbit moving according to the post-Newtonian equations of motion.

Let us associate each moving black hole with an accelerated frame and the center of mass with the global inertial frame. We denote $\lbrace X^b \rbrace$ as the coordinates of the accelerated frame and $\lbrace x^a \rbrace$ as the coordinates of the inertial frame. In the coordinates of the inertial frame, the trajectory of one of the BHs is given by 
$s^a(\tau) = \{t(\tau), \hat{s}^i(\tau)\}$ 
where $\tau$ is the proper time in the BH frame. The trajectories we are using are parameterized in terms of the global time, so we will consider $\hat{s}^a(\tau)=\hat{s}[\tau(t)] \equiv s^a(t_{\tau})$, where $\tau = \int^{t_{\tau}} dt \gamma^{-1}$. We now construct the transformation between these coordinate systems for each BH, i.e., $X^b(x^a)$, and use them to boost the BH terms, similar to an active Lorentz transformation. 

The local coordinates of a moving frame in a general spacetime are the so-called normal or Fermi coordinates \citep{fermi1962, Synge1960, manasse1963fermi, poisson2011motion}, in which the metric is locally flat.  We review how to build these coordinates, generalizing the methods in Refs. \cite{mashhoon2002length,combi2021superposed}. In Fermi coordinates, the time component at the origin of the moving frame is defined as the proper time of the trajectory, $T=\tau$, while the space coordinates of a given point $X^i$ are constructed by finding the unique spacelike geodesics passing through that point and orthogonal to the four-velocity $u^a = ds^a/d\tau$ \citep{poisson2011motion}. In a Minkowski background spacetime, the full transformation can be written as:
\begin{equation}
	x^{a} = s(t_{\tau})^{a} + X^i e^a_{i},
\end{equation}
where $e_{i} = e^a_{i} \partial_{a}$ is an orthonormal vector basis, Fermi-Walker transported by the accelerated frame and adapted to the local coordinates, i.e., the spatial axis of the frame \cite{poisson2011motion}. Because the tetrad field lives on the tangential space of a moving frame, for each time $t$ we can find the components of the basis in global coordinates $\lbrace x^a \rbrace$ by performing a Lorentz boost. This assumes that, for a fixed time, there is an inertial frame that is equivalent \textit{instantaneously} to an accelerated frame \citep{mashhoon2002length}.

At $X^i=0$, the orthonormal frame in Fermi normal coordinates is simply $e^{b}_{i}= \delta^{b}_{i}$ and thus applying a local boost, we have, in the global coordinate system,  $e^{a}_{i}= (\Lambda^{-1})^{a}_{i}$, where $(\Lambda^{-1})^{a}_{i}$ is the inverse of the Lorentz boost matrix, given by:
\begin{align}
\label{eq:lorentz}
& \Lambda({\beta}^i) = \exp{[\xi {n}^i (t) K^j \delta_{ij}]} =  \\
& \begin{pmatrix}
	  \gamma & -\gamma \beta n_{x} & -\gamma \beta n_y & -\gamma \beta n_{z} \\
	  -\gamma \beta n_{x} & 1+ (\gamma-1) n_{x}^2 & (\gamma-1) n_x n_y & (\gamma-1) n_x n_z \\
	  -\gamma \beta n_y & (\gamma-1) n_x n_y &  1+ (\gamma-1) n_{y}^2 & (\gamma-1) n_y n_z \\
	  -\gamma \beta n_z & (\gamma-1) n_x n_z & (\gamma-1) n_y n_z & 1+ (\gamma-1) n_z^2
  \end{pmatrix} \nonumber,
\end{align}
where $\xi= \tanh^{-1}{(\beta)}$ is the rapidity, $n^i =\beta^i/\beta = (n_x, n_y, n_z)$ is the unit direction of the velocity, and ${K}^i$ are the generators of the Lorentz group. Notice that $\Lambda^{-1}({\beta}^i) = \Lambda(-\beta^i)$.

Because for binary black holes $\gamma(t)$ changes slowly in time for most of the evolution, we approximate $t_{\tau} = \gamma \tau \equiv \gamma T$; The transformation from the inertial frame to the accelerated frame is then:
\begin{align}
        t = \gamma \: [T + &\beta( n_x X + n_y Y + n_z Z) ], \nonumber  \\
	x = s^x(t_{\tau}) &+ X\: [1 + n_x^2 (\gamma-1) ] \nonumber \\
	              &+ Y \: (\gamma-1) n_x n_y + Z \: (\gamma-1) n_x n_z,	\nonumber \\
	y = s^y(t_{\tau}) &+ Y \: [1 + n_y^2 (\gamma-1) ] \nonumber \\
	              &+ X \: (\gamma-1) n_x n_y + Z \: (\gamma-1) n_y n_z,	\nonumber 	\\
	z = s^z(t_{\tau}) &+ Z \: [1 + n_z^2 (\gamma-1) ] \nonumber \\
	              &+ Y \: (\gamma-1) n_z n_y + X \: (\gamma-1) n_x n_z.
\label{eq:fermicoord1}
\end{align}

The transformation reduces to a standard linear boost when the acceleration is zero; indeed, for a frame moving at uniform velocity, we have $s^i(t_\tau) = \beta n^i t_{\tau} =  \beta n^i \gamma T$, and the Jacobian of the transformation in Eqs.\ \eqref{eq:fermicoord1} is simply given by $\Lambda^a_{b}$. Let us note two issues with this time-dependent boost. First, the transformation to (and from) the accelerated frame is valid only in a neighborhood of the worldline. The transformation diverges when $ |X| \rightarrow c^2/\beta$; for a linearly accelerating frame, this is known as the Rindler horizon \citep{rindler1956visual}.
Beyond this horizon, the space-like geodesics used to construct the coordinates become time-like. 
A second issue is that to obtain $X^{b}(x^{a})$ from $x^a(X^{b})$ we need to invert Eqs.\ \eqref{eq:fermicoord1}, which is a set of non-linear equations for arbitrary trajectories because $s^i(t_\tau)$ is, in general, non-linear unless the velocity is constant. 

To avoid a non-linear inversion, we can expand the trajectory around the global time as $s^{i}(t_{\tau}) = s^i(t) + \beta^i (t) (t_{\tau}-t)$.
As we show in Appendix A, this approximation is good for a binary black hole and allows us to write the inverse as
\begin{align}
	T &= t \gamma^{-1} -\gamma \beta( n_x \bar{x} + n_y \bar{y} + n_z \bar{z} ), \nonumber  \\
	X &=  \bar{x}\: [1 + n_x^2 (\gamma-1) ]+ \bar{y} \: (\gamma-1) n_x n_y \nonumber \\
                &+ \bar{z}  (\gamma-1) n_x n_z,		\nonumber  \\
	Y &=  \bar{y} \: [1 + n_y^2 (\gamma-1) ]+ \bar{x} \: (\gamma-1) n_x n_y  \nonumber \\
	               &+ \bar{z} \: (\gamma-1) n_y n_z,	\nonumber \\
	Z &=  \bar{z} \: [1 + n_z^2 (\gamma-1) ]+ \bar{y} \: (\gamma-1) n_z n_y  \nonumber \\
	               &+ \bar{x} \: (\gamma-1) n_x n_z,
\label{eq:fermicoord2}
\end{align}
where we define
\begin{equation}
    \bar{x}^i =  x^i - s^i(t).
\end{equation}

To avoid coordinate pathologies at large distances, we also discard the acceleration term when calculating the Jacobian of Eqs.\ \eqref{eq:fermicoord2}. In this way, we obtain that the transformation is simply given by:
\begin{equation}
    \frac{\partial X^{a}}{\partial x^b} = \Lambda^{a}_{b}(\beta) + \mathcal{O}(\dot{\beta}),
\end{equation}
i.e., a simple Lorentz boost as defined before (see Appendix A for more details). The transformation is, in this way, well-defined at infinity. 

The full binary black hole metric, which we name Superposed Kerr-Schild (SKS), is then given by the boosted superposition:
\begin{equation}
	g_{ab} = \eta_{ab}  + [2H \Lambda^{d}_{a}l_{d} \Lambda^{c}_{b}l_{c}]_{(1)} +  
    [2H \Lambda^{c}_{a}  l_{c} 
    \Lambda^{d}_{b} l_{d}]_{(2)}
\label{eq:sks}
\end{equation}
where subscript $A=1,2$ refers to each BH; the null covectors, $l_{c}\,(X^{d})$, and $H(X^{d})$ function in the frame of the BH are given by Eqs.\ \eqref{eq:covectorBH} and \eqref{eq:hfun}, the coordinate transformation to the inertial frame is given by Eq.\ \eqref{eq:fermicoord2} and the Lorentz boost matrix $\Lambda^{d}_{a \,(A)}$ is given by Eq.\ \eqref{eq:lorentz}.

At large separations, the SKS metric tends naturally to an exact solution of Einstein's equations. We notice that even when the velocities are small (e.g., at very large separations), we still need to transform the superposition terms with the Lorentz boosts at first order in $c\beta$, i.e., we need to do the proper Galilean transformation of the metric. Otherwise, this solution does not have the right transformation properties in the Newtonian limit. In other words, even if $(1-\gamma) \sim 0$, there are still terms proportional to $c\beta$ in the transformation, so discarding them amounts to a first-order error in the solution. 


The free parameters of the metric are given by the spin ${a}^i_{(A)}$, the mass, $M_{(A)}$, the spatial trajectories, ${s}^i_{(A)}$, and three-velocities, ${v}^i_{(A)}$, for each black hole. 

\subsection{Black hole trajectories}

Although the metric represents a relativistic, strong-field spacetime, the motion of the black holes is computed using a post-Newtonian approximation{,} where the black holes are treated as point masses on a flat background and Einstein's equations are expanded in terms of $\epsilon \approx v^2 \approx M/r$ to compute corrections to Newtonian theory. Because we are interested in modeling the late stages of the inspiral where velocities reach tens of percent of the speed of light, we use a high-order expansion up to 4PN order.

Given the position of each black hole in the perturbed flat-background spacetime, ${s}^i_{(A)}$, we can eliminate the center of mass in the orbital equation, reducing the problem to an effective one-body equation of motion for ${x}^i = {s}^i_{(1)} - {s}^i_{(2)}$. This is given in Harmonic coordinates by:
\begin{align}
	\frac{d^2 {x}^i}{dt^2} &= {a}^i_{\rm N} + {a}^i_{\rm 1PN} + {a}^i_{\rm SO} + {a}^i_{\rm 2PN} + {a}^i_{\rm SS} + {a}^{i, \rm BT}_{\rm RR} \nonumber \\ 
	                         &+ {a}^i_{\rm PNSO} + {a}^i_{\rm 3PN} + {a}^i_{\rm RR1PN} + {a}^i_{\rm RRSO} \nonumber \\ 
				 & + {a}^i_{\rm RRSS} + {a}^i_{\rm 3.5PNSO} + {a}^i_{\rm 4PN},
\end{align}

where ${a}_{\rm N}$, ${a}_{\rm 1PN}$, ${a}_{\rm 2PN}$, ${a}_{\rm 3PN}$, and ${a}_{\rm 4PN}$ are the contributions to the acceleration due to different PN orders;
${a}_{\rm SO}$, ${a}_{\rm PNSO}$, and ${a}_{\rm 3.5PNSO}$, are the spin-orbit coupling contribution and its 1PN, and 3.5PN order corrections respectively; 
${a}_{\rm SS}$ is the spin-spin coupling term; 
and finally ${a}^{\rm BT}_{\rm RR}$, ${a}_{\rm RR1PN}$, ${a}_{\rm RRSO}$, ${a}_{\rm RRSS}$ are the radiation-reaction contribution, its 1PN correction, and contributions from spin-orbit and spin-spin interactions, respectively. Explicit expressions for all these terms can be found in Ref.~\cite{kidder1995coalescing}, in the Appendix of Ref.~\cite{csizmadia2012gravitational}, and explicitly in the \texttt{CBWaves} source code with corresponding references \footnote{\url{https://github.com/shinsei90/CBwaves}}.

The spin-orbit coupling term, which can induce orbital precession in the system, is not unique and depends on the choice of supplementary spin conditions. We fix these with the covariant expression $S^{ab} u_{a}=0$, where $u^a$ is the four-velocity of each black hole and $S^{ab}$ is the spin tensor \citep{blanchet2014gravitational}. The spin vector is defined as $S^i = (1/2) \epsilon^{i}_{jk} S^{jk}$, related to the spin parameter of Kerr-Schild as $S^i = M a^i$ and the dimensionless spin $S^i = \chi^i M^2$. 
Notice that the spin vector components are defined in the coordinate system associated with the moving BH.

Spins can precess due to spin-spin and spin-orbit coupling as the binary evolves. The equations of motion for the spins of one of the BHs are given by:
\begin{equation}
	\frac{d S^{i} }{dt} = \epsilon^{i}_{jk} {\Omega^{j}}  {S^{k}},
\end{equation}
where $\epsilon^{i}_{jk}$ is the Levi-Civita symbol, and the precession vector ${\Omega^{j}}$ depends on the angular momentum of the binary and {the} spins of both black holes. We use the expression from Equation (3) in Ref.~\cite{kacskovics2022orbital} containing terms up to 3.5PN order (see also references therein for explicit formulae).

We solve the coupled equation of motions for ${x}^i$, ${S}^i_{(1)}$, and ${S}^i_{(2)}$ with a fourth-order Runge-Kutta scheme using a version of the public code \texttt{CBWaves}, a modular C code that solves the PN equations efficiently.
Initial conditions are set by choosing the semi-major axis, $r_{12}$, spin directions, and eccentricity, defined as $e = (r_{\rm max} - r_{\rm min})/(r_{\rm max} + r_{\rm min})$, where $r_{\rm max}$ and $r_{\rm min}$ are the apocenter and pericenter of the orbit, respectively.
Once we obtain ${x}^i(t)$, we compute the individual trajectories ${s}^i_{(A)}$ and then the velocities ${v}^i_{(A)}$. Because our metric construction was formulated in Kerr-Schild coordinates, we must, in principle, transform the trajectories from Harmonic to Kerr-Schild using, e.g., the equations in Appendix A of Ref.~\cite{ma2021extending}. However, we have verified that this transformation usually involves a small offset for the trajectories and barely changes the constraints.

\subsection{Transition to merger}
\label{sec:tranmerger}

When the binary approaches merger, the dynamics of the spacetime near the black hole horizon become highly non-linear; accurate evolution during merger and gravitational wave emission can only be predicted by solving Einstein's equations numerically.
Nevertheless, when the binary separation is small $r_{12} \lesssim 5\,M$, the characteristic timescales from inspiral to merger \citep{Peters:1963ux}, $t_{\rm insp} \sim [r_{12}/(2M)]^4 \, M$, are comparable or smaller than the dynamical timescale of the gas surrounding the binary, $t_{\rm dyn} \gtrsim  r/v_{\rm dyn} \sim (r/M)^{3/2} \, M$, with $r \gtrsim r_{12}$. Moreover, we know that shortly after the merger and ringdown phase, the remnant will settle into a Kerr black hole, with properties that we can estimate well following analytical arguments \cite{buonanno2008estimating}, together with fitting formulae from NR \citep{lehner2014numerical}.
This motivates us to approximate the merger process by interpolating the superposed binary metric to a single black hole metric. We argue that this is a reasonable approximation for the binary spacetime if we only care about the behavior of matter surrounding the merger. 

We interpolate the mass and spin of the black holes between inspiral and post-merger in the following way. During binary evolution, the mass, spin, and velocity of each black hole are defined as:
\begin{equation}
M_{(A)}(t) = M_{0\, (A)} \left[1-\mathcal{W}(t)\right] +  \frac{1}{2} M_{\rm f} \mathcal{W}(t),
\end{equation}
\begin{equation}
{a}^{i}_{(A)}(t) = {a}^i_{0\, (A)} \left[1-\mathcal{W}(t)\right] +  {a}^i_{\rm f} \mathcal{W}(t),
\end{equation}
\begin{equation}
{v}^i_{(A)}(t)= {v}^i_{0\, (A)} \left[1-\mathcal{W}(t)\right] + {v}^i_{\rm f}\mathcal{W}(t),
\end{equation}
where $M_{0\, (A)}$, ${a}^i_{0\, (A)}$, ${v}^i_{0\, (A)}$ are the mass, spin, and velocity during inspiral and $M_{\rm f}$, ${a}^i_{\rm f}= \chi^i_{\rm f} M_{\rm f}$, ${v}^i_{\rm f}$ are the corresponding properties of the remnant black hole after merger. We use an interpolating function, $\mathcal{W}(t)$, that is $0$ for $t < t_{\rm merger} - dt_{\rm buffer}$, $1$ for $t > t_{\rm merger} + dt_{\rm buffer}$, and smoothly interpolates from 0 to 1 in between (see Appendix B).  After $t>t_{\rm merger} + dt_{\rm buffer}$, the interpolation of BH masses and spin brings each superposed term to a single boosted Kerr black hole. 

The remnant properties are computed using fitting formulas from numerical relativity. Following Ref.~\cite{gerosa2016precession}, we use Eq.\ (18) from Ref.~\cite{barausse2012mass} to compute the final remnant mass as
\begin{equation}
M_{\rm f}= \left( M_{0\,(1)}+M_{0\,(2)} \right) \left(1-E_{\rm rad}\right),
\end{equation}
where $E_{\rm rad}$ is the radiated energy. To compute ${a}^i_{\rm f}$ we calculate the dimensionless spin $|{\chi}_{\rm f}|={a}_{\rm f}/M_{\rm f}$ and the direction of the spin with respect to the total angular momentum using Eqs.\ (12) and (19) from Ref.~\cite{hofmann2016final}. 
To compute the BH kick velocity ${v}^i_{\rm f}$, we use the compiled methods in Ref.~\cite{gerosa2016precession}, which in turn, make use of formulas in Ref.~\cite{kesden2010final}.

The BH remnant properties depend on the mass-ratio $q=M_{(1)}/M_{(2)}$, the dimensionless spins $\chi^i_{(A)}$, the angle between spins and orbital angular momentum, $\cos{\theta_{(A)}} = \hat{S}^{(A)} \cdot \hat{L}$,  and the azimuthal angle between the projections of spins onto the orbital plane, defined as $\cos{\Delta \Phi}= \Big(( \hat{S}_{(1)} \times \hat{L} )/| \hat{S}_{(1)} \times \hat{L}| \Big) \cdot \Big (( \hat{S}_{(2)} \times \hat{L} )/| \hat{S}_{(2)} \times \hat{L}| \Big)$. Here, ${L}$ is the orbital angular momentum and hats indicate unit vectors. These quantities are calculated at $t_{\rm merger} - dt_{\rm buffer}$.

\begin{figure*}[ht!]
    \centering
    \includegraphics[width=0.9\columnwidth]{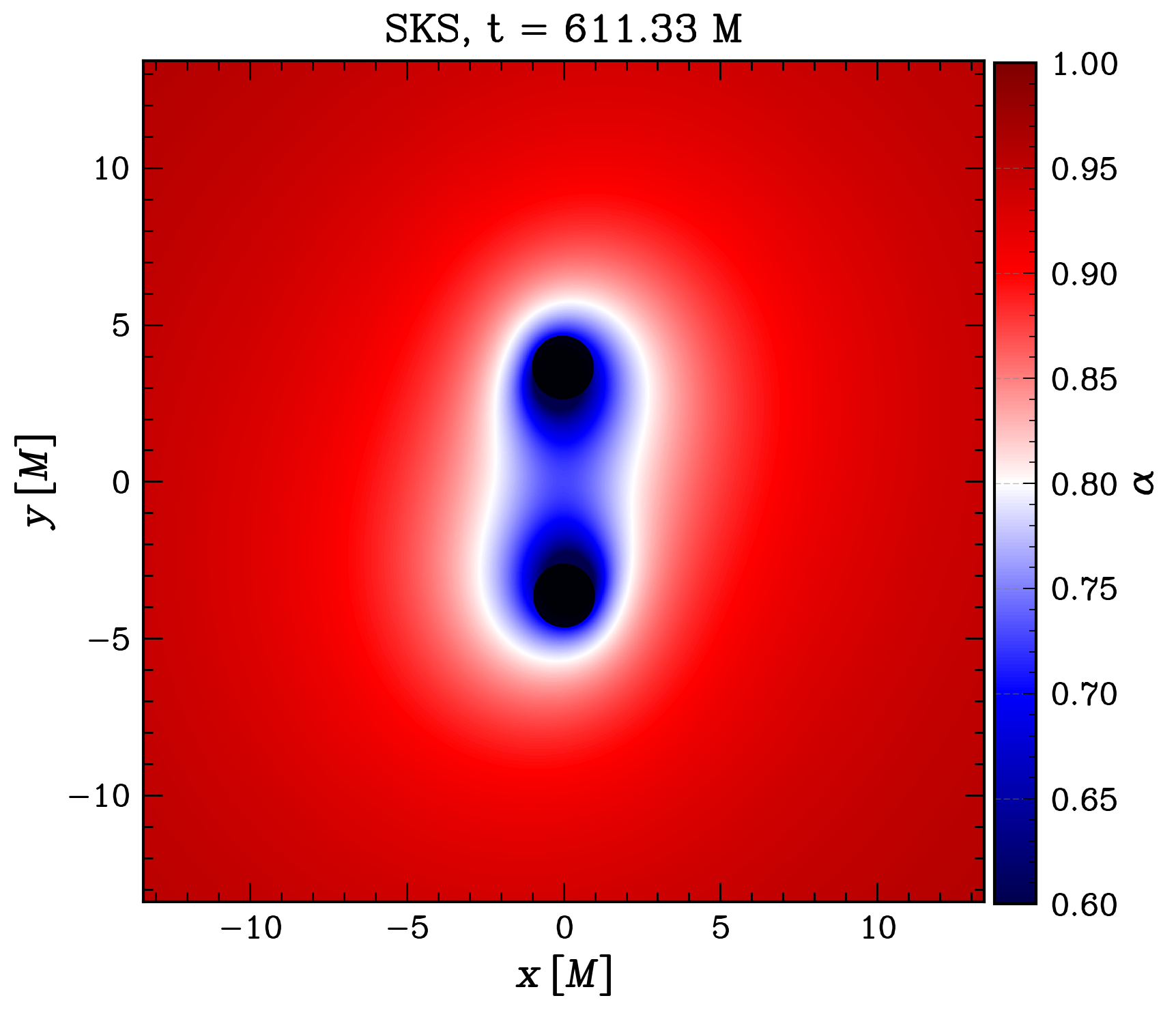}
    \includegraphics[width=0.9\columnwidth]{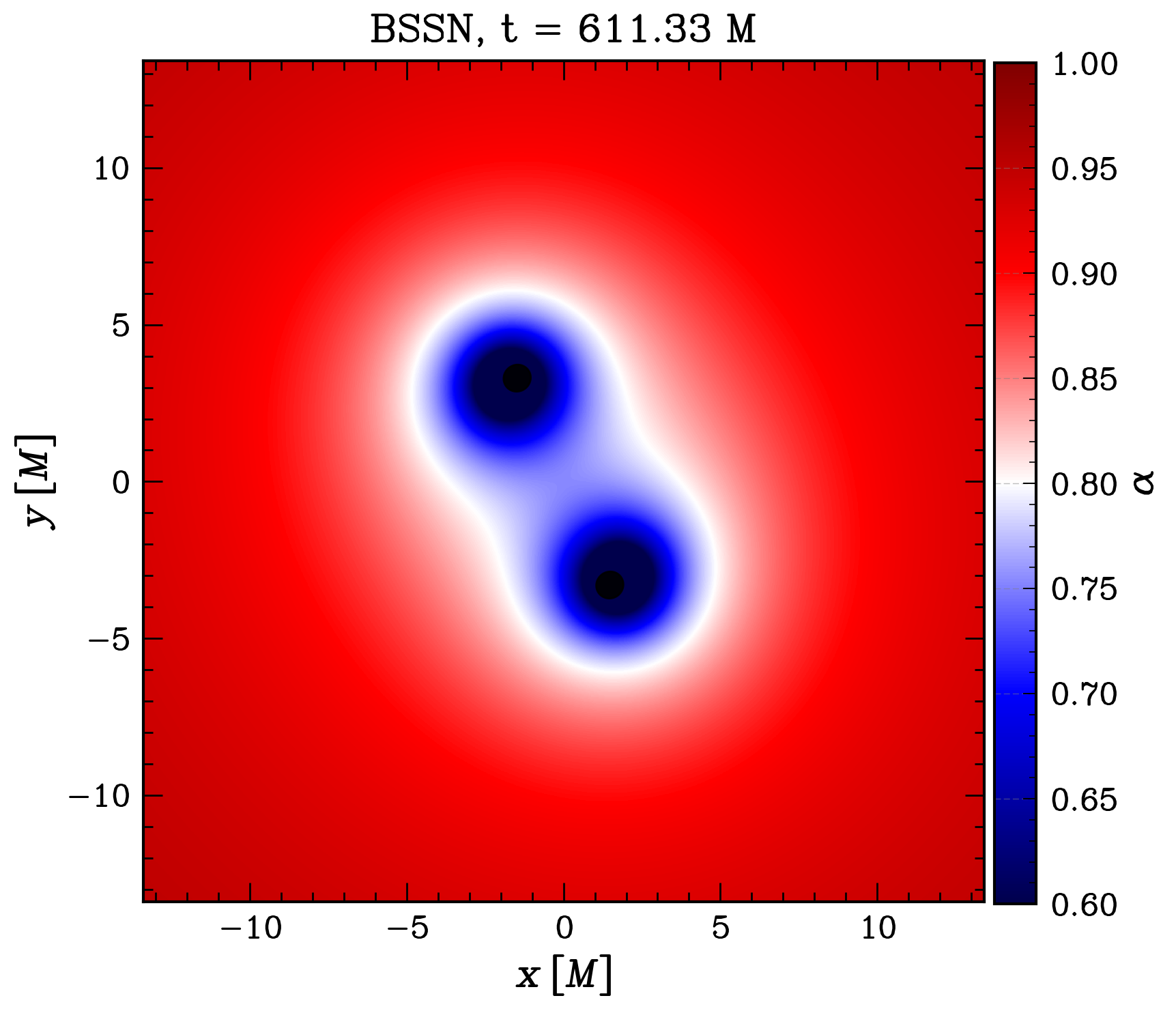}
    \caption{Lapse function in the equatorial plane for \texttt{Run-SKS} (left) and \texttt{Run-BSSN} (right). The SKS metric shows the lapse deformed by the boost while the puncture gauge maintains its symmetry.}
    \label{fig:eqalp}
\end{figure*}

\section{Comparing numerical relativity and the SKS approximation}
\label{sec:numrel}

\rvw{In the next sections, we present a series of tests for the SKS metric, analyzing its spacetime properties, such as metric components, Hamiltonian constraints, and apparent horizons. We also analyze the behavior of matter around the black holes by performing GRMHD simulations on this background spacetime.  We compare these results with a full numerical evolution of Einstein's equations using a $3+1$ decomposition in the Baumgarte-Shapiro-Shibata-Nakamura (BSSN) formalism \cite{nakamura1987General, shibata1995Evolution, baumgarte1999Numerical} together with the puncture gauge evolution for the coordinates. We carry on this comparison in two regimes, considering first (a) merging binary black holes with initial separation of $r_{12}= 10\,M$ surrounded by uniform gas, and (b) inspiralling binary black holes with initial separation of $r_{12}= 20\,M$ surrounded by a magnetized circumbinary disk. The first case is designed to test the robustness and stability of matter evolution onto the approximated spacetime as the two black holes merge, see Sec.~\ref{sec:results_10}; the second case, evolved for much longer ($\gtrsim 10000 \,M$), is designed to compare the numerical and approximated metric in a situation where magnetic fields and turbulence are involved, see Sec.~\ref{sec:results_20}.}

\subsection{Numerical implementation of the SKS metric}

The full expression of the metric given in Eq.\ \eqref{eq:sks} is written first in symbolic form using a {\tt Mathematica} notebook. This allows us to perform controlled checks and avoid typographical errors in writing large expressions. 
The covariant metric, $g_{ab}$, as a function returning a 4x4 matrix, is then exported to C language in an optimized way as a function of coordinates and trajectories; in particular, we optimize algebraic operations before exporting to avoid multiple evaluations of expressions when possible.
The black hole trajectories are calculated using a modified version of the public software {\texttt{CBwave}}, which is written in C and executed via a Python wrapper. The transition from inspiral to merger is calculated after this step in post-processing.
The trajectories are then exported to an HDF5 table containing 1D arrays of each property as a function of time. We use linear interpolation to obtain the orbital parameters at any given time.

The implementation of the metric is now available in a public repository \citep{softcombi} which includes {\tt Mathematica} notebooks, the PN solver (our version of {\texttt CBWave}), and explicit C functions to use for numerical computations. The metric has been implemented, tested, and used in production in the public GRMHD codes \texttt{Athena++} and within the framework of the \texttt{EinsteinToolkit} using either \texttt{GRHydro} and \texttt{IllinoisGRMHD} codes. These implementations will become publicly available soon.

\subsection{Numerical implementation of Einstein's equations}

In the 3+1 decomposition of spacetime \citep{lichnerowicz1944espaces, arnowitt2008Dynamics}, the metric is written as:
\begin{equation}
	g_{ab} dx^a dx^b = -\alpha^2 dt^2 + \gamma_{ij} (dx^i + \beta^i dt) (dx^j + \beta^j dt),
\label{eq:3p1}
\end{equation}
where $\alpha$ is the lapse, $\beta^i$ is the shift, and $\gamma_{ij}$ is the 3-dimensional metric. With these variables, Einstein's equations can be decomposed into evolution and constraint equations \citep{baumgarte2010Numerical}. The constraint equations are given by the Hamiltonian constraint
\begin{equation}
\label{eq:Hamiltonian}
	\mathcal{H}: = R - K_{ij} K^{ij} + K^2 - 16 \pi n_{a}n_{b} T^{ab} =0, 
\end{equation}
and the momentum constraints
\begin{equation}
	\mathcal{M}_i: = D_j K^j_i -D_i K + \gamma^j_i n^a  T_{ja} =0, 
\end{equation}
where $T_{ab}$ is the energy-momentum tensor, $R$ is the three-dimensional Ricci scalar, $K_{ij}$ is the extrinsic curvature, $K$ its trace, $n_a = \delta^{0}_{a} \alpha$ is the unit vector normal to the spatial slice, and $D_{i}$ is the covariant derivative associated with the 3-dimensional metric. In the BSSN formalism, to achieve stable evolution, we split the extrinsic curvature into a trace-free part ${A}_{ij}$ and its trace $K$; then, we conformally rescale the extrinsic curvature and the spatial metric with a factor $\phi$, and finally, we introduce the spatial projected Christoffel symbol $\Gamma^i$. The evolution equations in BSSN form can be found, e.g., in Ref.~\cite{baumgarte2010Numerical}. To fix the gauge (coordinate) freedom given by the variables $\alpha$ and $\beta^i$, we use the puncture evolution, consisting of the Bonna-Mass\'o equations for $\alpha$ \citep{bona1995New}  and the gamma-driver shift for $\beta^i$ \citep{alcubierre2003Gauge}, with an extra evolution variable, $B^i$. In particular, we use the $1+\log$ variant of these equations and we choose $\eta=2/M$ in the constant for the gamma-driver equations. This choice is particularly useful for dealing with black holes without the need to excise the singularity. 


For evolving these equations, we use the {\tt EinsteinToolkit} framework \citep{goodale2003Cactus, loffler2012Einstein, babiuc2019einstein}, which makes use of the {\tt Carpet} code \citep{schnetter2004Evolutions} for handling adaptive mesh-refinement grids with sub-cycling in time. To analyze various aspects of the spacetime and matter fields in our simulation, we make use of the \textit{thorns} {\tt AHFinderDirect}\citep{thornburg2004Fast}, {\tt QuasiLocalMeasures}, and {\tt Outflow}, which are all part of the Toolkit. To evolve the gas on top of the dynamically curved spacetime, we use our version of {\tt GRHydro}, which evolves the ideal MHD equations \citep{mosta2013GRHydro, combi2023grmhd, baiotti2005Threedimensional}. To evolve Einstein's equations, we use the BSSN solver in the {\tt McLahan} code \citep{Brown:2008sb} and we use the {\tt TwoPuncture} \citep{Ansorg:2004ds} code to set up initial data. To use the analytical metric within the Toolkit, we wrote a new thorn, {\tt AnalyticalSpacetime}, which set the metric components in {\tt ADMBase}.

\section{Merging black holes on a uniform unmagnetized gas}\label{sec:results_10}

\rvw{We evolve an equal-mass non-spinning binary on a quasi-circular orbit starting at an initial separation of $r_{12} = 10\,M$ surrounded by a uniform gas with {very} low magnetization. We denote as \texttt{Run-BSSN} the simulation using full numerical relativity coupled to GRMHD and as \texttt{Run-SKS} the simulation evolving the GRMHD equations with the SKS metric as a background spacetime.}

\subsection{Initial data}

For \texttt{Run-BSSN} we use puncture initial data to set an equal-mass binary at a coordinate separation of $10\,M$. This method solves the constraint equations using the so-called conformal-transverse-traceless decomposition \citep{brandt1997simple,gourgoulhon2007construction} where we choose a conformally flat metric $\gamma_{ij}= \psi^4 \delta_{ij}$ with $\psi = 1 + 2 m_{(1)}/r_1 + 2m_{(2)}/r_2 +u$; here, $m_{(A)}$ is the puncture parameter (or bare mass) of the A-th black hole and $u$ is a function that must be solved through the constraint equations. The extrinsic curvature is chosen to incorporate the initial momentum of the BH. We use $m_{(1)} = m_{(2)} = 0.48595$ and an initial momentum in the $y$ direction of $P_{(1)\,y} = 0.095433\,M$ and $P_{(2)\,y} = -0.095433\,M$ for each BH, which sets the binary on quasi-circular orbit, yielding a total ADM mass of $M \sim 1$.

For \texttt{Run-SKS}, we first solve the PN equations of two equal-mass black holes on a quasi-circular orbit ({set} via eccentricity reduction), choosing a coordinate separation of $10\,M$, and a total mass of $M=1$. We transition from inspiral to merger when the separation of the point masses is $r_{12}\sim3\,M$; we interpolate the spin, velocity, and mass of both black holes to the remnant properties given in this case by $M_{\rm f} = 0.95$, zero kick velocity, and spin given by $\chi= 0.66$, using a buffer time of $dt_{\rm buffer}= 0.01\:M$ (see Section \ref{sec:tranmerger}).

For the fluid part, we set up a static uniform ideal gas of $\rho_0 = 10^{-2}\,M^{-2}$   all over the domain, assuming an adiabatic equation of state (EOS) with an adiabatic index $\Gamma =4/3$ to calculate the pressure and internal energy. We note that the total density is decoupled from {the} spacetime, so the choice of $\rho_0$ is arbitrary.  For this test, we set the magnetic field to a very low value, but we effectively solve the full system of MHD equations.

\subsubsection{Simulation set-up}

For \texttt{Run-BSSN} we evolve Einstein's equations using the BSSN formulation and puncture gauge as explained in Sec.~\ref{sec:numrel}. We neglect the energy-momentum tensor of matter, assuming that the gas does not affect the spacetime. For \texttt{Run-SKS}, we implement the metric in the framework of {the} {\tt EinsteinToolkit}{,} first decomposing the 4D metric \eqref{eq:sks} in 3+1 form as in Eq.\ \eqref{eq:3p1}. We select the natural slicing adapted to the coordinates given by $\alpha^2=-1/g^{tt}$, $\beta_i = g_{ti}$, and $ \gamma_{ij}=g_{ij}$. We also need to compute the extrinsic curvature, $K_{ab}$, for finding the apparent horizon and its quasilocal properties; we compute derivatives using a finite-difference stencil of order two for the spatial and time derivatives.

With the dynamical spacetime as a background, we evolve the finite-volume GRMHD equations in the Valencia formulation \citep{font2002Threedimensional} using our version of the public code \texttt{GRHydro}. We solve the Riemann problem using an HLLE solver \citep{harten1987Uniformly}, fifth-order WENO-Z for reconstruction \citep{tchekhovskoy2007wham}, and a scheme for transforming conservative to primitive variables following Ref.~\citep{siegel2018Recovery}. We use an ideal EOS with an adiabatic index of $\Gamma =4/3$. The magnetic field is evolved using a vector potential in the generalized Lorenz gauge \citep{farris2012binary} with an upwind constraint transport scheme \cite{del2007echo}. 

The grid structure consists of two moving refinement hierarchies with 7 (6) levels each following the black holes, with a domain $-260\,M<x,y,z<260\,M$ for \texttt{Run-BSSN} (\texttt{Run-SKS}). The finer resolution for \texttt{Run-BSSN} is $dx=2/2^6 \, M=M/32 $ and for \texttt{Run-SKS} is $dx=2/2^5 \: M=M/16$. In both cases, the horizon is covered by 16 points, which is on the lower end of what is recommended for having a stable evolution. For \texttt{Run-BSSN} we evolve in time using a method of lines and a fourth-order Runge-Kutta with a Courant factor of $0.3$. For \texttt{Run-SKS}, because the spacetime is fixed and the convergence properties of the plasma are less strict than the spacetime evolution \citep{white2016extension}, we use a lower-order time integrator given by a second-order Runge-Kutta with a Courant factor of $0.3$. Since the velocity of the BHs is $v\sim 0.1\,c-0.2\,c$, the metric changes on a cell crossing time given by $dt \sim 10\,dx/c$ where $dx$ is the size of the smallest cell in the simulation grid. The Courant condition for the plasma, however, requires a much shorter time-step of the order of $dt \sim 0.3\,dx/c$ {(set by the Courant number times the cell crossing time for speeds $\sim$ $c$)}. In this way, we can safely update the metric every $\sim 10$ time-steps, saving additional computational time; see Sec.~\ref{sec:comper} {and Appendix A of Ref.\ \citep{ressler2024}}.


\begin{figure*}[ht!]
    \centering
    \includegraphics[width=1\columnwidth]{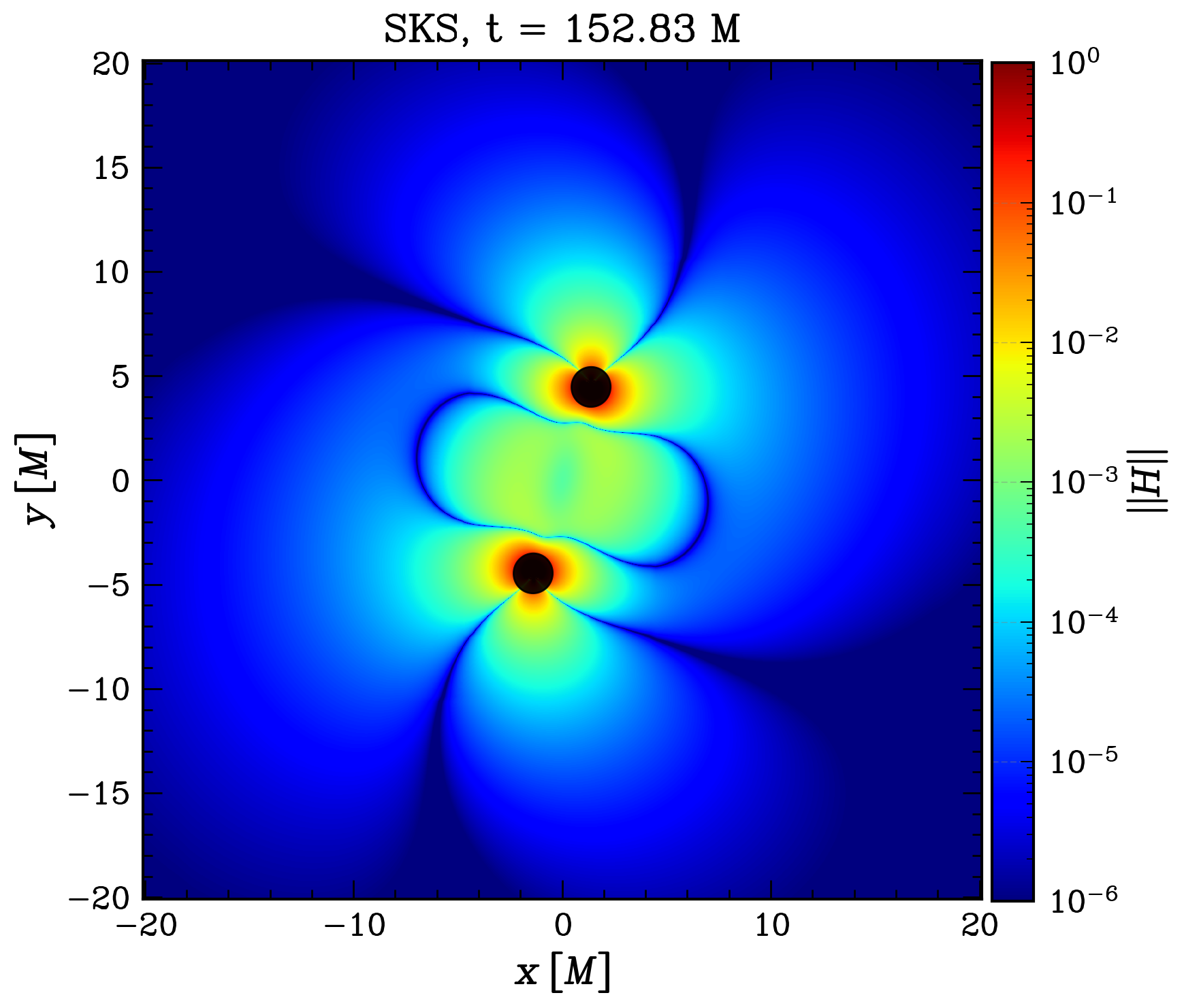}
    \includegraphics[width=1\columnwidth]{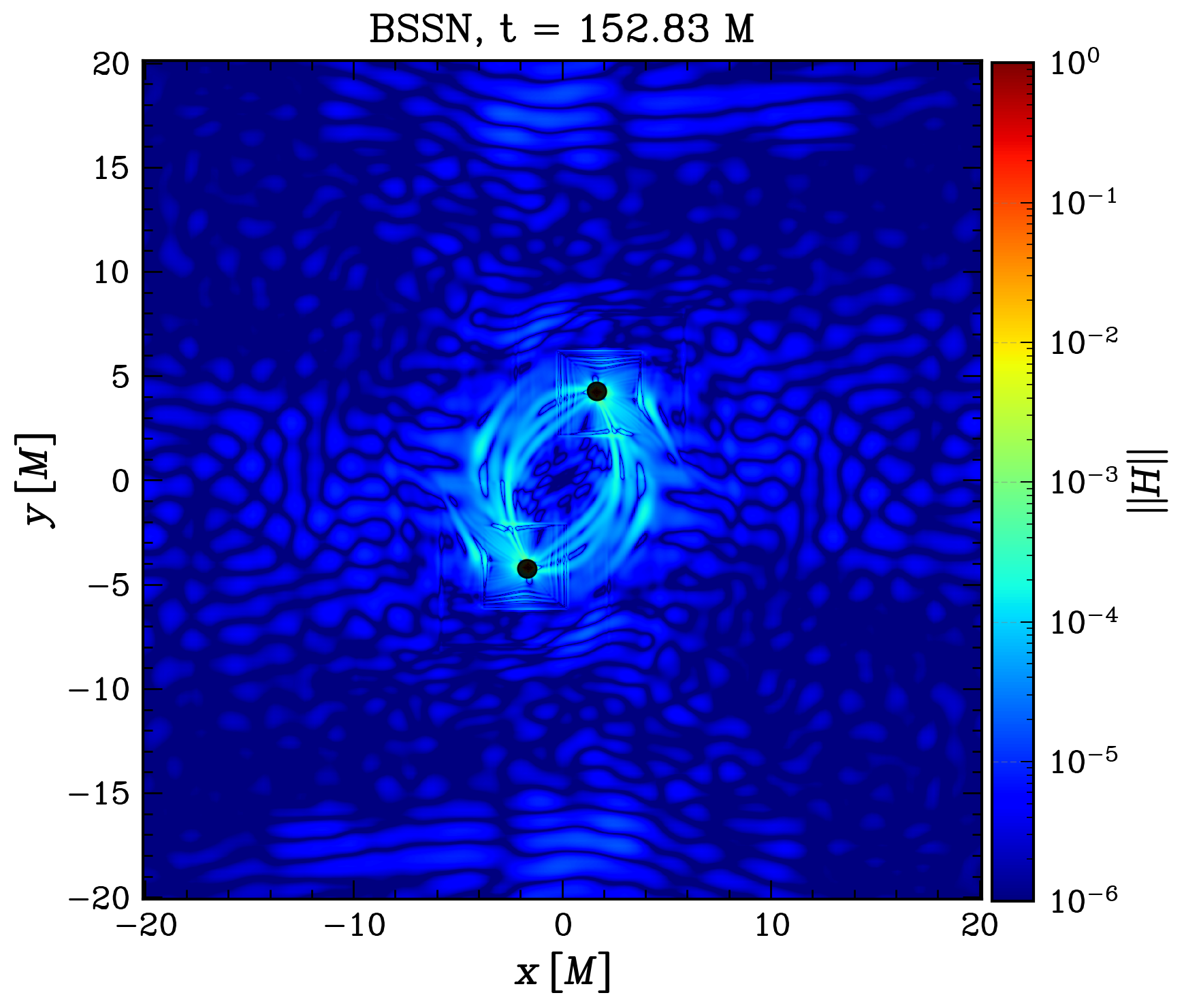}
    \caption{Absolute value of the constraint violations $\mathcal{H}$ in the equatorial plane for \texttt{Run-SKS} (left) and \texttt{Run-BSSN} (right) }
    \label{fig:eqham}
\end{figure*}

\begin{figure}[ht!]
    \centering
    \includegraphics[width=\columnwidth]{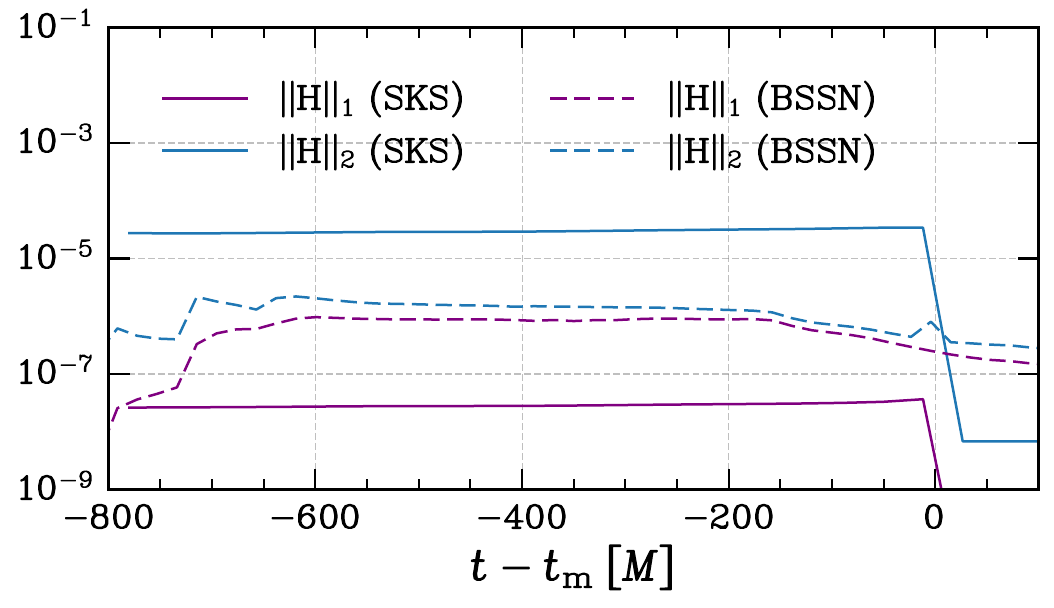}
    \caption{Evolution of Hamiltonian constraint for the superposed metric (SKS) and the numerically evolved metric (BSSN).}
    \label{fig:hamvst}
\end{figure}

\subsubsection{Spacetime properties}

\begin{figure}[ht!]
    \centering
    \includegraphics[width=\columnwidth]{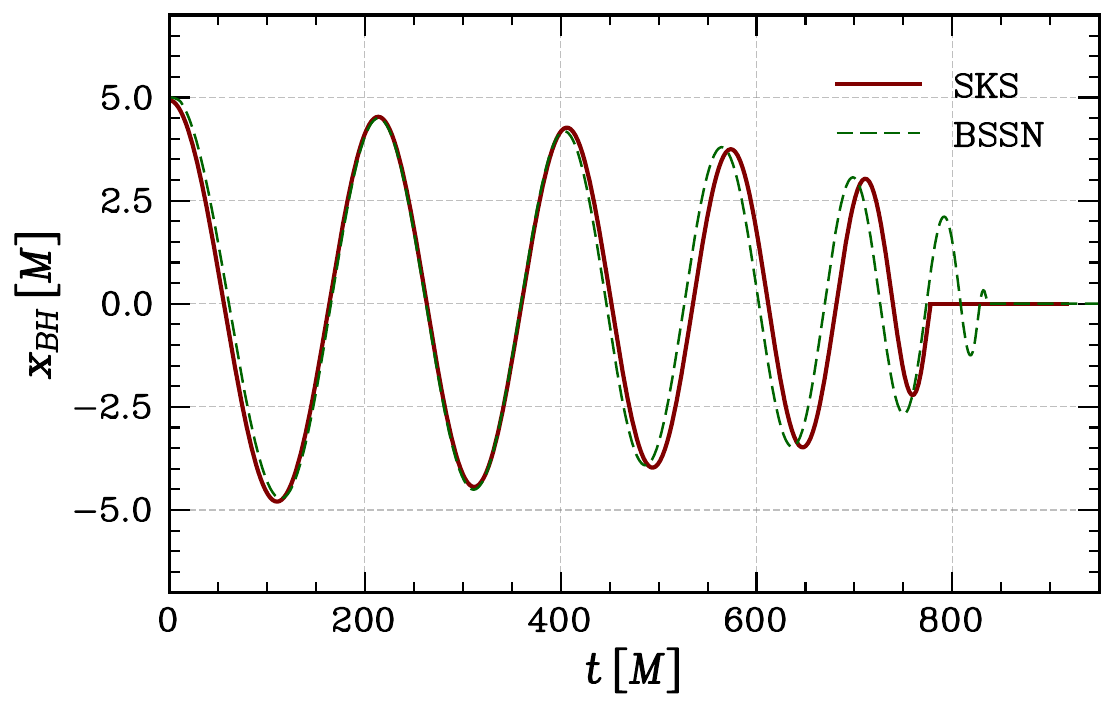} 
    \caption{Evolution of the $x$ component of a BH trajectory for \texttt{Run-SKS} and \texttt{Run-BSSN}. The SKS run starts dephasing on the last orbits as expected, given that the BH trajectories are obtained from a PN evolution.}
    \label{fig:traj}
\end{figure}

\begin{figure*}[ht!]
    \centering
    \includegraphics[width=0.9\columnwidth]{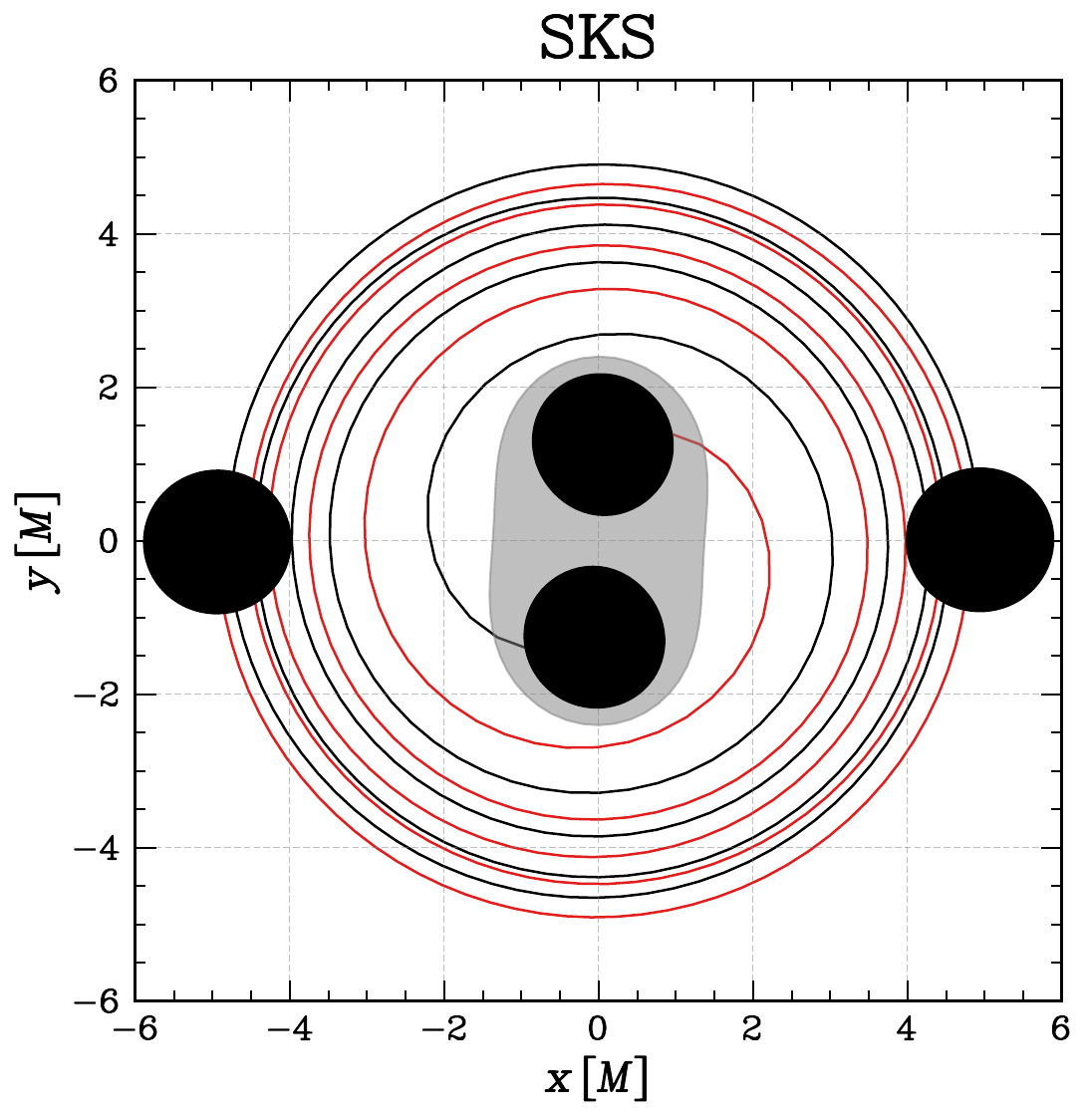}
    \includegraphics[width=0.9\columnwidth]{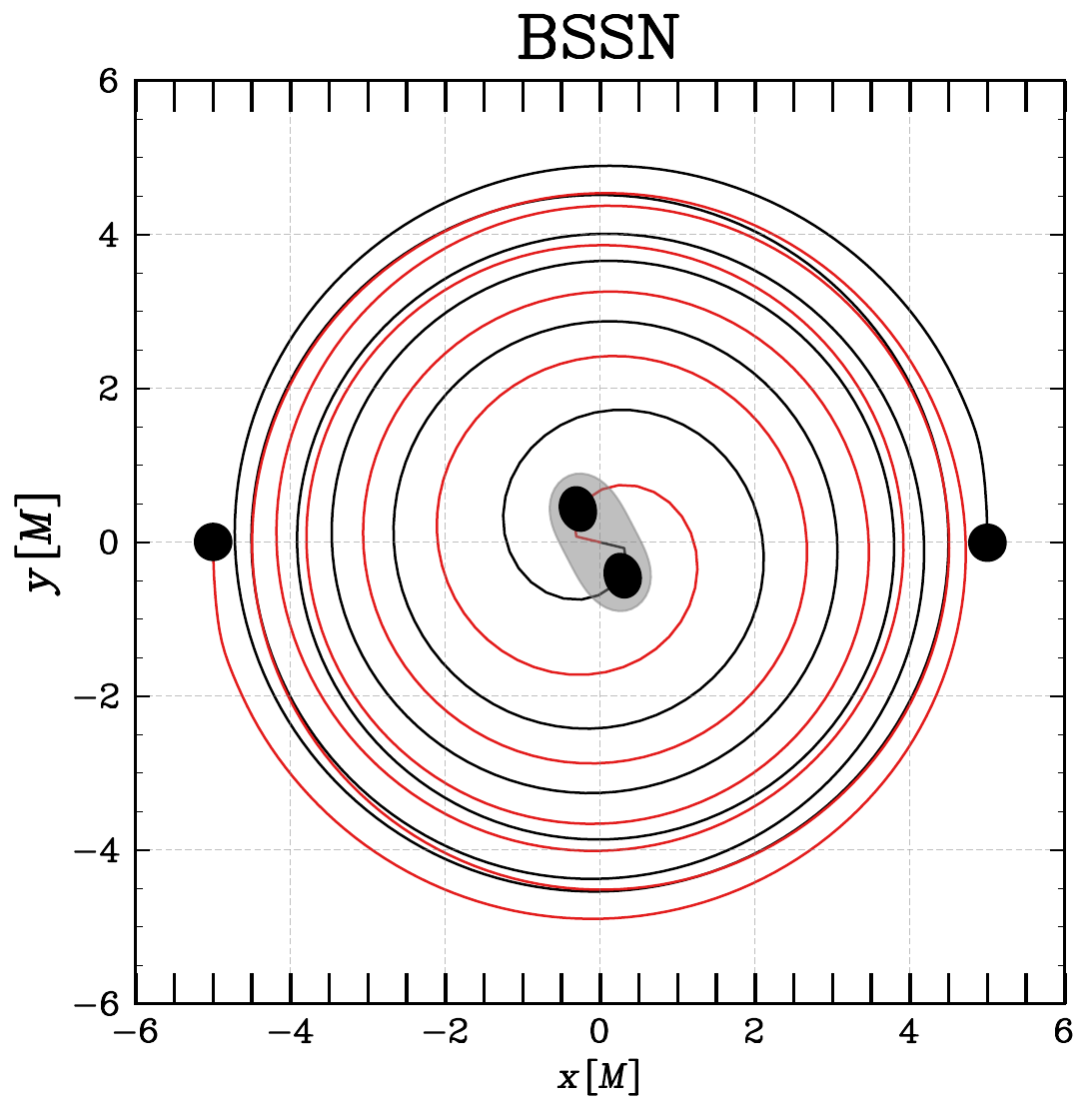}
    \caption{Apparent horizons (black surfaces) and their trajectory at the beginning of the simulation and at merger, where a final common horizon is formed (grey surface) for \texttt{Run-SKS} (left) and \texttt{Run-BSSN} (right).}
    \label{fig:eqah}
\end{figure*}

Because the gauge evolves in a full numerical relativistic evolution, it is hard to make a one-to-one comparison with our SKS metric (see Ref.~\citep{sadiq2018}). Moreover, the puncture gauge used in the BSSN evolution is initially different from the Kerr-Schild superposition, even at large separations. For a single non-spinning BH in puncture coordinates, the radial coordinate of the horizon is $r \sim 0.75\, M$ \citep{baumgarte2007analytical}. In Kerr-Schild coordinates, the radial coordinate of the horizon is instead $r = 2\,M$. While both coordinate systems agree far away from the BH, the horizon can be more easily resolved in Kerr-Schild coordinates; this is especially useful for avoiding numerical artifacts and the necessity of using more resolution. Nevertheless, this coordinate difference, noticeable for instance in the lapse in Fig.~\ref{fig:eqalp} {(plotted for both simulations at $t\sim600\,M$)}, could be problematic for comparing diagnostics. As we will show, these differences amount to a fixed constant offset in some quantities.  

\subsubsection{Hamiltonian constraints and metric}

In solving Einstein's equations, different sources of numerical error introduce violations to the constraints of Einstein's equations, which can spoil the solution if they are not controlled \citep{alic2013Constraint}. For an exact solution, the values of the constraint calculated numerically depend on truncation errors due to the resolution. {In contrast,} because of the approximated nature of our SKS metric, the constraints will not converge to zero with resolution. A local measure of $\mathcal{H}$ {(Eq.~\ref{eq:Hamiltonian})} indicates the relative quality of the approximation for different points in space and time.
Violations of the constraints, $\mathcal{H} \neq 0$, can be interpreted as deviations from a vacuum solution of GR by the presence of a spurious rest-mass density.
To quantify the global quality of the metric, we use the global 1-norm, $||\mathcal{H}||_1$, and 2-norm, $||\mathcal{H}||_2$, of the Hamiltonian constraint. The 1-norm of the constraints can be interpreted as the `fake' mass introduced by the approximation. These quantities should be negligible compared to the total mass of the spacetime. One can also compare these numbers with the constraint violations for an exact solution at the same resolution; in our case, the norm of the constraint for the exact Kerr metric is given by $||\mathcal{H}||_{2} \sim 10^{-8}$) at these resolutions.

In Fig.~\ref{fig:eqham} we show an equatorial snapshot of the Hamiltonian constraints {at a particular time} for \texttt{Run-SKS} and \texttt{Run-BSSN}. 
For \texttt{Run-SKS}, the highest values of the constraints violations $(\sim 10^{-1}-10^{-2})$ are located near the black hole horizons, as well as the region between them ($\sim 10^{-3}$). In this region, the values of the constraint increase as the BHs approach each other. The bulk of the constraint violations is restricted to the orbital region and decreases rapidly outwards as $\propto r^{-4}$. For \texttt{Run-BSSN}, the constraint violations are also concentrated in the orbital region but are more diffuse, and the highest constraint violations are $\sim 10^{-3}-10^{-4}$, lower in magnitude than those in \texttt{Run-SKS}.

In Fig.~\ref{fig:hamvst}, we compare the 1-norm and 2-norm of the {Hamiltonian constraints for} \texttt{Run-SKS} and \texttt{Run-BSSN}.
Both $||\mathcal{H}||_1$ and $||\mathcal{H}||_{2}$ remain almost constant for \texttt{Run-SKS} during inspiral up to the merger, where it drops orders of magnitude as the metric transitions to the exact solution of the remnant black hole. This final value is determined by resolution even for the exact Kerr metric we use in the post-merger SKS, because we use a finite difference method to calculate it. For \texttt{Run-BSSN}, $||\mathcal{H}||_1$ and $||\mathcal{H}||_{2}$ increase a few orders of magnitude with respect to the initial data settling down after $\sim 200\,M$. 
We find that $||\mathcal{H}||_2$ for the SKS metric is higher than the BSSN metric by one order of magnitude, showing that locally, on average, constraints are lower for the BSSN metric. \rvw{The total amount of constraints, quantified by $||\mathcal{H}||_1$, is higher for the BSSN metric because numerical violations propagate through the entire domain, but this does not imply that the numerical evolution is a worse solution.}

\subsubsection{Apparent horizon formation and quasi-local properties}

\begin{figure}[ht!]
    \centering
    \includegraphics[width=\columnwidth]{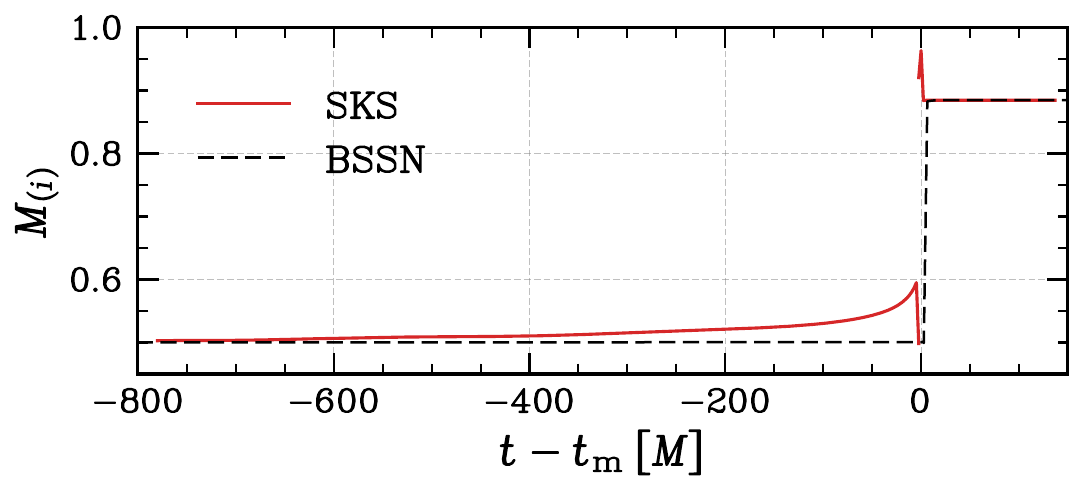} 
    \caption{Evolution of irreducible mass for one black hole and the remnant black hole after merger of \texttt{Run-SKS} and \texttt{Run-BSSN}.}
    \label{fig:mass}
\end{figure}

The SKS metric approximation is a strong field representation of the BH, e.g. different from a PN metric approximation \citep{zilhao2014pn}. 
To characterize the BHs and their properties (quasi-)locally, we resort to the concept of apparent horizon.
An apparent horizon is defined as the outermost marginally trapped surface in a spatial slice; a marginally trapped surface is a closed 2-surface with future-pointing outgoing null geodesics having zero expansion:
\begin{equation}
	\Theta =D_{a} R^{a} + K_{ab} R^a R^b -K =0.
\label{eq:ah}
\end{equation}

We use the methods developed in Ref.\cite{thornburg2004Fast} and implemented in the {\tt AHFinderDirect} code to find and characterize the horizons in \texttt{Run-SKS} and \texttt{Run-BSSN}.  In Fig.\ \ref{fig:traj} we plot the (coordinate) separation of the binary measured from the centroid of the apparent horizons as a function of time for both \texttt{Run-SKS} and \texttt{Run-BSSN}. For \texttt{Run-SKS}, the apparent horizons follow naturally the PN trajectories we use for the boost. Besides the small eccentricity in the initial data, we see a very good agreement with \texttt{Run-BSSN} all the way up to the merger. They differ only in the last orbit, where \texttt{Run-BSSN} has one more orbital cycle; this is expected since the size of the horizon is larger for \texttt{Run-SKS} and the PN approximation is less accurate. 

Because of the gauge differences that we pointed out before, the coordinate radius of the horizon in\texttt{Run-SKS} and \texttt{Run-BSSN} are different during evolution, as can be seen in the equatorial plane in Fig.~\ref{fig:eqah}, where \texttt{Run-SKS} horizons are bigger. The shape of the horizons found self-consistently through Eq.\ \eqref{eq:ah}, minimally changes during the inspiral. Remarkably, in \texttt{Run-SKS} we found a common apparent horizon once the separation of the black holes is sufficiently small ($r_{12} \sim 3\,M$) just before the interpolation to merger. The emergence of a third common apparent horizon is an expected feature of a binary merger \cite{hawking2023large, pook2021happens} and is also found in our \texttt{Run-BSSN} simulation. During the transition to the final remnant BH in \texttt{Run-SKS}, this common horizon evolves smoothly to the horizon of a Kerr-Schild black hole, as we interpolate the mass and spin.

As the separation of the binary gets smaller and smaller, the SKS approximation becomes less accurate. The expected properties of the binary merger are, nevertheless, reproduced by the approximation, i.e., two BH horizons evolve up to merger and form a common apparent horizon which relaxes to a final remnant Kerr BH. The quasi-local properties of the horizon are also well-behaved during evolution. To see this, we plot in Fig.~\ref{fig:mass} the irreducible mass of one of the BHs for \texttt{Run-SKS} and \texttt{Run-BSSN}, defined simply as the $M_{(i)} = R/2$ where $R$ is the radius of the apparent horizon. For \texttt{Run-BSSN}, the mass of one black hole is constant to $\sim 0.5\,M$ during the entire simulation. In \texttt{Run-SKS}, the mass increases as the approximation worsens but only on $10-15\%$ level before merger. At merger, the mass settles to a final value of $M_{\rm f} = 0.8844$ for \texttt{Run-SKS} (calculated from numerical relativity fitting formulae) and $M_{\rm f} = 0.8847$ for \texttt{Run-BSSN}. The final spin measured at the horizon \citep{Dreyer:2002mx,schnetter2006introduction} $\chi_{\rm f} = 0.61$ for \texttt{Run-SKS} and $\chi_{\rm f} = 0.68$ for \texttt{Run-BSSN}.

\subsubsection{Uniform plasma evolution onto binary black holes}

\begin{figure*}[ht!]
    \centering
    \includegraphics[width=0.9\columnwidth]{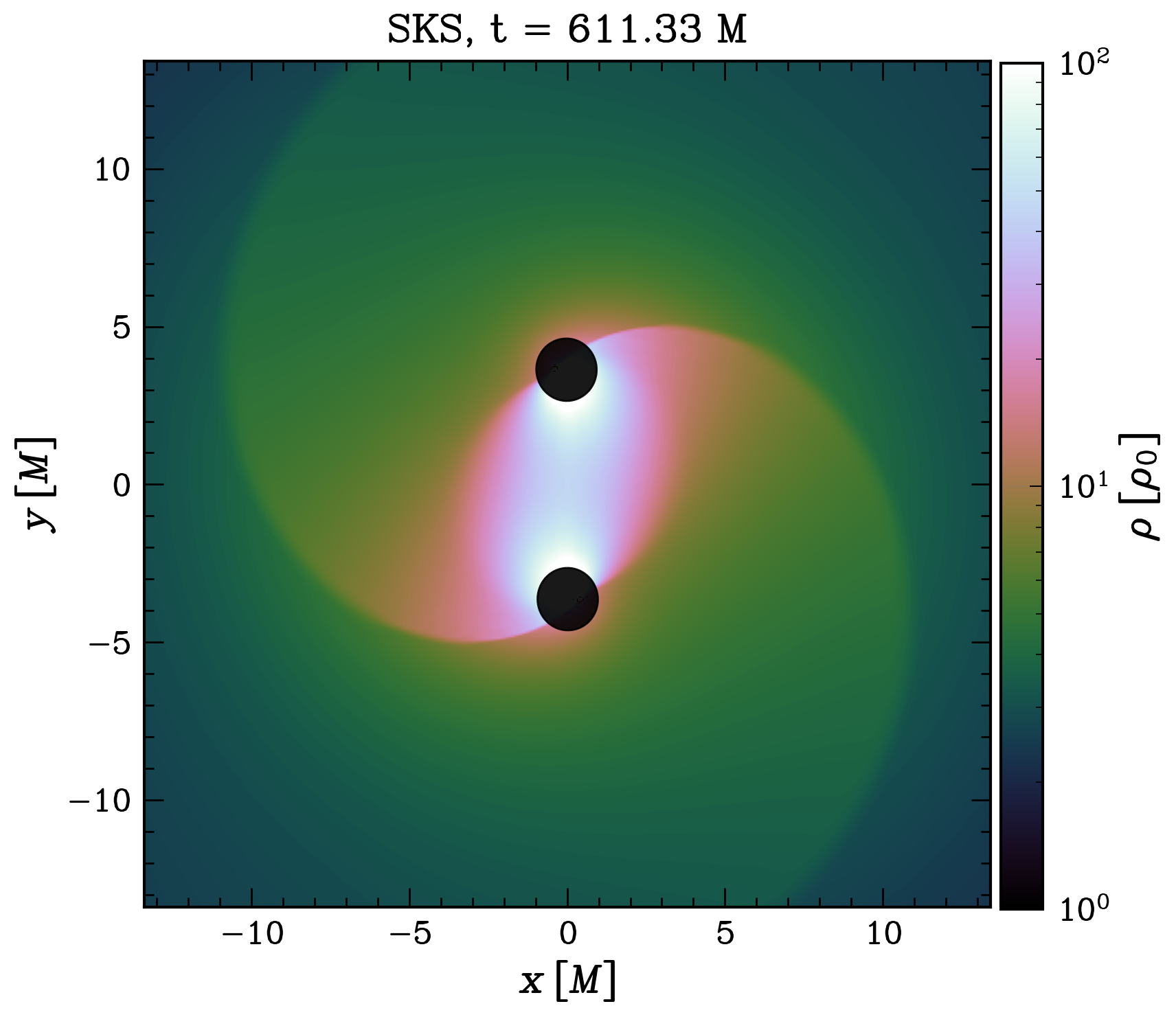}
    \includegraphics[width=0.9\columnwidth]{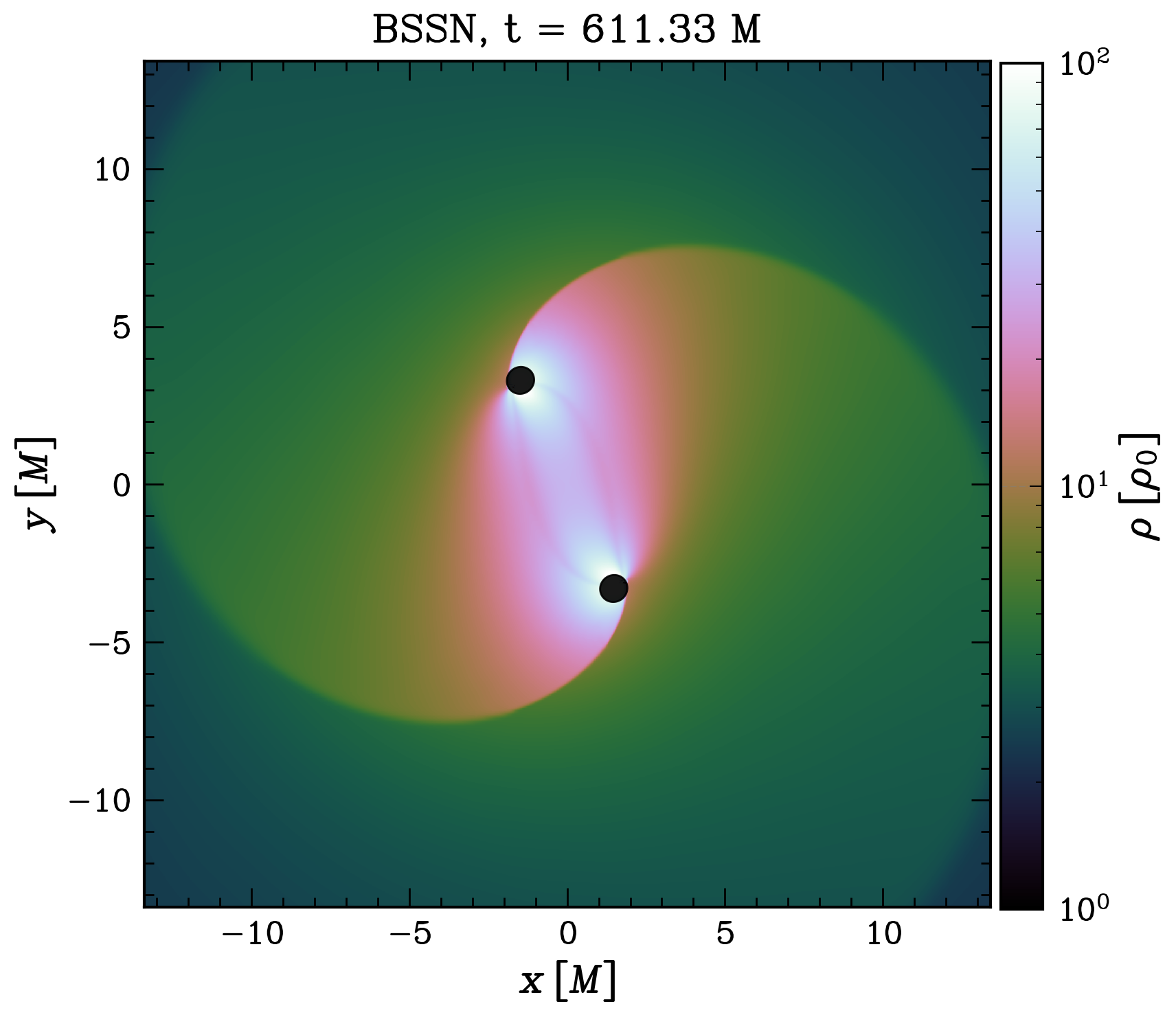}
    \caption{Rest-mass density in the equatorial plane normalized with the initial uniform density $\rho_0 = 10^{-2}$ for \texttt{Run-SKS} (left) and \texttt{Run-BSSN} (right). The apparent horizon of the black holes is represented by black circles. We observe spiral waves traveling through the fluid and a tidal bridge forming between the black holes, qualitatively similar in both simulations.}
    \label{fig:eqrho}
\end{figure*}

To show that the SKS metric is a good approximation for accretion physics even at very small separations, we evolve an initially static uniform gas onto the spacetime. This resembles a Bondi flow problem, which has been investigated for BH binaries in NR \cite{farris2010binary, giacomazzo2012, cattorini2021fully}. For this system, the Bondi radius is given by $R_{\rm b} = 2M/c^2_{\rm s, \infty} = 2M/(\Gamma P_{\infty}/\rho_{\infty}) \approx  25\,M$, where $rho_{\infty}$, $P_{\infty}$ and $c^2_{\rm s, \infty}$ are the density, pressure, and sound speed at infinity.

\begin{figure}[ht!]
    \centering
    \includegraphics[width=\columnwidth]{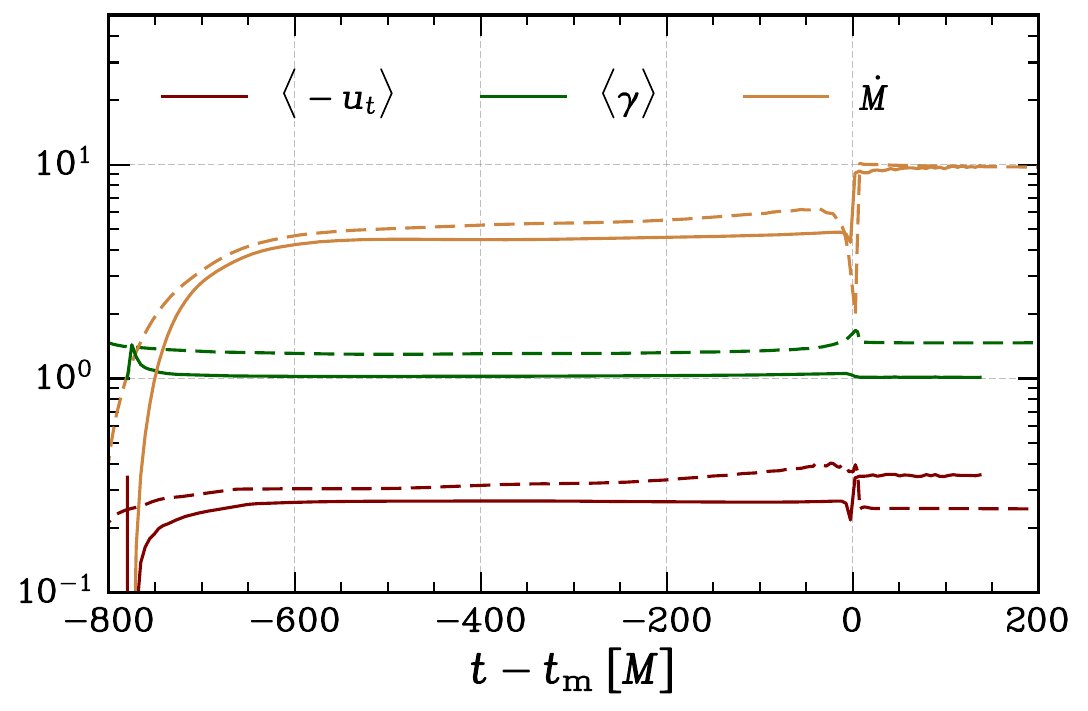} 
    \caption{Evolution of accretion rate onto the BHs, $\dot{M}$, average Lorentz factor, $\langle \gamma \rangle$, and average specific energy, $\langle -u_t \rangle$, for \texttt{Run-SKS} (dashed lines) and \texttt{Run-BSSN} (thick lines).}
    \label{fig:accrate10}
\end{figure}

After a short radially infalling phase, the flow reaches a quasi-steady state, as can be seen in the accretion rate evolution onto the BHs in Fig.~\ref{fig:accrate10}. Close to the black holes, the flow resembles a Bondi-Hoyle-Lyttleton fluid solution \citep{edgar2004review}, with a shock front that propagates in the direction of motion of the black holes, creating spiral waves due to the orbital motion of the binary. In between the black holes, a tidal bridge is formed, creating a stationary high-density region. 

All these features are noticeably similar for \texttt{Run-SKS} and \texttt{Run-BSSN} even one orbit before the merger, as shown, e.g., in the equatorial density in Fig.~\ref{fig:eqrho}. We find that \texttt{Run-SKS} has higher density regions near the horizon, by a factor of a few, while the flow velocity in \texttt{Run-BSSN} is slightly higher (see Fig.~\ref{fig:accrate10}). The accretion rate {measured at the event horizon} $\dot{M}=\int \sqrt{-g} d\Omega u^r \rho$, where $u^r$ is the radial four-velocity in the BH frame, however, follows the same behavior in both simulations: it rises in the first $\sim 100M$, reaching a steady state that lasts until merger, when it settles into a new steady state very quickly. We observe only tens of percent differences between the accretion rates in each simulation. The specific energy of the flow averaged over the horizon differs by a factor of $\sim 2$, although this quantity is strongly gauge-dependent, especially near the horizon. The average velocity of the flow onto the horizon follows the same behavior in each run, but differs throughout the initial evolution by a fixed constant offset that could be attributed to gauge differences in how we measure the velocity.

\section{Inspiralling black holes accreting from a magnetized torus}\label{sec:results_20}

\rvw{
We also perform GRMHD simulations of non-spinning black holes starting at a separation of $r\approx 20\,M$ surrounded by an initially weakly-magnetized torus, comparing BSSN (\texttt{Run-BSSN-20}) and SKS (\texttt{Run-SKS-20}) metric approaches.
}
\subsection{Initial data}

\rvw{
For \texttt{Run-BSSN-20}, as in the previous Sec.~\ref{sec:results_10}, we use puncture initial data for a non-spinning, equal-mass binary at a separation of $20\,M$. We use $m_{(1)} = m_{(2)} = 0.48595$ and an initial momentum in the $y$ direction of $P_{(1)\,y} = 0.095433 M$ and $P_{(2)\,y} = -0.095433 M$ for each BH, with a total ADM mass of $M \approx 1$. For \texttt{Run-SKS-20}, we solve the PN equations of two equal-mass black holes on a quasi-circular orbit ({set} via eccentricity reduction), choosing a coordinate separation of $20\,M$, and a total mass of $M=1$. 
For the MHD sector, in both simulations, we start with a gas torus in hydrodynamic equilibrium with constant angular momentum around the binary, fixing the inner edge at $r_{\rm in}=18\,M$, and maximum pressure at $r_{\rm  p,max}=29\,M$. We endow the torus with a weak seed magnetic field contained in a single poloidal loop{,}
fixing the strength of the field to obtain a density-averaged plasma-$\beta$ of $\beta=2p/b^2=100$ in the disk.  The solution assumes that the spacetime is of a single BH in Kerr-Schild coordinates, and thus it quickly goes out of equilibrium. A quasi-steady state is established near the black holes after a transient that lasts $\approx 3500\,M$.
}

\subsection{Numerical setup}

We use a similar setup for evolving spacetime and MHD as in the previous section, with a larger grid consisting of 10 refinement levels following the BHs with a global domain $-1000\,M<x,y,z<1000\,M$ and finest resolution block of $dx=M/64$ covering a box of radius $r=1\,M$ centered on the BH. We use the exact same grid for each simulation.

\subsection{Circumbinary accretion onto inspiralling black holes}

The strong binary torques perturb the initially axisymmetric magnetized torus, and the gas quickly falls into the black holes through dual streams. In the first $\approx 3500\,M$ of evolution, the magnetic field is amplified by the magnetorotational instability near the black hole in combination with tidal forces that compress the field. Rotationally supported mini-disks form around each black hole, accreting through MHD and tidally-induced stresses.

\begin{figure*}
    \centering
    \includegraphics[width=0.49\linewidth]{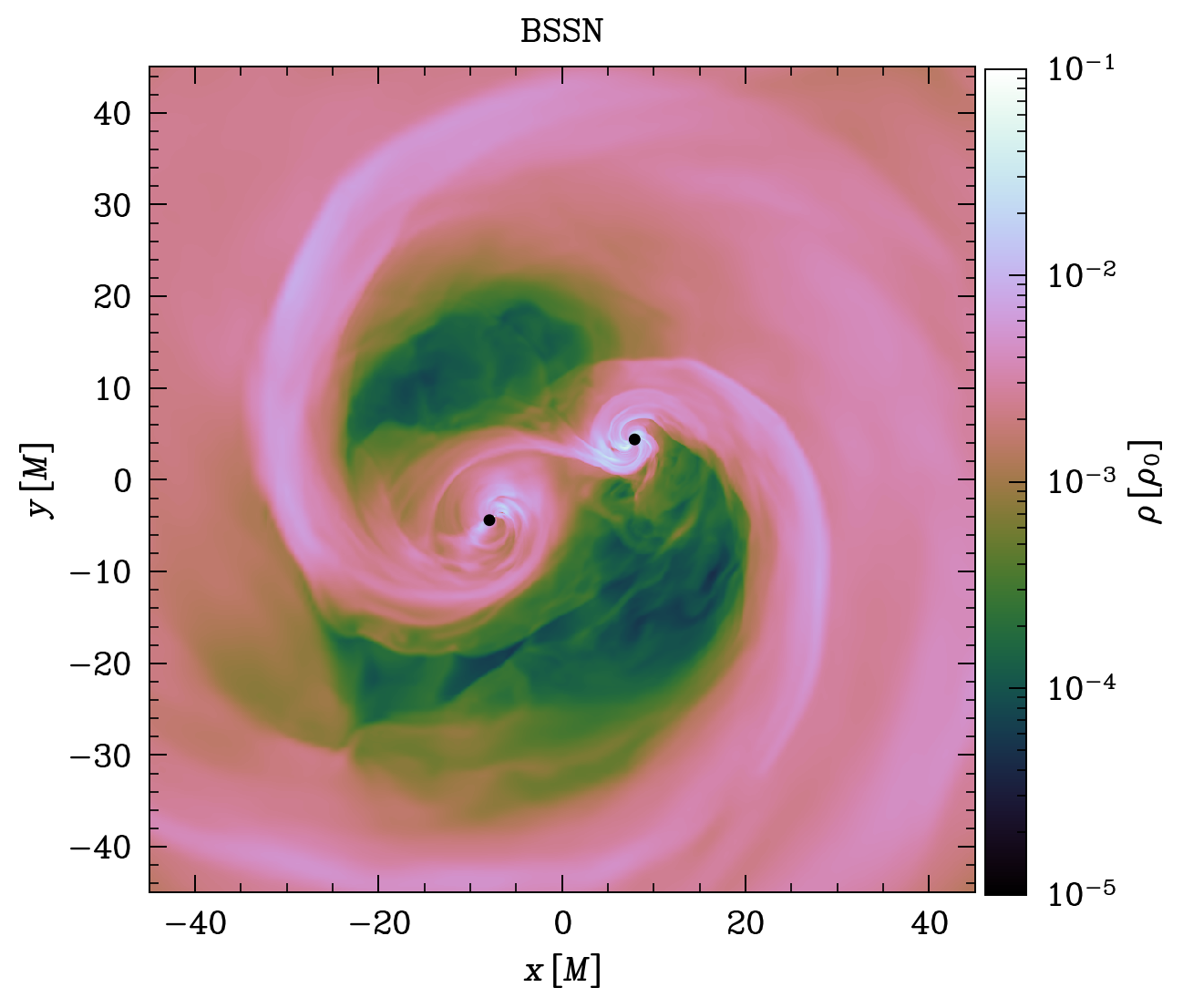}
    \includegraphics[width=0.49\linewidth]{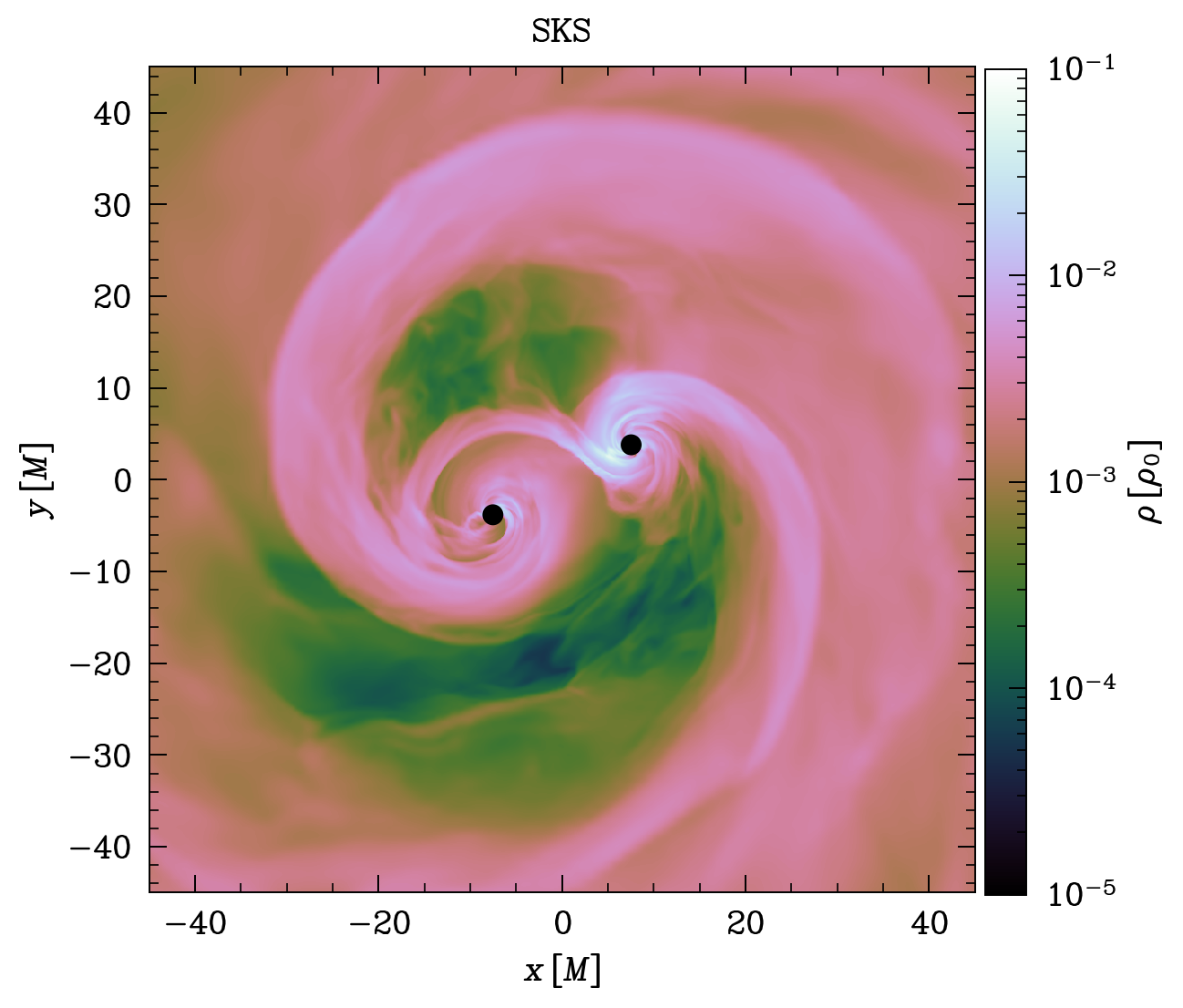}
    \caption{Rest-mass density in the equatorial plane showing the circumbianry disk and mini-disks structure around each black hole for \texttt{Run-BSSN-20} (left) and \texttt{Run-SKS-20} (right), where their coordinate horizons are indicated by black circles. Notice that the size of mini-disks, the shape of tidal streams, and the low-density gap agree well between each simulation.}
    \label{fig:cbd}
\end{figure*}

\begin{figure}
    \centering
    \includegraphics[width=1\linewidth]{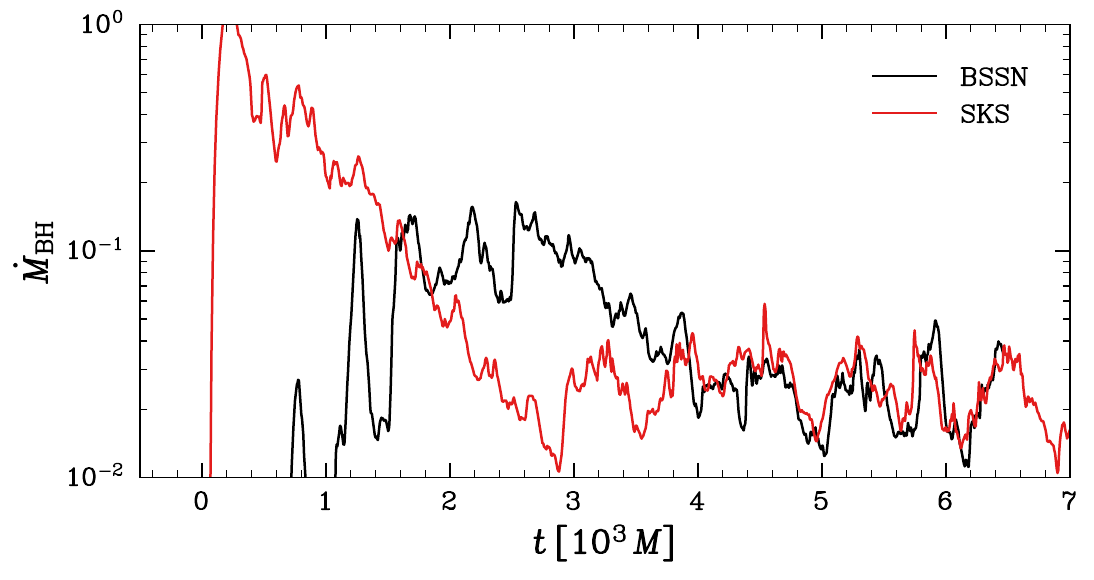}
    \caption{\rvw{Total accretion rate onto one black hole as a function of time for each run. The initial transient is different in each simulation due to the initial conditions, but remarkably, $\dot{M}_{\rm BH}$ converges to roughly the same value after $t \approx 3500\,M$, exhibiting, in particular, the same periodicity and modulation.}}
    \label{fig:accrate}
\end{figure}

\begin{figure}
    \centering
    \includegraphics[width=1.0\linewidth]{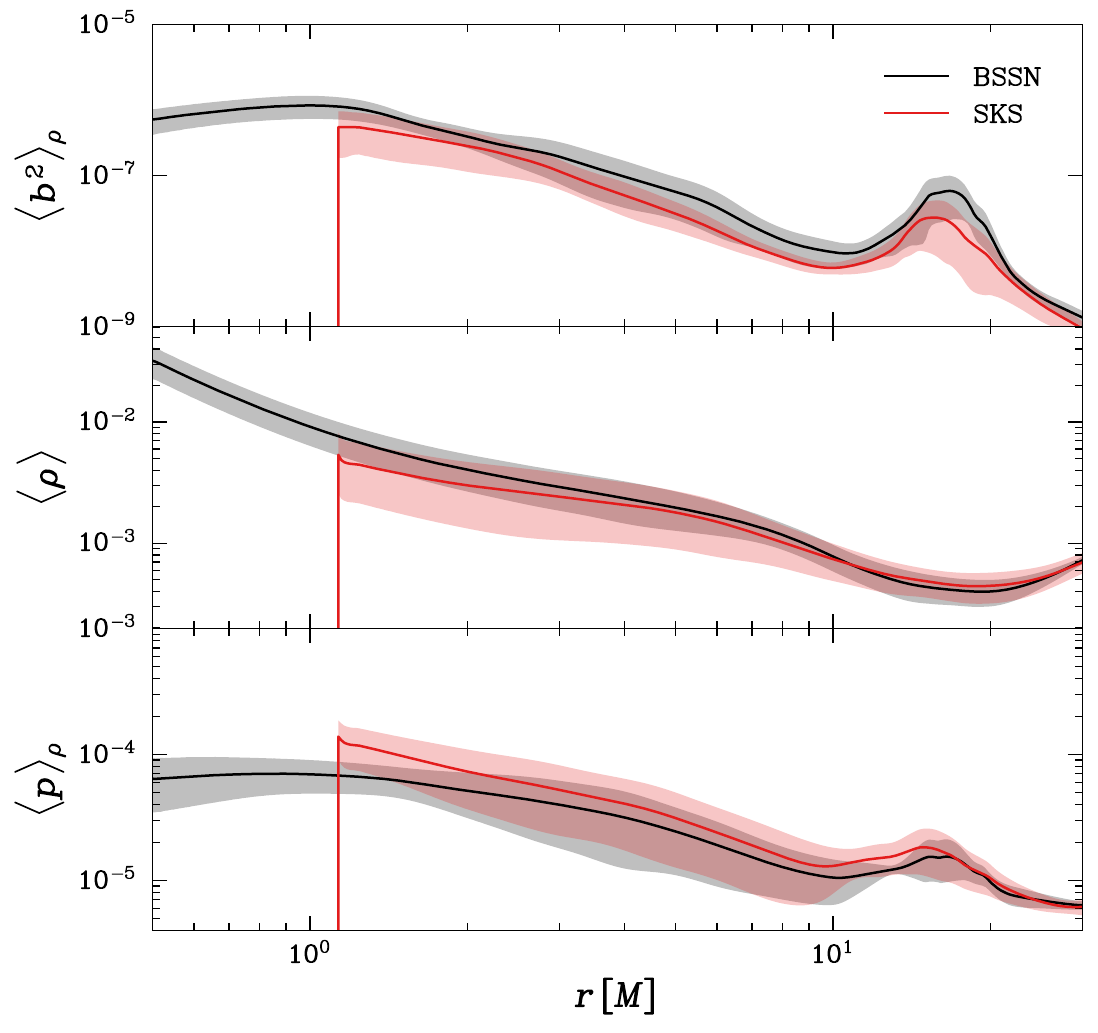}
    \caption{\rvw{MHD properties of the flow in the rest frame of one black hole, time-averaged over a binary period, where $\langle b^2\rangle_{\rho}$ (top panel), is the density-weighted, spherical-averaged comoving magnetic energy,  $\langle \rho \rangle$ (middle panel), is the spherical-averaged rest-mass density, and $\langle p \rangle_{\rho}$ (bottom panel), is the density-weighted, spherical-averaged pressure. Quantities in \texttt{Run-SKS-20} (SKS) are multiplied by a global constant of $1.5$. Thick lines show the time-averaged quantity, and the shadowed areas the mean deviations. The coordinate radius of the horizon in the SKS run is larger and thus the red curves cut off at $r\approx 1\,M$.}}
    \label{fig:mhdq}
\end{figure}

A low-density, eccentric gap forms around the binary due to binary torques ({as seen in the midplane slices of mass density shown in }Fig.~\ref{fig:cbd}), establishing an asymmetric accretion cycle between each BH. In particular,  the accretion rate onto each BH is modulated by the beat frequency once it reaches a quasi-steady state ({as seen in the accretion rate vs.\ time in} Fig.~\ref{fig:accrate}). These features, including the periodicity of the accretion rate and the size of the circumbinary disk gap, agree remarkably well between SKS and BSSN simulations. The time-averaged MHD properties of the mini-disk measured in the rest-frame of the BH also agree well between the two approaches within the intrinsic hydrodynamical variations of the flow (Fig.~\ref{fig:mhdq}) up to a small multiplicative constant which appears due to gauge differences. Indeed, we find that the accretion rate is (nearly) gauge invariant when measured at the horizon, while volume-averaged quantities near the BHs, such as density ($\rho$), differ by a multiplicative constant ($\sim 1.5$) when measured at the same coordinate radius.
{We reiterate here that the MHD fluid variables, as defined here, are gauge-dependent; there is no straightforward way to precisely compare these quantities in a local context without explicit coordinate transformations. 
These transformations are difficult to obtain because in NR the gauge itself is evolving. We can, however, confirm aposteriori which quantities remain gauge invariant.}

\section{Computational Performance}
\label{sec:comper}
\label{sec:results_comp}

\rvw{Using a prescribed metric offers several computational advantages compared to a full numerical evolution. Most importantly, evolving MHD equations on a semi-analytical metric background can be cheaper than performing a full NR simulation for several reasons.}

\subsection{Synchronization}
{
A numerical scheme for solving Einstein's equations, such as BSSN or Z4c, requires solving equations for many additional variables than are used in MHD. For BSSN, there is a set of 24 variables, $\lbrace \phi, \gamma_{ij},K, A_{ij}, \Gamma^i, \alpha, \beta^i,B^i \rbrace$, {that} we need to evolve using equations with lengthy source terms, while communicating information among possibly many CPU cores. The cost of our analytical metric, on the other hand, comes from calculating all terms in Eq.\ \eqref{eq:sks}, the metric inverse and its derivatives, and doing a binary search in a table to interpolate and obtain the position and velocity of BHs; all these operations are local and computationally inexpensive. With a large number of CPUs $(\gtrsim1000)$, the inter-processor and inter-mesh communication for the numerous extra spacetime variables severely impacts the performance of the code; a detailed account of this problem can be found in Refs.~\cite{reisswig2013ThreeDimensional, ng2024hybrid}).
}

\rvw
{ 
Using the same fiducial setup of $100^3$ cells with 10 refinement levels in each simulation and a total of 1920 CPU cores (20 nodes in the Lise cluster), we measure the code speed (in units of $M$ per runtime hour) averaged over the first 256 iterations. We observe that the speed-up of the analytical spacetime run compared with the BSSN run is a factor of $\gtrsim 5$, see Table \ref{tab}.
}

\subsection{Metric update}
\rvw{
Because truncation errors are dominated by the MHD sector, a further boost in efficiency can be obtained by updating the metric every $N$ iterations, where $N$ is constrained by the metric rate of change. For black holes moving at $\sim 0.1c$ and a timestep of $dt=0.25dx/c$ {(set by the Courant number times the cell crossing time for speeds $\sim$ $c$)}, we can safely update the metric every $\sim 10$ iterations. In Ref.~\cite{ressler2024}, we showed explicitly that the hydrodynamical convergence order is maintained when updating every 10 iterations instead of every timestep. Updating the metric every 10 iterations gives us only an improvement of $10\%-15 \%$ in our code, but an improvement of order 2 in Ref.~\cite{ressler2024}  where we used \texttt{Athena++} {and did not optimize the metric computation with {\tt Mathematica}}. 
}

\subsection{Gauge differences}

\rvw{The coordinate radius of the horizon is smaller in the puncture gauge $(r^{\rm NR}_{\rm BH}\lesssim 0.5\,M)$ compared to SKS $(r^{\rm SKS}_{\rm BH}\approx 1\,M)$. On the other hand, the binary separation can be set consistently among coordinates through the gauge invariant quantity $x=(M\Omega_{\rm orb})^{2/3}$, where $\Omega_{\rm orb}$ is the angular velocity of the binary. For the same value of $\Omega$ (set at $r\gtrsim 10\,M$), the resulting binary separation for each gauge is approximately invariant $r^{\rm SKS}_{12} \approx r^{\rm NR}_{\rm 12} =r_{12}$ (see, for instance, the excellent matching in trajectories in Fig.~\ref{fig:traj}). This means that, given a fixed resolution at the binary separation, we can fit more cells within the horizon of the SKS metric. More importantly, the distance that a fluid parcel needs to cross to go from one horizon to the other is then better resolved in the SKS runs. It is common practice in numerical relativity to find gauge prescriptions that maximize the ratio $r_{\rm BH}/r_{12}$, to alleviate the resolution requirements \cite{ma2021extending}.}

\rvw{
Having an adequate gauge is particularly important for spinning black holes, since the coordinate horizon in the puncture (and harmonic) gauge shrinks substantially as the dimensionless spin approaches $a=1$ \cite{ma2021extending, combi2021superposed}. Kerr-Schild-like coordinates, as used in our SKS metric, have the advantage that the horizon radius only shrinks moderately for high spin, and thus are commonly used in accretion physics for this reason. Fixing the same number of cells within the black hole horizons for each gauge amounts to deactivating one refinement level in the analytical SKS metric run (using non-spinning black holes), which gives us a total speed-up of order $\approx 8${, including the gain from reduced variable communication and less frequent metric updates}. This is summarized in Table~\ref{tab}, where we list the computational speed of our fiducial runs with the same number of grid cells ({\texttt{BSSN-fidu} }and {\texttt{SKS-fidu}}) as well as an SKS run with the metric only updated every 10 timesteps ({\texttt{SKS-it10}}) and an SKS run with one less level of mesh refinement ({\texttt{SKS-rL9}}})

\begin{table}[h]
    \centering
    \begin{tabular}{lccc}
        \toprule
        \textbf{Name of Run} & \textbf{Description} & \textbf{Speed} & \textbf{CPU} \\
        \midrule
        \texttt{BSSN-fidu}     & 10 RL,  $N=125^3$ & $10\,M/{\rm hr}$ & $1920$ \\
        \texttt{SKS-fidu}      & 10 RL,  $N=125^3$ & $45\,M/{\rm hr}$ & $1920$ \\
        \texttt{SKS-it10} & 10 RL,  $N=125^3$ & $50\,M/{\rm hr}$ & $1920$ \\
        \texttt{SKS-rL9}  & 9 RL,   $N=125^3$ & $76\,M/{\rm hr}$ & $1920$ \\
        \bottomrule
    \end{tabular}
    \caption{Table with different runs testing the performance of the metric against the numerical relativity evolution, where \texttt{SKS} denotes GRMHD simulations using the SKS metric and \texttt{BSSN} denotes simulations done evolving Einstein's equations. {Here, RL stands for refinement levels.} }
    \label{tab}
\end{table}

\section{Conclusions}
\label{sec:con}

We have presented a new general approximation for a binary black hole spacetime that accounts for arbitrary spins, eccentricities, and mass ratios. The new approximation is an excellent approximation for the inspiral regime and can be used even through merger, where we interpolate the metric to a BH remnant with properties obtained from NR fittings.  We tested this approximation by analyzing its spacetime properties and performing GRMHD simulations of uniform gas and a magnetized disk evolving on this dynamical metric. We compare these results with a full numerical relativity simulation from small separations up to the merger. Our main results can be summarized as follows:

\begin{itemize}
    
	\item The SKS approximation is well-behaved through inspiral and merger. The norm of the Hamiltonian constraint violations remains constant during the entire evolution (Fig.~\ref{fig:hamvst}), increasing locally between the black holes as the separation shortens (Fig.\ref{fig:hamvst}). We show that the black hole trajectories, calculated using a 4PN approximation, follow the numerical relativity trajectory of the BHs quite well up to merger, see Fig.~\ref{fig:traj}.

    \item When the BHs reach separations of $\sim 3\,M$, we show that the SKS metric develops a common apparent horizon that relaxes smoothly to the remnant Kerr black hole horizon (Fig.~\ref{fig:eqah}). The mass of the BHs in \texttt{Run-SKS}, measured as the irreducible mass of the horizons, remains almost constant except in the last orbit before merger when it increases $15\%$ (Fig.~\ref{fig:mass}). 

	\item We perform GRMHD simulations using the SKS metric and a full numerical relativity evolution of Einstein's equations with the BSSN formalism and puncture gauge. We tested two scenarios: merging black holes on a uniform gas, and inspiralling black holes accreting from a magnetized disk. We show that the gas flow behaves qualitatively and quantitatively similarly in both cases. We compare time-averaged properties such as density, magnetic field energy density, and pressure, as well as time-dependent properties such as accretion rate, finding exceptional agreement between both approaches {when considering the gauge-dependent nature of most MHD quantities}.

     \item We show that GRMHD simulations using the SKS metric as a background spacetime are much more efficient than a full numerical relativity evolution {with performance gains up to a factor of  $\sim$ 8 for non-spinning black holes in the most optimistic case}.{The performance boost could be even larger for rapidly spinning black holes, where the puncture gauge requires particularly high resolution within the event horizons. }
     
\end{itemize}

Our analytical metric can be useful to explore regimes that are prohibitively expensive in numerical relativity, such as small-mass ratios and large separations. The numerical implementation of the SKS is public and can be found in a public repository  \citep{softcombi} with the PN solver and {\tt Mathematica} notebooks included. We are also planning to make public our implementation of the metric in the GR version of the {\tt Athena++} code as well as a module (thorn) for the {\tt EinsteinToolkit} (a preliminary version can be found in \footnote{\url{https://gitlab.com/combi.luciano/analyticalbbh}}). We welcome bug reports from the community.


\acknowledgments

LC would like to thank Manuela Campanelli, Carlos Lousto, and Luis Lehner for their support and valuable ideas that contributed to this work. We thank Raphael Mignon-Risse for discussions, testing the metric, and identifying bugs. We also thank Sizheng Ma, Geoff Ryan, Edu Gutierrez, Scott Noble, Will East, Erik Schnetter, and Joaquin Pelle for useful conversations, and the reviewer for their constructive criticism and suggestions, which significantly improved the manuscript. This research was enabled in part by support provided by SciNet (www.scinethpc.ca) and Compute Canada (www.computecanada.ca).
We also acknowledge the support of the Natural Sciences and Engineering Research Council of Canada (NSERC), [funding reference number 568580]
Cette recherche a \'et\'e financ\'ee par le Conseil de recherches en sciences naturelles et en g\'enie du Canada (CRSNG), [num\'ero de r\'ef\'erence 568580]. Research at Perimeter Institute is supported in part by the Government of Canada through the Department of Innovation, Science and Economic Development Canada and by the Province of Ontario through the Ministry of Colleges and Universities.  The authors gratefully acknowledge the computing time made available to them on the high-performance computer ``Lise'' at the NHR Center NHR@ZIB. This center is jointly supported by the German Federal Ministry of Education and Research and the state governments participating in the NHR (www.nhr-verein.de/unsere-partner).

\appendix

\section{Time-dependent boost transformation}

To transform our BH-superposed terms, we need to invert the Fermi coordinate transformation, which is defined from global (inertial) coordinates $x^a = (t,x,y,z)$ to BH (accelerated) coordinates  $X^a = (T,X,Y,Z)$. If the frame were moving on a uniform trajectory, this transformation and its inverse would be simply a Lorentz boost. For an arbitrary accelerated frame, we need, in principle, to solve a non-linear equation to obtain the inverse since the trajectory $s^a(t_{\tau})$ can depend non-linearly on the global time. We can, however, take the following approximation: expanding the trajectory around a time $t$, we obtain $s^{i}(t_{\tau}) \approx s^i(t) + \beta^i (t) (t_{\tau}-t)$. This approximation is good if $\beta |t_{\tau} - t|$ is small. Using the first Equation in \eqref{eq:fermicoord1} we have that $\beta |t_{\tau} - t| = \beta^2 \gamma (n_x X + n_y Y +n_z Z)$. For a fixed position in space, the error is then proportional to $\beta^2 \gamma$, which is small for a binary black hole orbit ($\beta \sim 0.1$). On the other hand, even though this term grows as we move away from the origin, the actual position of the black holes becomes irrelevant for the transformation at larger distances given that the orbit is bound; in other words, we have that $x^a \sim s^a + X^a \sim  X^a(1+\beta^2\gamma)$ as $|x^i| \rightarrow \infty$, 

Now, using this assumption and rewriting things as $s^{i}(t_{\tau}) \approx s^i(t) + \beta^i (t) (t_{\tau}-t) \approx \delta x^i + \beta^i(t) \gamma T$, with $\delta x^i = [s^i(t)-\beta^i t]$, we can write the  transformation in Eq.\ \eqref{eq:fermicoord1} in a form very similar to linear Lorentz boost plus a spatial displacement
\begin{align}
        t = \gamma \: [T + &\beta( n_x X + n_y Y + n_z Z) ], \\
	(x-\delta x) &= \beta^x(t) \gamma T + X\: [1 + n_x^2 (\gamma-1) ] \nonumber \\
	              &+ Y \: (\gamma-1) n_x n_y + Z \: (\gamma-1) n_x n_z,		\\
	(y - \delta y) &= \beta^z(t) \gamma T + Y \: [1 + n_y^2 (\gamma-1) ] \nonumber \\
	              &+ X \: (\gamma-1) n_x n_y + Z \: (\gamma-1) n_y n_z,		\\
	(z - \delta z) &=\beta^z(t) \gamma T + Z \: [1 + n_z^2 (\gamma-1) ] \nonumber \\
	              &+ Y \: (\gamma-1) n_z n_y + X \: (\gamma-1) n_x n_z.
\end{align}

In this form, we can invert the system of equations by applying the inverse Lorentz matrix to obtain, in the spatial part:
\begin{align*}
	X = & -\beta^x \gamma t + (x-\delta x)\: [1 + n_x^2 (\gamma-1) ] \\
            &+ (y-\delta y) \: (\gamma-1) n_x n_y + (z-\delta {z}) (\gamma-1) n_x n_z,	\\
        Y = &-\beta^y \gamma t  + (y-\delta y)\: [1 + n_y^2 (\gamma-1) ] \\ 
              &+ (x-\delta x) \: (\gamma-1) n_x n_y + (z-\delta {z}) (\gamma-1) n_y n_z,		\\
        Z = &-\beta^z \gamma t + (z-\delta z)\: [1 + n_z^2 (\gamma-1) ] \\ 
              &+ (y-\delta y) \: (\gamma-1) n_y n_z + (x-\delta {x})  (\gamma-1) n_x n_z.		\\
\end{align*}

The time part is more complicated but not needed given that we use this tranformation in stationary metrics (i.e.\ single Kerr black holes). To find the Jacobian of this transformation, if we discard all acceleration terms $(\dot{\beta}^i\sim 0)$, we have that $d (x^i-\delta x^i) = dx^i$, since $d(\delta x^i)=d(s^i(t) -\beta^i t) = \beta^i dt - \beta^i dt + \dot{\beta}^i dt \sim 0$  and then it is easy to see that $\partial X^{a}/\partial x^{b} = \Lambda^{a}_{b} +\mathcal{O}(\dot{\beta})$. Also, notice that for a uniformly moving body, the transformation reduces directly to a linear boost since $\delta x^i = 0$ in that case. To obtain the explicit transformation given in Eq.\ \eqref{eq:fermicoord2}, we can expand $\delta x$, and show that all linear dependence in time disappears from the transformation.

\section{Interpolation function}

In Sec.~\ref{sec:tranmerger}, we use the smooth interpolation function $\mathcal{W}(t)$ adopted from Ref.~\cite{afterglowpy}. The function is defined as:
\begin{equation}
    \mathcal{W}(t) = \frac{E(\mathcal{T}(t))}{E(\mathcal{T}(t)
    )+E(\mathcal{T}(t)-1)},
\end{equation}
being
\begin{equation}
    E(t)= \lbrace \exp(-1/t) \: {\rm if} \:  t>0, \:  0 \: {\rm if} \:  t \leq 0 \rbrace,
\end{equation}
and the time parameter $\mathcal{T}$ is normalized as $\mathcal{T}(t)=[t-(t_{\rm merger}-dt_{\rm merger})]/dt_{\rm merger}$, so $\mathcal{T}(t_{\rm merger})=1$.

%


\begin{thebibliography}{117}%
\makeatletter
\providecommand \@ifxundefined [1]{%
 \@ifx{#1\undefined}
}%
\providecommand \@ifnum [1]{%
 \ifnum #1\expandafter \@firstoftwo
 \else \expandafter \@secondoftwo
 \fi
}%
\providecommand \@ifx [1]{%
 \ifx #1\expandafter \@firstoftwo
 \else \expandafter \@secondoftwo
 \fi
}%
\providecommand \natexlab [1]{#1}%
\providecommand \enquote  [1]{``#1''}%
\providecommand \bibnamefont  [1]{#1}%
\providecommand \bibfnamefont [1]{#1}%
\providecommand \citenamefont [1]{#1}%
\providecommand \href@noop [0]{\@secondoftwo}%
\providecommand \href [0]{\begingroup \@sanitize@url \@href}%
\providecommand \@href[1]{\@@startlink{#1}\@@href}%
\providecommand \@@href[1]{\endgroup#1\@@endlink}%
\providecommand \@sanitize@url [0]{\catcode `\\12\catcode `\$12\catcode
  `\&12\catcode `\#12\catcode `\^12\catcode `\_12\catcode `\%12\relax}%
\providecommand \@@startlink[1]{}%
\providecommand \@@endlink[0]{}%
\providecommand \url  [0]{\begingroup\@sanitize@url \@url }%
\providecommand \@url [1]{\endgroup\@href {#1}{\urlprefix }}%
\providecommand \urlprefix  [0]{URL }%
\providecommand \Eprint [0]{\href }%
\providecommand \doibase [0]{https://doi.org/}%
\providecommand \selectlanguage [0]{\@gobble}%
\providecommand \bibinfo  [0]{\@secondoftwo}%
\providecommand \bibfield  [0]{\@secondoftwo}%
\providecommand \translation [1]{[#1]}%
\providecommand \BibitemOpen [0]{}%
\providecommand \bibitemStop [0]{}%
\providecommand \bibitemNoStop [0]{.\EOS\space}%
\providecommand \EOS [0]{\spacefactor3000\relax}%
\providecommand \BibitemShut  [1]{\csname bibitem#1\endcsname}%
\let\auto@bib@innerbib\@empty
\bibitem [{\citenamefont {Abbott}\ \emph {et~al.}(2017)\citenamefont {Abbott}
  \emph {et~al.}}]{abbott2017gravitational}%
  \BibitemOpen
  \bibfield  {author} {\bibinfo {author} {\bibfnamefont {B.~P.}\ \bibnamefont
  {Abbott}} \emph {et~al.},\ }\href {https://doi.org/10/gcx9dc} {\bibfield
  {journal} {\bibinfo  {journal} {Astrophys. J. Lett.}\ }\textbf {\bibinfo
  {volume} {848}},\ \bibinfo {pages} {L13} (\bibinfo {year}
  {2017})}\BibitemShut {NoStop}%
\bibitem [{\citenamefont {{Kelly}}\ \emph {et~al.}(2021)\citenamefont
  {{Kelly}}, \citenamefont {{Etienne}}, \citenamefont {{Golomb}}, \citenamefont
  {{Schnittman}}, \citenamefont {{Baker}}, \citenamefont {{Noble}},\ and\
  \citenamefont {{Ryan}}}]{kelly2021electromagnetic}%
  \BibitemOpen
  \bibfield  {author} {\bibinfo {author} {\bibfnamefont {B.~J.}\ \bibnamefont
  {{Kelly}}}, \bibinfo {author} {\bibfnamefont {Z.~B.}\ \bibnamefont
  {{Etienne}}}, \bibinfo {author} {\bibfnamefont {J.}~\bibnamefont {{Golomb}}},
  \bibinfo {author} {\bibfnamefont {J.~D.}\ \bibnamefont {{Schnittman}}},
  \bibinfo {author} {\bibfnamefont {J.~G.}\ \bibnamefont {{Baker}}}, \bibinfo
  {author} {\bibfnamefont {S.~C.}\ \bibnamefont {{Noble}}},\ and\ \bibinfo
  {author} {\bibfnamefont {G.}~\bibnamefont {{Ryan}}},\ }\href
  {https://doi.org/10.1103/PhysRevD.103.063039} {\bibfield  {journal} {\bibinfo
   {journal} {\prd}\ }\textbf {\bibinfo {volume} {103}},\ \bibinfo {eid}
  {063039} (\bibinfo {year} {2021})},\ \Eprint
  {https://arxiv.org/abs/2010.11259} {arXiv:2010.11259 [astro-ph.HE]}
  \BibitemShut {NoStop}%
\bibitem [{\citenamefont {Bogdanovi{\'c}}\ \emph {et~al.}(2022)\citenamefont
  {Bogdanovi{\'c}}, \citenamefont {Miller},\ and\ \citenamefont
  {Blecha}}]{bogdanovic2022electromagnetic}%
  \BibitemOpen
  \bibfield  {author} {\bibinfo {author} {\bibfnamefont {T.}~\bibnamefont
  {Bogdanovi{\'c}}}, \bibinfo {author} {\bibfnamefont {M.~C.}\ \bibnamefont
  {Miller}},\ and\ \bibinfo {author} {\bibfnamefont {L.}~\bibnamefont
  {Blecha}},\ }\href@noop {} {\bibfield  {journal} {\bibinfo  {journal} {Living
  Rev. Relativ.}\ }\textbf {\bibinfo {volume} {25}},\ \bibinfo {pages} {3}
  (\bibinfo {year} {2022})}\BibitemShut {NoStop}%
\bibitem [{\citenamefont {{Begelman}}\ \emph {et~al.}(1980)\citenamefont
  {{Begelman}}, \citenamefont {{Blandford}},\ and\ \citenamefont
  {{Rees}}}]{Begelman1980}%
  \BibitemOpen
  \bibfield  {author} {\bibinfo {author} {\bibfnamefont {M.~C.}\ \bibnamefont
  {{Begelman}}}, \bibinfo {author} {\bibfnamefont {R.~D.}\ \bibnamefont
  {{Blandford}}},\ and\ \bibinfo {author} {\bibfnamefont {M.~J.}\ \bibnamefont
  {{Rees}}},\ }\href {https://doi.org/10.1038/287307a0} {\bibfield  {journal}
  {\bibinfo  {journal} {\nat}\ }\textbf {\bibinfo {volume} {287}},\ \bibinfo
  {pages} {307} (\bibinfo {year} {1980})}\BibitemShut {NoStop}%
\bibitem [{\citenamefont {{Artymowicz}}\ and\ \citenamefont
  {{Lubow}}(1996)}]{Artymowicz1996}%
  \BibitemOpen
  \bibfield  {author} {\bibinfo {author} {\bibfnamefont {P.}~\bibnamefont
  {{Artymowicz}}}\ and\ \bibinfo {author} {\bibfnamefont {S.~H.}\ \bibnamefont
  {{Lubow}}},\ }\href {https://doi.org/10.1086/310200} {\bibfield  {journal}
  {\bibinfo  {journal} {{Astrophys. J.}l}\ }\textbf {\bibinfo {volume} {467}},\
  \bibinfo {pages} {L77} (\bibinfo {year} {1996})}\BibitemShut {NoStop}%
\bibitem [{\citenamefont {Peters}\ and\ \citenamefont
  {Mathews}(1963)}]{Peters:1963ux}%
  \BibitemOpen
  \bibfield  {author} {\bibinfo {author} {\bibfnamefont {P.~C.}\ \bibnamefont
  {Peters}}\ and\ \bibinfo {author} {\bibfnamefont {J.}~\bibnamefont
  {Mathews}},\ }\href {https://doi.org/10.1103/PhysRev.131.435} {\bibfield
  {journal} {\bibinfo  {journal} {Phys. Rev.}\ }\textbf {\bibinfo {volume}
  {131}},\ \bibinfo {pages} {435} (\bibinfo {year} {1963})}\BibitemShut
  {NoStop}%
\bibitem [{\citenamefont {Peters}(1964)}]{Peters:1964zz}%
  \BibitemOpen
  \bibfield  {author} {\bibinfo {author} {\bibfnamefont {P.~C.}\ \bibnamefont
  {Peters}},\ }\href {https://doi.org/10.1103/PhysRev.136.B1224} {\bibfield
  {journal} {\bibinfo  {journal} {Phys. Rev.}\ }\textbf {\bibinfo {volume}
  {136}},\ \bibinfo {pages} {B1224} (\bibinfo {year} {1964})}\BibitemShut
  {NoStop}%
\bibitem [{\citenamefont {Afzal}\ \emph {et~al.}(2023)\citenamefont {Afzal},
  \citenamefont {Agazie}, \citenamefont {Anumarlapudi}, \citenamefont
  {Archibald}, \citenamefont {Arzoumanian}, \citenamefont {Baker},
  \citenamefont {B{\'e}csy}, \citenamefont {Blanco-Pillado}, \citenamefont
  {Blecha}, \citenamefont {Boddy} \emph {et~al.}}]{afzal2023nanograv}%
  \BibitemOpen
  \bibfield  {author} {\bibinfo {author} {\bibfnamefont {A.}~\bibnamefont
  {Afzal}}, \bibinfo {author} {\bibfnamefont {G.}~\bibnamefont {Agazie}},
  \bibinfo {author} {\bibfnamefont {A.}~\bibnamefont {Anumarlapudi}}, \bibinfo
  {author} {\bibfnamefont {A.~M.}\ \bibnamefont {Archibald}}, \bibinfo {author}
  {\bibfnamefont {Z.}~\bibnamefont {Arzoumanian}}, \bibinfo {author}
  {\bibfnamefont {P.~T.}\ \bibnamefont {Baker}}, \bibinfo {author}
  {\bibfnamefont {B.}~\bibnamefont {B{\'e}csy}}, \bibinfo {author}
  {\bibfnamefont {J.~J.}\ \bibnamefont {Blanco-Pillado}}, \bibinfo {author}
  {\bibfnamefont {L.}~\bibnamefont {Blecha}}, \bibinfo {author} {\bibfnamefont
  {K.~K.}\ \bibnamefont {Boddy}}, \emph {et~al.},\ }\href@noop {} {\bibfield
  {journal} {\bibinfo  {journal} {Astrophys.\ J.\ Lett.}\ }\textbf {\bibinfo
  {volume} {951}},\ \bibinfo {pages} {L11} (\bibinfo {year}
  {2023})}\BibitemShut {NoStop}%
\bibitem [{\citenamefont {{Agazie}}\ \emph {et~al.}(2023)\citenamefont
  {{Agazie}}, \citenamefont {{Anumarlapudi}}, \citenamefont {{Archibald}},
  \citenamefont {{Baker}}, \citenamefont {{B{\'e}csy}}, \citenamefont
  {{Blecha}}, \citenamefont {{Bonilla}}, \citenamefont {{Brazier}},
  \citenamefont {{Brook}}, \citenamefont {{Burke-Spolaor}}, \citenamefont
  {{Burnette}}, \citenamefont {{Case}}, \citenamefont {{Casey-Clyde}},
  \citenamefont {{Charisi}}, \citenamefont {{Chatterjee}}, \citenamefont
  {{Chatziioannou}}, \citenamefont {{Cheeseboro}}, \citenamefont {{Chen}},
  \citenamefont {{Cohen}}, \citenamefont {{Cordes}}, \citenamefont {{Cornish}},
  \citenamefont {{Crawford}}, \citenamefont {{Cromartie}}, \citenamefont
  {{Crowter}}, \citenamefont {{Cutler}}, \citenamefont {{D'Orazio}},
  \citenamefont {{Decesar}}, \citenamefont {{Degan}}, \citenamefont
  {{Demorest}}, \citenamefont {{Deng}}, \citenamefont {{Dolch}}, \citenamefont
  {{Drachler}}, \citenamefont {{Ferrara}}, \citenamefont {{Fiore}},
  \citenamefont {{Fonseca}}, \citenamefont {{Freedman}}, \citenamefont
  {{Gardiner}}, \citenamefont {{Garver-Daniels}}, \citenamefont {{Gentile}},
  \citenamefont {{Gersbach}}, \citenamefont {{Glaser}}, \citenamefont {{Good}},
  \citenamefont {{G{\"u}ltekin}}, \citenamefont {{Hazboun}}, \citenamefont
  {{Hourihane}}, \citenamefont {{Islo}}, \citenamefont {{Jennings}},
  \citenamefont {{Johnson}}, \citenamefont {{Jones}}, \citenamefont {{Kaiser}},
  \citenamefont {{Kaplan}}, \citenamefont {{Kelley}}, \citenamefont {{Kerr}},
  \citenamefont {{Key}}, \citenamefont {{Laal}}, \citenamefont {{Lam}},
  \citenamefont {{Lamb}}, \citenamefont {{Lazio}}, \citenamefont
  {{Lewandowska}}, \citenamefont {{Littenberg}}, \citenamefont {{Liu}},
  \citenamefont {{Luo}}, \citenamefont {{Lynch}}, \citenamefont {{Ma}},
  \citenamefont {{Madison}}, \citenamefont {{McEwen}}, \citenamefont {{McKee}},
  \citenamefont {{McLaughlin}}, \citenamefont {{McMann}}, \citenamefont
  {{Meyers}}, \citenamefont {{Meyers}}, \citenamefont {{Mingarelli}},
  \citenamefont {{Mitridate}}, \citenamefont {{Natarajan}}, \citenamefont
  {{Ng}}, \citenamefont {{Nice}}, \citenamefont {{Ocker}}, \citenamefont
  {{Olum}}, \citenamefont {{Pennucci}}, \citenamefont {{Perera}}, \citenamefont
  {{Petrov}}, \citenamefont {{Pol}}, \citenamefont {{Radovan}}, \citenamefont
  {{Ransom}}, \citenamefont {{Ray}}, \citenamefont {{Romano}}, \citenamefont
  {{Runnoe}}, \citenamefont {{Sardesai}}, \citenamefont {{Schmiedekamp}},
  \citenamefont {{Schmiedekamp}}, \citenamefont {{Schmitz}}, \citenamefont
  {{Schult}}, \citenamefont {{Shapiro-Albert}}, \citenamefont {{Siemens}},
  \citenamefont {{Simon}}, \citenamefont {{Siwek}}, \citenamefont {{Stairs}},
  \citenamefont {{Stinebring}}, \citenamefont {{Stovall}}, \citenamefont
  {{Sun}}, \citenamefont {{Susobhanan}}, \citenamefont {{Swiggum}},
  \citenamefont {{Taylor}}, \citenamefont {{Taylor}}, \citenamefont {{Turner}},
  \citenamefont {{Unal}}, \citenamefont {{Vallisneri}}, \citenamefont
  {{Vigeland}}, \citenamefont {{Wachter}}, \citenamefont {{Wahl}},
  \citenamefont {{Wang}}, \citenamefont {{Witt}}, \citenamefont {{Wright}},
  \citenamefont {{Young}},\ and\ \citenamefont {{Nanograv
  Collaboration}}}]{agazie2023nanograv}%
  \BibitemOpen
  \bibfield  {author} {\bibinfo {author} {\bibfnamefont {G.}~\bibnamefont
  {{Agazie}}}, \bibinfo {author} {\bibfnamefont {A.}~\bibnamefont
  {{Anumarlapudi}}}, \bibinfo {author} {\bibfnamefont {A.~M.}\ \bibnamefont
  {{Archibald}}}, \bibinfo {author} {\bibfnamefont {P.~T.}\ \bibnamefont
  {{Baker}}}, \bibinfo {author} {\bibfnamefont {B.}~\bibnamefont
  {{B{\'e}csy}}}, \bibinfo {author} {\bibfnamefont {L.}~\bibnamefont
  {{Blecha}}}, \bibinfo {author} {\bibfnamefont {A.}~\bibnamefont {{Bonilla}}},
  \bibinfo {author} {\bibfnamefont {A.}~\bibnamefont {{Brazier}}}, \bibinfo
  {author} {\bibfnamefont {P.~R.}\ \bibnamefont {{Brook}}}, \bibinfo {author}
  {\bibfnamefont {S.}~\bibnamefont {{Burke-Spolaor}}}, \bibinfo {author}
  {\bibfnamefont {R.}~\bibnamefont {{Burnette}}}, \bibinfo {author}
  {\bibfnamefont {R.}~\bibnamefont {{Case}}}, \bibinfo {author} {\bibfnamefont
  {J.~A.}\ \bibnamefont {{Casey-Clyde}}}, \bibinfo {author} {\bibfnamefont
  {M.}~\bibnamefont {{Charisi}}}, \bibinfo {author} {\bibfnamefont
  {S.}~\bibnamefont {{Chatterjee}}}, \bibinfo {author} {\bibfnamefont
  {K.}~\bibnamefont {{Chatziioannou}}}, \bibinfo {author} {\bibfnamefont
  {B.~D.}\ \bibnamefont {{Cheeseboro}}}, \bibinfo {author} {\bibfnamefont
  {S.}~\bibnamefont {{Chen}}}, \bibinfo {author} {\bibfnamefont
  {T.}~\bibnamefont {{Cohen}}}, \bibinfo {author} {\bibfnamefont {J.~M.}\
  \bibnamefont {{Cordes}}}, \bibinfo {author} {\bibfnamefont {N.~J.}\
  \bibnamefont {{Cornish}}}, \bibinfo {author} {\bibfnamefont {F.}~\bibnamefont
  {{Crawford}}}, \bibinfo {author} {\bibfnamefont {H.~T.}\ \bibnamefont
  {{Cromartie}}}, \bibinfo {author} {\bibfnamefont {K.}~\bibnamefont
  {{Crowter}}}, \bibinfo {author} {\bibfnamefont {C.~J.}\ \bibnamefont
  {{Cutler}}}, \bibinfo {author} {\bibfnamefont {D.~J.}\ \bibnamefont
  {{D'Orazio}}}, \bibinfo {author} {\bibfnamefont {M.~E.}\ \bibnamefont
  {{Decesar}}}, \bibinfo {author} {\bibfnamefont {D.}~\bibnamefont {{Degan}}},
  \bibinfo {author} {\bibfnamefont {P.~B.}\ \bibnamefont {{Demorest}}},
  \bibinfo {author} {\bibfnamefont {H.}~\bibnamefont {{Deng}}}, \bibinfo
  {author} {\bibfnamefont {T.}~\bibnamefont {{Dolch}}}, \bibinfo {author}
  {\bibfnamefont {B.}~\bibnamefont {{Drachler}}}, \bibinfo {author}
  {\bibfnamefont {E.~C.}\ \bibnamefont {{Ferrara}}}, \bibinfo {author}
  {\bibfnamefont {W.}~\bibnamefont {{Fiore}}}, \bibinfo {author} {\bibfnamefont
  {E.}~\bibnamefont {{Fonseca}}}, \bibinfo {author} {\bibfnamefont {G.~E.}\
  \bibnamefont {{Freedman}}}, \bibinfo {author} {\bibfnamefont
  {E.}~\bibnamefont {{Gardiner}}}, \bibinfo {author} {\bibfnamefont
  {N.}~\bibnamefont {{Garver-Daniels}}}, \bibinfo {author} {\bibfnamefont
  {P.~A.}\ \bibnamefont {{Gentile}}}, \bibinfo {author} {\bibfnamefont {K.~A.}\
  \bibnamefont {{Gersbach}}}, \bibinfo {author} {\bibfnamefont
  {J.}~\bibnamefont {{Glaser}}}, \bibinfo {author} {\bibfnamefont {D.~C.}\
  \bibnamefont {{Good}}}, \bibinfo {author} {\bibfnamefont {K.}~\bibnamefont
  {{G{\"u}ltekin}}}, \bibinfo {author} {\bibfnamefont {J.~S.}\ \bibnamefont
  {{Hazboun}}}, \bibinfo {author} {\bibfnamefont {S.}~\bibnamefont
  {{Hourihane}}}, \bibinfo {author} {\bibfnamefont {K.}~\bibnamefont {{Islo}}},
  \bibinfo {author} {\bibfnamefont {R.~J.}\ \bibnamefont {{Jennings}}},
  \bibinfo {author} {\bibfnamefont {A.}~\bibnamefont {{Johnson}}}, \bibinfo
  {author} {\bibfnamefont {M.~L.}\ \bibnamefont {{Jones}}}, \bibinfo {author}
  {\bibfnamefont {A.~R.}\ \bibnamefont {{Kaiser}}}, \bibinfo {author}
  {\bibfnamefont {D.~L.}\ \bibnamefont {{Kaplan}}}, \bibinfo {author}
  {\bibfnamefont {L.~Z.}\ \bibnamefont {{Kelley}}}, \bibinfo {author}
  {\bibfnamefont {M.}~\bibnamefont {{Kerr}}}, \bibinfo {author} {\bibfnamefont
  {J.~S.}\ \bibnamefont {{Key}}}, \bibinfo {author} {\bibfnamefont
  {N.}~\bibnamefont {{Laal}}}, \bibinfo {author} {\bibfnamefont {M.~T.}\
  \bibnamefont {{Lam}}}, \bibinfo {author} {\bibfnamefont {W.~G.}\ \bibnamefont
  {{Lamb}}}, \bibinfo {author} {\bibfnamefont {T.~J.~W.}\ \bibnamefont
  {{Lazio}}}, \bibinfo {author} {\bibfnamefont {N.}~\bibnamefont
  {{Lewandowska}}}, \bibinfo {author} {\bibfnamefont {T.~B.}\ \bibnamefont
  {{Littenberg}}}, \bibinfo {author} {\bibfnamefont {T.}~\bibnamefont {{Liu}}},
  \bibinfo {author} {\bibfnamefont {J.}~\bibnamefont {{Luo}}}, \bibinfo
  {author} {\bibfnamefont {R.~S.}\ \bibnamefont {{Lynch}}}, \bibinfo {author}
  {\bibfnamefont {C.-P.}\ \bibnamefont {{Ma}}}, \bibinfo {author}
  {\bibfnamefont {D.~R.}\ \bibnamefont {{Madison}}}, \bibinfo {author}
  {\bibfnamefont {A.}~\bibnamefont {{McEwen}}}, \bibinfo {author}
  {\bibfnamefont {J.~W.}\ \bibnamefont {{McKee}}}, \bibinfo {author}
  {\bibfnamefont {M.~A.}\ \bibnamefont {{McLaughlin}}}, \bibinfo {author}
  {\bibfnamefont {N.}~\bibnamefont {{McMann}}}, \bibinfo {author}
  {\bibfnamefont {B.~W.}\ \bibnamefont {{Meyers}}}, \bibinfo {author}
  {\bibfnamefont {P.~M.}\ \bibnamefont {{Meyers}}}, \bibinfo {author}
  {\bibfnamefont {C.~M.~F.}\ \bibnamefont {{Mingarelli}}}, \bibinfo {author}
  {\bibfnamefont {A.}~\bibnamefont {{Mitridate}}}, \bibinfo {author}
  {\bibfnamefont {P.}~\bibnamefont {{Natarajan}}}, \bibinfo {author}
  {\bibfnamefont {C.}~\bibnamefont {{Ng}}}, \bibinfo {author} {\bibfnamefont
  {D.~J.}\ \bibnamefont {{Nice}}}, \bibinfo {author} {\bibfnamefont {S.~K.}\
  \bibnamefont {{Ocker}}}, \bibinfo {author} {\bibfnamefont {K.~D.}\
  \bibnamefont {{Olum}}}, \bibinfo {author} {\bibfnamefont {T.~T.}\
  \bibnamefont {{Pennucci}}}, \bibinfo {author} {\bibfnamefont {B.~B.~P.}\
  \bibnamefont {{Perera}}}, \bibinfo {author} {\bibfnamefont {P.}~\bibnamefont
  {{Petrov}}}, \bibinfo {author} {\bibfnamefont {N.~S.}\ \bibnamefont {{Pol}}},
  \bibinfo {author} {\bibfnamefont {H.~A.}\ \bibnamefont {{Radovan}}}, \bibinfo
  {author} {\bibfnamefont {S.~M.}\ \bibnamefont {{Ransom}}}, \bibinfo {author}
  {\bibfnamefont {P.~S.}\ \bibnamefont {{Ray}}}, \bibinfo {author}
  {\bibfnamefont {J.~D.}\ \bibnamefont {{Romano}}}, \bibinfo {author}
  {\bibfnamefont {J.~C.}\ \bibnamefont {{Runnoe}}}, \bibinfo {author}
  {\bibfnamefont {S.~C.}\ \bibnamefont {{Sardesai}}}, \bibinfo {author}
  {\bibfnamefont {A.}~\bibnamefont {{Schmiedekamp}}}, \bibinfo {author}
  {\bibfnamefont {C.}~\bibnamefont {{Schmiedekamp}}}, \bibinfo {author}
  {\bibfnamefont {K.}~\bibnamefont {{Schmitz}}}, \bibinfo {author}
  {\bibfnamefont {L.}~\bibnamefont {{Schult}}}, \bibinfo {author}
  {\bibfnamefont {B.~J.}\ \bibnamefont {{Shapiro-Albert}}}, \bibinfo {author}
  {\bibfnamefont {X.}~\bibnamefont {{Siemens}}}, \bibinfo {author}
  {\bibfnamefont {J.}~\bibnamefont {{Simon}}}, \bibinfo {author} {\bibfnamefont
  {M.~S.}\ \bibnamefont {{Siwek}}}, \bibinfo {author} {\bibfnamefont {I.~H.}\
  \bibnamefont {{Stairs}}}, \bibinfo {author} {\bibfnamefont {D.~R.}\
  \bibnamefont {{Stinebring}}}, \bibinfo {author} {\bibfnamefont
  {K.}~\bibnamefont {{Stovall}}}, \bibinfo {author} {\bibfnamefont {J.~P.}\
  \bibnamefont {{Sun}}}, \bibinfo {author} {\bibfnamefont {A.}~\bibnamefont
  {{Susobhanan}}}, \bibinfo {author} {\bibfnamefont {J.~K.}\ \bibnamefont
  {{Swiggum}}}, \bibinfo {author} {\bibfnamefont {J.}~\bibnamefont {{Taylor}}},
  \bibinfo {author} {\bibfnamefont {S.~R.}\ \bibnamefont {{Taylor}}}, \bibinfo
  {author} {\bibfnamefont {J.~E.}\ \bibnamefont {{Turner}}}, \bibinfo {author}
  {\bibfnamefont {C.}~\bibnamefont {{Unal}}}, \bibinfo {author} {\bibfnamefont
  {M.}~\bibnamefont {{Vallisneri}}}, \bibinfo {author} {\bibfnamefont {S.~J.}\
  \bibnamefont {{Vigeland}}}, \bibinfo {author} {\bibfnamefont {J.~M.}\
  \bibnamefont {{Wachter}}}, \bibinfo {author} {\bibfnamefont {H.~M.}\
  \bibnamefont {{Wahl}}}, \bibinfo {author} {\bibfnamefont {Q.}~\bibnamefont
  {{Wang}}}, \bibinfo {author} {\bibfnamefont {C.~A.}\ \bibnamefont {{Witt}}},
  \bibinfo {author} {\bibfnamefont {D.}~\bibnamefont {{Wright}}}, \bibinfo
  {author} {\bibfnamefont {O.}~\bibnamefont {{Young}}},\ and\ \bibinfo {author}
  {\bibnamefont {{Nanograv Collaboration}}},\ }\href
  {https://doi.org/10.3847/2041-8213/ace18b} {\bibfield  {journal} {\bibinfo
  {journal} {Astrophys. J. Lett.}\ }\textbf {\bibinfo {volume} {952}},\
  \bibinfo {eid} {L37} (\bibinfo {year} {2023})},\ \Eprint
  {https://arxiv.org/abs/2306.16220} {arXiv:2306.16220 [astro-ph.HE]}
  \BibitemShut {NoStop}%
\bibitem [{\citenamefont {{Amaro-Seoane}}\ \emph {et~al.}(2017)\citenamefont
  {{Amaro-Seoane}}, \citenamefont {{Audley}}, \citenamefont {{Babak}},
  \citenamefont {{Baker}}, \citenamefont {{Barausse}}, \citenamefont
  {{Bender}}, \citenamefont {{Berti}}, \citenamefont {{Binetruy}},
  \citenamefont {{Born}}, \citenamefont {{Bortoluzzi}}, \citenamefont {{Camp}},
  \citenamefont {{Caprini}}, \citenamefont {{Cardoso}}, \citenamefont
  {{Colpi}}, \citenamefont {{Conklin}}, \citenamefont {{Cornish}},
  \citenamefont {{Cutler}}, \citenamefont {{Danzmann}}, \citenamefont
  {{Dolesi}}, \citenamefont {{Ferraioli}}, \citenamefont {{Ferroni}},
  \citenamefont {{Fitzsimons}}, \citenamefont {{Gair}}, \citenamefont {{Gesa
  Bote}}, \citenamefont {{Giardini}}, \citenamefont {{Gibert}}, \citenamefont
  {{Grimani}}, \citenamefont {{Halloin}}, \citenamefont {{Heinzel}},
  \citenamefont {{Hertog}}, \citenamefont {{Hewitson}}, \citenamefont
  {{Holley-Bockelmann}}, \citenamefont {{Hollington}}, \citenamefont
  {{Hueller}}, \citenamefont {{Inchauspe}}, \citenamefont {{Jetzer}},
  \citenamefont {{Karnesis}}, \citenamefont {{Killow}}, \citenamefont
  {{Klein}}, \citenamefont {{Klipstein}}, \citenamefont {{Korsakova}},
  \citenamefont {{Larson}}, \citenamefont {{Livas}}, \citenamefont {{Lloro}},
  \citenamefont {{Man}}, \citenamefont {{Mance}}, \citenamefont {{Martino}},
  \citenamefont {{Mateos}}, \citenamefont {{McKenzie}}, \citenamefont
  {{McWilliams}}, \citenamefont {{Miller}}, \citenamefont {{Mueller}},
  \citenamefont {{Nardini}}, \citenamefont {{Nelemans}}, \citenamefont
  {{Nofrarias}}, \citenamefont {{Petiteau}}, \citenamefont {{Pivato}},
  \citenamefont {{Plagnol}}, \citenamefont {{Porter}}, \citenamefont
  {{Reiche}}, \citenamefont {{Robertson}}, \citenamefont {{Robertson}},
  \citenamefont {{Rossi}}, \citenamefont {{Russano}}, \citenamefont {{Schutz}},
  \citenamefont {{Sesana}}, \citenamefont {{Shoemaker}}, \citenamefont
  {{Slutsky}}, \citenamefont {{Sopuerta}}, \citenamefont {{Sumner}},
  \citenamefont {{Tamanini}}, \citenamefont {{Thorpe}}, \citenamefont
  {{Troebs}}, \citenamefont {{Vallisneri}}, \citenamefont {{Vecchio}},
  \citenamefont {{Vetrugno}}, \citenamefont {{Vitale}}, \citenamefont
  {{Volonteri}}, \citenamefont {{Wanner}}, \citenamefont {{Ward}},
  \citenamefont {{Wass}}, \citenamefont {{Weber}}, \citenamefont {{Ziemer}},\
  and\ \citenamefont {{Zweifel}}}]{LISA2017}%
  \BibitemOpen
  \bibfield  {author} {\bibinfo {author} {\bibfnamefont {P.}~\bibnamefont
  {{Amaro-Seoane}}}, \bibinfo {author} {\bibfnamefont {H.}~\bibnamefont
  {{Audley}}}, \bibinfo {author} {\bibfnamefont {S.}~\bibnamefont {{Babak}}},
  \bibinfo {author} {\bibfnamefont {J.}~\bibnamefont {{Baker}}}, \bibinfo
  {author} {\bibfnamefont {E.}~\bibnamefont {{Barausse}}}, \bibinfo {author}
  {\bibfnamefont {P.}~\bibnamefont {{Bender}}}, \bibinfo {author}
  {\bibfnamefont {E.}~\bibnamefont {{Berti}}}, \bibinfo {author} {\bibfnamefont
  {P.}~\bibnamefont {{Binetruy}}}, \bibinfo {author} {\bibfnamefont
  {M.}~\bibnamefont {{Born}}}, \bibinfo {author} {\bibfnamefont
  {D.}~\bibnamefont {{Bortoluzzi}}}, \bibinfo {author} {\bibfnamefont
  {J.}~\bibnamefont {{Camp}}}, \bibinfo {author} {\bibfnamefont
  {C.}~\bibnamefont {{Caprini}}}, \bibinfo {author} {\bibfnamefont
  {V.}~\bibnamefont {{Cardoso}}}, \bibinfo {author} {\bibfnamefont
  {M.}~\bibnamefont {{Colpi}}}, \bibinfo {author} {\bibfnamefont
  {J.}~\bibnamefont {{Conklin}}}, \bibinfo {author} {\bibfnamefont
  {N.}~\bibnamefont {{Cornish}}}, \bibinfo {author} {\bibfnamefont
  {C.}~\bibnamefont {{Cutler}}}, \bibinfo {author} {\bibfnamefont
  {K.}~\bibnamefont {{Danzmann}}}, \bibinfo {author} {\bibfnamefont
  {R.}~\bibnamefont {{Dolesi}}}, \bibinfo {author} {\bibfnamefont
  {L.}~\bibnamefont {{Ferraioli}}}, \bibinfo {author} {\bibfnamefont
  {V.}~\bibnamefont {{Ferroni}}}, \bibinfo {author} {\bibfnamefont
  {E.}~\bibnamefont {{Fitzsimons}}}, \bibinfo {author} {\bibfnamefont
  {J.}~\bibnamefont {{Gair}}}, \bibinfo {author} {\bibfnamefont
  {L.}~\bibnamefont {{Gesa Bote}}}, \bibinfo {author} {\bibfnamefont
  {D.}~\bibnamefont {{Giardini}}}, \bibinfo {author} {\bibfnamefont
  {F.}~\bibnamefont {{Gibert}}}, \bibinfo {author} {\bibfnamefont
  {C.}~\bibnamefont {{Grimani}}}, \bibinfo {author} {\bibfnamefont
  {H.}~\bibnamefont {{Halloin}}}, \bibinfo {author} {\bibfnamefont
  {G.}~\bibnamefont {{Heinzel}}}, \bibinfo {author} {\bibfnamefont
  {T.}~\bibnamefont {{Hertog}}}, \bibinfo {author} {\bibfnamefont
  {M.}~\bibnamefont {{Hewitson}}}, \bibinfo {author} {\bibfnamefont
  {K.}~\bibnamefont {{Holley-Bockelmann}}}, \bibinfo {author} {\bibfnamefont
  {D.}~\bibnamefont {{Hollington}}}, \bibinfo {author} {\bibfnamefont
  {M.}~\bibnamefont {{Hueller}}}, \bibinfo {author} {\bibfnamefont
  {H.}~\bibnamefont {{Inchauspe}}}, \bibinfo {author} {\bibfnamefont
  {P.}~\bibnamefont {{Jetzer}}}, \bibinfo {author} {\bibfnamefont
  {N.}~\bibnamefont {{Karnesis}}}, \bibinfo {author} {\bibfnamefont
  {C.}~\bibnamefont {{Killow}}}, \bibinfo {author} {\bibfnamefont
  {A.}~\bibnamefont {{Klein}}}, \bibinfo {author} {\bibfnamefont
  {B.}~\bibnamefont {{Klipstein}}}, \bibinfo {author} {\bibfnamefont
  {N.}~\bibnamefont {{Korsakova}}}, \bibinfo {author} {\bibfnamefont {S.~L.}\
  \bibnamefont {{Larson}}}, \bibinfo {author} {\bibfnamefont {J.}~\bibnamefont
  {{Livas}}}, \bibinfo {author} {\bibfnamefont {I.}~\bibnamefont {{Lloro}}},
  \bibinfo {author} {\bibfnamefont {N.}~\bibnamefont {{Man}}}, \bibinfo
  {author} {\bibfnamefont {D.}~\bibnamefont {{Mance}}}, \bibinfo {author}
  {\bibfnamefont {J.}~\bibnamefont {{Martino}}}, \bibinfo {author}
  {\bibfnamefont {I.}~\bibnamefont {{Mateos}}}, \bibinfo {author}
  {\bibfnamefont {K.}~\bibnamefont {{McKenzie}}}, \bibinfo {author}
  {\bibfnamefont {S.~T.}\ \bibnamefont {{McWilliams}}}, \bibinfo {author}
  {\bibfnamefont {C.}~\bibnamefont {{Miller}}}, \bibinfo {author}
  {\bibfnamefont {G.}~\bibnamefont {{Mueller}}}, \bibinfo {author}
  {\bibfnamefont {G.}~\bibnamefont {{Nardini}}}, \bibinfo {author}
  {\bibfnamefont {G.}~\bibnamefont {{Nelemans}}}, \bibinfo {author}
  {\bibfnamefont {M.}~\bibnamefont {{Nofrarias}}}, \bibinfo {author}
  {\bibfnamefont {A.}~\bibnamefont {{Petiteau}}}, \bibinfo {author}
  {\bibfnamefont {P.}~\bibnamefont {{Pivato}}}, \bibinfo {author}
  {\bibfnamefont {E.}~\bibnamefont {{Plagnol}}}, \bibinfo {author}
  {\bibfnamefont {E.}~\bibnamefont {{Porter}}}, \bibinfo {author}
  {\bibfnamefont {J.}~\bibnamefont {{Reiche}}}, \bibinfo {author}
  {\bibfnamefont {D.}~\bibnamefont {{Robertson}}}, \bibinfo {author}
  {\bibfnamefont {N.}~\bibnamefont {{Robertson}}}, \bibinfo {author}
  {\bibfnamefont {E.}~\bibnamefont {{Rossi}}}, \bibinfo {author} {\bibfnamefont
  {G.}~\bibnamefont {{Russano}}}, \bibinfo {author} {\bibfnamefont
  {B.}~\bibnamefont {{Schutz}}}, \bibinfo {author} {\bibfnamefont
  {A.}~\bibnamefont {{Sesana}}}, \bibinfo {author} {\bibfnamefont
  {D.}~\bibnamefont {{Shoemaker}}}, \bibinfo {author} {\bibfnamefont
  {J.}~\bibnamefont {{Slutsky}}}, \bibinfo {author} {\bibfnamefont {C.~F.}\
  \bibnamefont {{Sopuerta}}}, \bibinfo {author} {\bibfnamefont
  {T.}~\bibnamefont {{Sumner}}}, \bibinfo {author} {\bibfnamefont
  {N.}~\bibnamefont {{Tamanini}}}, \bibinfo {author} {\bibfnamefont
  {I.}~\bibnamefont {{Thorpe}}}, \bibinfo {author} {\bibfnamefont
  {M.}~\bibnamefont {{Troebs}}}, \bibinfo {author} {\bibfnamefont
  {M.}~\bibnamefont {{Vallisneri}}}, \bibinfo {author} {\bibfnamefont
  {A.}~\bibnamefont {{Vecchio}}}, \bibinfo {author} {\bibfnamefont
  {D.}~\bibnamefont {{Vetrugno}}}, \bibinfo {author} {\bibfnamefont
  {S.}~\bibnamefont {{Vitale}}}, \bibinfo {author} {\bibfnamefont
  {M.}~\bibnamefont {{Volonteri}}}, \bibinfo {author} {\bibfnamefont
  {G.}~\bibnamefont {{Wanner}}}, \bibinfo {author} {\bibfnamefont
  {H.}~\bibnamefont {{Ward}}}, \bibinfo {author} {\bibfnamefont
  {P.}~\bibnamefont {{Wass}}}, \bibinfo {author} {\bibfnamefont
  {W.}~\bibnamefont {{Weber}}}, \bibinfo {author} {\bibfnamefont
  {J.}~\bibnamefont {{Ziemer}}},\ and\ \bibinfo {author} {\bibfnamefont
  {P.}~\bibnamefont {{Zweifel}}},\ }\href@noop {} {\bibfield  {journal}
  {\bibinfo  {journal} {arXiv e-prints}\ ,\ \bibinfo {eid} {arXiv:1702.00786}}
  (\bibinfo {year} {2017})},\ \Eprint {https://arxiv.org/abs/1702.00786}
  {arXiv:1702.00786 [astro-ph.IM]} \BibitemShut {NoStop}%
\bibitem [{\citenamefont {Graham}\ \emph {et~al.}(2015)\citenamefont {Graham},
  \citenamefont {Djorgovski}, \citenamefont {Stern}, \citenamefont {Drake},
  \citenamefont {Mahabal}, \citenamefont {Donalek}, \citenamefont {Glikman},
  \citenamefont {Larson},\ and\ \citenamefont
  {Christensen}}]{graham2015systematic}%
  \BibitemOpen
  \bibfield  {author} {\bibinfo {author} {\bibfnamefont {M.~J.}\ \bibnamefont
  {Graham}}, \bibinfo {author} {\bibfnamefont {S.}~\bibnamefont {Djorgovski}},
  \bibinfo {author} {\bibfnamefont {D.}~\bibnamefont {Stern}}, \bibinfo
  {author} {\bibfnamefont {A.~J.}\ \bibnamefont {Drake}}, \bibinfo {author}
  {\bibfnamefont {A.~A.}\ \bibnamefont {Mahabal}}, \bibinfo {author}
  {\bibfnamefont {C.}~\bibnamefont {Donalek}}, \bibinfo {author} {\bibfnamefont
  {E.}~\bibnamefont {Glikman}}, \bibinfo {author} {\bibfnamefont
  {S.}~\bibnamefont {Larson}},\ and\ \bibinfo {author} {\bibfnamefont
  {E.}~\bibnamefont {Christensen}},\ }\href@noop {} {\bibfield  {journal}
  {\bibinfo  {journal} {Mon. Not. R. Astron. Soc.}\ }\textbf {\bibinfo {volume}
  {453}},\ \bibinfo {pages} {1562} (\bibinfo {year} {2015})}\BibitemShut
  {NoStop}%
\bibitem [{\citenamefont {{Charisi}}\ \emph {et~al.}(2016)\citenamefont
  {{Charisi}}, \citenamefont {{Bartos}}, \citenamefont {{Haiman}},
  \citenamefont {{Price-Whelan}}, \citenamefont {{Graham}}, \citenamefont
  {{Bellm}}, \citenamefont {{Laher}},\ and\ \citenamefont
  {{M{\'a}rka}}}]{Charisi2016}%
  \BibitemOpen
  \bibfield  {author} {\bibinfo {author} {\bibfnamefont {M.}~\bibnamefont
  {{Charisi}}}, \bibinfo {author} {\bibfnamefont {I.}~\bibnamefont {{Bartos}}},
  \bibinfo {author} {\bibfnamefont {Z.}~\bibnamefont {{Haiman}}}, \bibinfo
  {author} {\bibfnamefont {A.~M.}\ \bibnamefont {{Price-Whelan}}}, \bibinfo
  {author} {\bibfnamefont {M.~J.}\ \bibnamefont {{Graham}}}, \bibinfo {author}
  {\bibfnamefont {E.~C.}\ \bibnamefont {{Bellm}}}, \bibinfo {author}
  {\bibfnamefont {R.~R.}\ \bibnamefont {{Laher}}},\ and\ \bibinfo {author}
  {\bibfnamefont {S.}~\bibnamefont {{M{\'a}rka}}},\ }\href
  {https://doi.org/10.1093/mnras/stw1838} {\bibfield  {journal} {\bibinfo
  {journal} {Mon. Not. R. Astron. Soc.}\ }\textbf {\bibinfo {volume} {463}},\
  \bibinfo {pages} {2145} (\bibinfo {year} {2016})},\ \Eprint
  {https://arxiv.org/abs/1604.01020} {arXiv:1604.01020 [astro-ph.GA]}
  \BibitemShut {NoStop}%
\bibitem [{\citenamefont {Roedig}\ \emph {et~al.}(2014)\citenamefont {Roedig},
  \citenamefont {Krolik},\ and\ \citenamefont {Miller}}]{Roedig2014}%
  \BibitemOpen
  \bibfield  {author} {\bibinfo {author} {\bibfnamefont {C.}~\bibnamefont
  {Roedig}}, \bibinfo {author} {\bibfnamefont {J.~H.}\ \bibnamefont {Krolik}},\
  and\ \bibinfo {author} {\bibfnamefont {M.~C.}\ \bibnamefont {Miller}},\
  }\href {https://doi.org/10.1088/0004-637x/785/2/115} {\bibfield  {journal}
  {\bibinfo  {journal} {Astrophys.\ J.}\ }\textbf {\bibinfo {volume} {785}},\
  \bibinfo {pages} {115} (\bibinfo {year} {2014})}\BibitemShut {NoStop}%
\bibitem [{\citenamefont {{MacFadyen}}\ and\ \citenamefont
  {{Milosavljevi{\'c}}}(2008)}]{MacFadyen2008}%
  \BibitemOpen
  \bibfield  {author} {\bibinfo {author} {\bibfnamefont {A.~I.}\ \bibnamefont
  {{MacFadyen}}}\ and\ \bibinfo {author} {\bibfnamefont {M.}~\bibnamefont
  {{Milosavljevi{\'c}}}},\ }\href {https://doi.org/10.1086/523869} {\bibfield
  {journal} {\bibinfo  {journal} {{Astrophys. J.}}\ }\textbf {\bibinfo {volume}
  {672}},\ \bibinfo {pages} {83} (\bibinfo {year} {2008})},\ \Eprint
  {https://arxiv.org/abs/astro-ph/0607467} {arXiv:astro-ph/0607467 [astro-ph]}
  \BibitemShut {NoStop}%
\bibitem [{\citenamefont {Noble}\ \emph
  {et~al.}(2012{\natexlab{a}})\citenamefont {Noble}, \citenamefont {Mundim},
  \citenamefont {Nakano}, \citenamefont {Krolik}, \citenamefont {Campanelli},
  \citenamefont {Zlochower},\ and\ \citenamefont {Yunes}}]{Noble12}%
  \BibitemOpen
  \bibfield  {author} {\bibinfo {author} {\bibfnamefont {S.~C.}\ \bibnamefont
  {Noble}}, \bibinfo {author} {\bibfnamefont {B.~C.}\ \bibnamefont {Mundim}},
  \bibinfo {author} {\bibfnamefont {H.}~\bibnamefont {Nakano}}, \bibinfo
  {author} {\bibfnamefont {J.~H.}\ \bibnamefont {Krolik}}, \bibinfo {author}
  {\bibfnamefont {M.}~\bibnamefont {Campanelli}}, \bibinfo {author}
  {\bibfnamefont {Y.}~\bibnamefont {Zlochower}},\ and\ \bibinfo {author}
  {\bibfnamefont {N.}~\bibnamefont {Yunes}},\ }\href
  {http://stacks.iop.org/0004-637X/755/i=1/a=51} {\bibfield  {journal}
  {\bibinfo  {journal} {Astrophys.\ J.}\ }\textbf {\bibinfo {volume} {755}},\
  \bibinfo {pages} {51} (\bibinfo {year} {2012}{\natexlab{a}})}\BibitemShut
  {NoStop}%
\bibitem [{\citenamefont {{Shi}}\ \emph {et~al.}(2012)\citenamefont {{Shi}},
  \citenamefont {{Krolik}}, \citenamefont {{Lubow}},\ and\ \citenamefont
  {{Hawley}}}]{Shi:2012ApJ}%
  \BibitemOpen
  \bibfield  {author} {\bibinfo {author} {\bibfnamefont {J.-M.}\ \bibnamefont
  {{Shi}}}, \bibinfo {author} {\bibfnamefont {J.~H.}\ \bibnamefont {{Krolik}}},
  \bibinfo {author} {\bibfnamefont {S.~H.}\ \bibnamefont {{Lubow}}},\ and\
  \bibinfo {author} {\bibfnamefont {J.~F.}\ \bibnamefont {{Hawley}}},\ }\href
  {https://doi.org/10.1088/0004-637X/749/2/118} {\bibfield  {journal} {\bibinfo
   {journal} {{Astrophys. J.}}\ }\textbf {\bibinfo {volume} {749}},\ \bibinfo
  {eid} {118} (\bibinfo {year} {2012})},\ \Eprint
  {https://arxiv.org/abs/1110.4866} {arXiv:1110.4866 [astro-ph.HE]}
  \BibitemShut {NoStop}%
\bibitem [{\citenamefont {{Bowen}}\ \emph {et~al.}(2017)\citenamefont
  {{Bowen}}, \citenamefont {{Campanelli}}, \citenamefont {{Krolik}},
  \citenamefont {{Mewes}},\ and\ \citenamefont {{Noble}}}]{bowen2017}%
  \BibitemOpen
  \bibfield  {author} {\bibinfo {author} {\bibfnamefont {D.~B.}\ \bibnamefont
  {{Bowen}}}, \bibinfo {author} {\bibfnamefont {M.}~\bibnamefont
  {{Campanelli}}}, \bibinfo {author} {\bibfnamefont {J.~H.}\ \bibnamefont
  {{Krolik}}}, \bibinfo {author} {\bibfnamefont {V.}~\bibnamefont {{Mewes}}},\
  and\ \bibinfo {author} {\bibfnamefont {S.~C.}\ \bibnamefont {{Noble}}},\
  }\href {https://doi.org/10.3847/1538-4357/aa63f3} {\bibfield  {journal}
  {\bibinfo  {journal} {{Astrophys. J.}}\ }\textbf {\bibinfo {volume} {838}},\
  \bibinfo {eid} {42} (\bibinfo {year} {2017})},\ \Eprint
  {https://arxiv.org/abs/1612.02373} {arXiv:1612.02373 [astro-ph.HE]}
  \BibitemShut {NoStop}%
\bibitem [{\citenamefont {{Ryan}}\ and\ \citenamefont
  {{MacFadyen}}(2017)}]{RyanMacFadyen17}%
  \BibitemOpen
  \bibfield  {author} {\bibinfo {author} {\bibfnamefont {G.}~\bibnamefont
  {{Ryan}}}\ and\ \bibinfo {author} {\bibfnamefont {A.}~\bibnamefont
  {{MacFadyen}}},\ }\href {https://doi.org/10.3847/1538-4357/835/2/199}
  {\bibfield  {journal} {\bibinfo  {journal} {{Astrophys. J.}}\ }\textbf
  {\bibinfo {volume} {835}},\ \bibinfo {eid} {199} (\bibinfo {year} {2017})},\
  \Eprint {https://arxiv.org/abs/1611.00341} {arXiv:1611.00341 [astro-ph.HE]}
  \BibitemShut {NoStop}%
\bibitem [{\citenamefont {{Westernacher-Schneider}}\ \emph
  {et~al.}(2022)\citenamefont {{Westernacher-Schneider}}, \citenamefont
  {{Zrake}}, \citenamefont {{MacFadyen}},\ and\ \citenamefont
  {{Haiman}}}]{westernacher2021}%
  \BibitemOpen
  \bibfield  {author} {\bibinfo {author} {\bibfnamefont {J.~R.}\ \bibnamefont
  {{Westernacher-Schneider}}}, \bibinfo {author} {\bibfnamefont
  {J.}~\bibnamefont {{Zrake}}}, \bibinfo {author} {\bibfnamefont
  {A.}~\bibnamefont {{MacFadyen}}},\ and\ \bibinfo {author} {\bibfnamefont
  {Z.}~\bibnamefont {{Haiman}}},\ }\href
  {https://doi.org/10.1103/PhysRevD.106.103010} {\bibfield  {journal} {\bibinfo
   {journal} {\prd}\ }\textbf {\bibinfo {volume} {106}},\ \bibinfo {eid}
  {103010} (\bibinfo {year} {2022})},\ \Eprint
  {https://arxiv.org/abs/2111.06882} {arXiv:2111.06882 [astro-ph.HE]}
  \BibitemShut {NoStop}%
\bibitem [{\citenamefont {{D'Orazio}}\ \emph {et~al.}(2013)\citenamefont
  {{D'Orazio}}, \citenamefont {{Haiman}},\ and\ \citenamefont
  {{MacFadyen}}}]{DOrazio13}%
  \BibitemOpen
  \bibfield  {author} {\bibinfo {author} {\bibfnamefont {D.~J.}\ \bibnamefont
  {{D'Orazio}}}, \bibinfo {author} {\bibfnamefont {Z.}~\bibnamefont
  {{Haiman}}},\ and\ \bibinfo {author} {\bibfnamefont {A.}~\bibnamefont
  {{MacFadyen}}},\ }\href {https://doi.org/10.1093/mnras/stt1787} {\bibfield
  {journal} {\bibinfo  {journal} {Mon. Not. R. Astron. Soc.}\ }\textbf
  {\bibinfo {volume} {436}},\ \bibinfo {pages} {2997} (\bibinfo {year}
  {2013})},\ \Eprint {https://arxiv.org/abs/1210.0536} {arXiv:1210.0536
  [astro-ph.GA]} \BibitemShut {NoStop}%
\bibitem [{\citenamefont {{Komossa}}(2006)}]{Komossa2006}%
  \BibitemOpen
  \bibfield  {author} {\bibinfo {author} {\bibfnamefont {S.}~\bibnamefont
  {{Komossa}}},\ }\href@noop {} {\bibfield  {journal} {\bibinfo  {journal}
  {MmSAI}\ }\textbf {\bibinfo {volume} {77}},\ \bibinfo {pages} {733} (\bibinfo
  {year} {2006})}\BibitemShut {NoStop}%
\bibitem [{\citenamefont {{Sukov{\'a}}}\ \emph {et~al.}(2021)\citenamefont
  {{Sukov{\'a}}}, \citenamefont {{Zaja{\v{c}}ek}}, \citenamefont {{Witzany}},\
  and\ \citenamefont {{Karas}}}]{Sukova2021}%
  \BibitemOpen
  \bibfield  {author} {\bibinfo {author} {\bibfnamefont {P.}~\bibnamefont
  {{Sukov{\'a}}}}, \bibinfo {author} {\bibfnamefont {M.}~\bibnamefont
  {{Zaja{\v{c}}ek}}}, \bibinfo {author} {\bibfnamefont {V.}~\bibnamefont
  {{Witzany}}},\ and\ \bibinfo {author} {\bibfnamefont {V.}~\bibnamefont
  {{Karas}}},\ }\href {https://doi.org/10.3847/1538-4357/ac05c6} {\bibfield
  {journal} {\bibinfo  {journal} {Astrophys. J.}\ }\textbf {\bibinfo {volume}
  {917}},\ \bibinfo {eid} {43} (\bibinfo {year} {2021})},\ \Eprint
  {https://arxiv.org/abs/2102.08135} {arXiv:2102.08135 [astro-ph.HE]}
  \BibitemShut {NoStop}%
\bibitem [{\citenamefont {{Lehto}}\ and\ \citenamefont
  {{Valtonen}}(1996)}]{Lehto1996}%
  \BibitemOpen
  \bibfield  {author} {\bibinfo {author} {\bibfnamefont {H.~J.}\ \bibnamefont
  {{Lehto}}}\ and\ \bibinfo {author} {\bibfnamefont {M.~J.}\ \bibnamefont
  {{Valtonen}}},\ }\href {https://doi.org/10.1086/176962} {\bibfield  {journal}
  {\bibinfo  {journal} {Astrophys. J.}\ }\textbf {\bibinfo {volume} {460}},\
  \bibinfo {pages} {207} (\bibinfo {year} {1996})}\BibitemShut {NoStop}%
\bibitem [{\citenamefont {Britzen}\ \emph {et~al.}(2023)\citenamefont
  {Britzen}, \citenamefont {Zaja{\v{c}}ek}, \citenamefont {Fendt},
  \citenamefont {Kun}, \citenamefont {Jaron}, \citenamefont
  {Sillanp{\"a}{\"a}}, \citenamefont {Eckart} \emph
  {et~al.}}]{britzen2023precession}%
  \BibitemOpen
  \bibfield  {author} {\bibinfo {author} {\bibfnamefont {S.}~\bibnamefont
  {Britzen}}, \bibinfo {author} {\bibfnamefont {M.}~\bibnamefont
  {Zaja{\v{c}}ek}}, \bibinfo {author} {\bibfnamefont {C.}~\bibnamefont
  {Fendt}}, \bibinfo {author} {\bibfnamefont {E.}~\bibnamefont {Kun}}, \bibinfo
  {author} {\bibfnamefont {F.}~\bibnamefont {Jaron}}, \bibinfo {author}
  {\bibfnamefont {A.}~\bibnamefont {Sillanp{\"a}{\"a}}}, \bibinfo {author}
  {\bibfnamefont {A.}~\bibnamefont {Eckart}}, \emph {et~al.},\ }\href@noop {}
  {\bibfield  {journal} {\bibinfo  {journal} {Astrophys. J.}\ }\textbf
  {\bibinfo {volume} {951}},\ \bibinfo {pages} {106} (\bibinfo {year}
  {2023})}\BibitemShut {NoStop}%
\bibitem [{\citenamefont {{Dittmann}}\ \emph {et~al.}(2024)\citenamefont
  {{Dittmann}}, \citenamefont {{Dempsey}},\ and\ \citenamefont
  {{Li}}}]{dittmann2023evolution}%
  \BibitemOpen
  \bibfield  {author} {\bibinfo {author} {\bibfnamefont {A.~J.}\ \bibnamefont
  {{Dittmann}}}, \bibinfo {author} {\bibfnamefont {A.~M.}\ \bibnamefont
  {{Dempsey}}},\ and\ \bibinfo {author} {\bibfnamefont {H.}~\bibnamefont
  {{Li}}},\ }\href {https://doi.org/10.3847/1538-4357/ad23ce} {\bibfield
  {journal} {\bibinfo  {journal} {{Astrophys. J.}}\ }\textbf {\bibinfo {volume}
  {964}},\ \bibinfo {eid} {61} (\bibinfo {year} {2024})},\ \Eprint
  {https://arxiv.org/abs/2310.03832} {arXiv:2310.03832 [astro-ph.HE]}
  \BibitemShut {NoStop}%
\bibitem [{\citenamefont {Franchini}\ \emph {et~al.}(2023)\citenamefont
  {Franchini}, \citenamefont {Bonetti}, \citenamefont {Lupi}, \citenamefont
  {Miniutti}, \citenamefont {Bortolas}, \citenamefont {Giustini}, \citenamefont
  {Dotti}, \citenamefont {Sesana}, \citenamefont {Arcodia},\ and\ \citenamefont
  {Ryu}}]{franchini2023quasi}%
  \BibitemOpen
  \bibfield  {author} {\bibinfo {author} {\bibfnamefont {A.}~\bibnamefont
  {Franchini}}, \bibinfo {author} {\bibfnamefont {M.}~\bibnamefont {Bonetti}},
  \bibinfo {author} {\bibfnamefont {A.}~\bibnamefont {Lupi}}, \bibinfo {author}
  {\bibfnamefont {G.}~\bibnamefont {Miniutti}}, \bibinfo {author}
  {\bibfnamefont {E.}~\bibnamefont {Bortolas}}, \bibinfo {author}
  {\bibfnamefont {M.}~\bibnamefont {Giustini}}, \bibinfo {author}
  {\bibfnamefont {M.}~\bibnamefont {Dotti}}, \bibinfo {author} {\bibfnamefont
  {A.}~\bibnamefont {Sesana}}, \bibinfo {author} {\bibfnamefont
  {R.}~\bibnamefont {Arcodia}},\ and\ \bibinfo {author} {\bibfnamefont
  {T.}~\bibnamefont {Ryu}},\ }\href@noop {} {\bibfield  {journal} {\bibinfo
  {journal} {Astronomy \& Astrophysics}\ }\textbf {\bibinfo {volume} {675}},\
  \bibinfo {pages} {A100} (\bibinfo {year} {2023})}\BibitemShut {NoStop}%
\bibitem [{\citenamefont {Mu{\~{n}}oz}\ \emph {et~al.}(2020)\citenamefont
  {Mu{\~{n}}oz}, \citenamefont {Lai}, \citenamefont {Kratter},\ and\
  \citenamefont {Miranda}}]{Munoz2020}%
  \BibitemOpen
  \bibfield  {author} {\bibinfo {author} {\bibfnamefont {D.~J.}\ \bibnamefont
  {Mu{\~{n}}oz}}, \bibinfo {author} {\bibfnamefont {D.}~\bibnamefont {Lai}},
  \bibinfo {author} {\bibfnamefont {K.}~\bibnamefont {Kratter}},\ and\ \bibinfo
  {author} {\bibfnamefont {R.}~\bibnamefont {Miranda}},\ }\href
  {https://doi.org/10.3847/1538-4357/ab5d33} {\bibfield  {journal} {\bibinfo
  {journal} {Astrophys.\ J.}\ }\textbf {\bibinfo {volume} {889}},\ \bibinfo
  {pages} {114} (\bibinfo {year} {2020})}\BibitemShut {NoStop}%
\bibitem [{\citenamefont {{Miranda}}\ \emph {et~al.}(2017)\citenamefont
  {{Miranda}}, \citenamefont {{Mu{\~n}oz}},\ and\ \citenamefont
  {{Lai}}}]{Miranda2017}%
  \BibitemOpen
  \bibfield  {author} {\bibinfo {author} {\bibfnamefont {R.}~\bibnamefont
  {{Miranda}}}, \bibinfo {author} {\bibfnamefont {D.~J.}\ \bibnamefont
  {{Mu{\~n}oz}}},\ and\ \bibinfo {author} {\bibfnamefont {D.}~\bibnamefont
  {{Lai}}},\ }\href {https://doi.org/10.1093/mnras/stw3189} {\bibfield
  {journal} {\bibinfo  {journal} {Mon. Not. R. Astron. Soc.}\ }\textbf
  {\bibinfo {volume} {466}},\ \bibinfo {pages} {1170} (\bibinfo {year}
  {2017})},\ \Eprint {https://arxiv.org/abs/1610.07263} {arXiv:1610.07263
  [astro-ph.SR]} \BibitemShut {NoStop}%
\bibitem [{\citenamefont {{Duffell}}\ \emph {et~al.}(2024)\citenamefont
  {{Duffell}}, \citenamefont {{Dittmann}}, \citenamefont {{D'Orazio}},
  \citenamefont {{Franchini}}, \citenamefont {{Kratter}}, \citenamefont
  {{Penzlin}}, \citenamefont {{Ragusa}}, \citenamefont {{Siwek}}, \citenamefont
  {{Tiede}}, \citenamefont {{Wang}}, \citenamefont {{Zrake}}, \citenamefont
  {{Dempsey}}, \citenamefont {{Haiman}}, \citenamefont {{Lupi}}, \citenamefont
  {{Pirog}},\ and\ \citenamefont {{Ryan}}}]{duffell2024santa}%
  \BibitemOpen
  \bibfield  {author} {\bibinfo {author} {\bibfnamefont {P.~C.}\ \bibnamefont
  {{Duffell}}}, \bibinfo {author} {\bibfnamefont {A.~J.}\ \bibnamefont
  {{Dittmann}}}, \bibinfo {author} {\bibfnamefont {D.~J.}\ \bibnamefont
  {{D'Orazio}}}, \bibinfo {author} {\bibfnamefont {A.}~\bibnamefont
  {{Franchini}}}, \bibinfo {author} {\bibfnamefont {K.~M.}\ \bibnamefont
  {{Kratter}}}, \bibinfo {author} {\bibfnamefont {A.~B.~T.}\ \bibnamefont
  {{Penzlin}}}, \bibinfo {author} {\bibfnamefont {E.}~\bibnamefont {{Ragusa}}},
  \bibinfo {author} {\bibfnamefont {M.}~\bibnamefont {{Siwek}}}, \bibinfo
  {author} {\bibfnamefont {C.}~\bibnamefont {{Tiede}}}, \bibinfo {author}
  {\bibfnamefont {H.}~\bibnamefont {{Wang}}}, \bibinfo {author} {\bibfnamefont
  {J.}~\bibnamefont {{Zrake}}}, \bibinfo {author} {\bibfnamefont {A.~M.}\
  \bibnamefont {{Dempsey}}}, \bibinfo {author} {\bibfnamefont {Z.}~\bibnamefont
  {{Haiman}}}, \bibinfo {author} {\bibfnamefont {A.}~\bibnamefont {{Lupi}}},
  \bibinfo {author} {\bibfnamefont {M.}~\bibnamefont {{Pirog}}},\ and\ \bibinfo
  {author} {\bibfnamefont {G.}~\bibnamefont {{Ryan}}},\ }\href
  {https://doi.org/10.3847/1538-4357/ad5a7e} {\bibfield  {journal} {\bibinfo
  {journal} {{Astrophys. J.}}\ }\textbf {\bibinfo {volume} {970}},\ \bibinfo
  {eid} {156} (\bibinfo {year} {2024})},\ \Eprint
  {https://arxiv.org/abs/2402.13039} {arXiv:2402.13039 [astro-ph.SR]}
  \BibitemShut {NoStop}%
\bibitem [{\citenamefont {Cattorini}\ and\ \citenamefont
  {Giacomazzo}(2023)}]{cattorini2023grmhd}%
  \BibitemOpen
  \bibfield  {author} {\bibinfo {author} {\bibfnamefont {F.}~\bibnamefont
  {Cattorini}}\ and\ \bibinfo {author} {\bibfnamefont {B.}~\bibnamefont
  {Giacomazzo}},\ }\href@noop {} {\bibfield  {journal} {\bibinfo  {journal}
  {Astropart. Phys.}\ ,\ \bibinfo {pages} {102892}} (\bibinfo {year}
  {2023})}\BibitemShut {NoStop}%
\bibitem [{\citenamefont {{Gold}}(2019)}]{Gold2019}%
  \BibitemOpen
  \bibfield  {author} {\bibinfo {author} {\bibfnamefont {R.}~\bibnamefont
  {{Gold}}},\ }\href {https://doi.org/10.3390/galaxies7020063} {\bibfield
  {journal} {\bibinfo  {journal} {Galaxies}\ }\textbf {\bibinfo {volume} {7}},\
  \bibinfo {eid} {63} (\bibinfo {year} {2019})}\BibitemShut {NoStop}%
\bibitem [{\citenamefont {Noble}\ \emph
  {et~al.}(2012{\natexlab{b}})\citenamefont {Noble}, \citenamefont {Mundim},
  \citenamefont {Nakano}, \citenamefont {Krolik}, \citenamefont {Campanelli},
  \citenamefont {Zlochower},\ and\ \citenamefont {Yunes}}]{noble2012x}%
  \BibitemOpen
  \bibfield  {author} {\bibinfo {author} {\bibfnamefont {S.~C.}\ \bibnamefont
  {Noble}}, \bibinfo {author} {\bibfnamefont {B.~C.}\ \bibnamefont {Mundim}},
  \bibinfo {author} {\bibfnamefont {H.}~\bibnamefont {Nakano}}, \bibinfo
  {author} {\bibfnamefont {J.~H.}\ \bibnamefont {Krolik}}, \bibinfo {author}
  {\bibfnamefont {M.}~\bibnamefont {Campanelli}}, \bibinfo {author}
  {\bibfnamefont {Y.}~\bibnamefont {Zlochower}},\ and\ \bibinfo {author}
  {\bibfnamefont {N.}~\bibnamefont {Yunes}},\ }\href
  {https://doi.org/10.1088/0004-637X/755/1/51} {\bibfield  {journal} {\bibinfo
  {journal} {Astrophys. J.}\ }\textbf {\bibinfo {volume} {755}},\ \bibinfo
  {pages} {51} (\bibinfo {year} {2012}{\natexlab{b}})},\ \Eprint
  {https://arxiv.org/abs/1204.1073} {arXiv:1204.1073 [astro-ph.HE]}
  \BibitemShut {NoStop}%
\bibitem [{\citenamefont {{Zilh{\~a}o}}\ \emph {et~al.}(2015)\citenamefont
  {{Zilh{\~a}o}}, \citenamefont {{Noble}}, \citenamefont {{Campanelli}},\ and\
  \citenamefont {{Zlochower}}}]{zilhao2014pn}%
  \BibitemOpen
  \bibfield  {author} {\bibinfo {author} {\bibfnamefont {M.}~\bibnamefont
  {{Zilh{\~a}o}}}, \bibinfo {author} {\bibfnamefont {S.~C.}\ \bibnamefont
  {{Noble}}}, \bibinfo {author} {\bibfnamefont {M.}~\bibnamefont
  {{Campanelli}}},\ and\ \bibinfo {author} {\bibfnamefont {Y.}~\bibnamefont
  {{Zlochower}}},\ }\href {https://doi.org/10.1103/PhysRevD.91.024034}
  {\bibfield  {journal} {\bibinfo  {journal} {\prd}\ }\textbf {\bibinfo
  {volume} {91}},\ \bibinfo {eid} {024034} (\bibinfo {year} {2015})},\ \Eprint
  {https://arxiv.org/abs/1409.4787} {arXiv:1409.4787 [gr-qc]} \BibitemShut
  {NoStop}%
\bibitem [{\citenamefont {{Noble}}\ \emph {et~al.}(2021)\citenamefont
  {{Noble}}, \citenamefont {{Krolik}}, \citenamefont {{Campanelli}},
  \citenamefont {{Zlochower}}, \citenamefont {{Mundim}}, \citenamefont
  {{Nakano}},\ and\ \citenamefont {{Zilh{\~a}o}}}]{noble2021}%
  \BibitemOpen
  \bibfield  {author} {\bibinfo {author} {\bibfnamefont {S.~C.}\ \bibnamefont
  {{Noble}}}, \bibinfo {author} {\bibfnamefont {J.~H.}\ \bibnamefont
  {{Krolik}}}, \bibinfo {author} {\bibfnamefont {M.}~\bibnamefont
  {{Campanelli}}}, \bibinfo {author} {\bibfnamefont {Y.}~\bibnamefont
  {{Zlochower}}}, \bibinfo {author} {\bibfnamefont {B.~C.}\ \bibnamefont
  {{Mundim}}}, \bibinfo {author} {\bibfnamefont {H.}~\bibnamefont {{Nakano}}},\
  and\ \bibinfo {author} {\bibfnamefont {M.}~\bibnamefont {{Zilh{\~a}o}}},\
  }\href {https://doi.org/10.3847/1538-4357/ac2229} {\bibfield  {journal}
  {\bibinfo  {journal} {{Astrophys. J.}}\ }\textbf {\bibinfo {volume} {922}},\
  \bibinfo {eid} {175} (\bibinfo {year} {2021})},\ \Eprint
  {https://arxiv.org/abs/2103.12100} {arXiv:2103.12100 [astro-ph.HE]}
  \BibitemShut {NoStop}%
\bibitem [{\citenamefont {Mundim}\ \emph {et~al.}(2014)\citenamefont {Mundim},
  \citenamefont {Nakano}, \citenamefont {Yunes}, \citenamefont {Campanelli},
  \citenamefont {Noble},\ and\ \citenamefont {Zlochower}}]{Mundim2014}%
  \BibitemOpen
  \bibfield  {author} {\bibinfo {author} {\bibfnamefont {B.~C.}\ \bibnamefont
  {Mundim}}, \bibinfo {author} {\bibfnamefont {H.}~\bibnamefont {Nakano}},
  \bibinfo {author} {\bibfnamefont {N.}~\bibnamefont {Yunes}}, \bibinfo
  {author} {\bibfnamefont {M.}~\bibnamefont {Campanelli}}, \bibinfo {author}
  {\bibfnamefont {S.~C.}\ \bibnamefont {Noble}},\ and\ \bibinfo {author}
  {\bibfnamefont {Y.}~\bibnamefont {Zlochower}},\ }\href
  {https://doi.org/10.1103/PhysRevD.89.084008} {\bibfield  {journal} {\bibinfo
  {journal} {Phys. Rev. D}\ }\textbf {\bibinfo {volume} {89}},\ \bibinfo
  {pages} {084008} (\bibinfo {year} {2014})}\BibitemShut {NoStop}%
\bibitem [{\citenamefont {Ireland}\ \emph {et~al.}(2016)\citenamefont
  {Ireland}, \citenamefont {Mundim}, \citenamefont {Nakano},\ and\
  \citenamefont {Campanelli}}]{ireland16}%
  \BibitemOpen
  \bibfield  {author} {\bibinfo {author} {\bibfnamefont {B.}~\bibnamefont
  {Ireland}}, \bibinfo {author} {\bibfnamefont {B.~C.}\ \bibnamefont {Mundim}},
  \bibinfo {author} {\bibfnamefont {H.}~\bibnamefont {Nakano}},\ and\ \bibinfo
  {author} {\bibfnamefont {M.}~\bibnamefont {Campanelli}},\ }\href@noop {}
  {\bibfield  {journal} {\bibinfo  {journal} {Phys. Rev. D}\ }\textbf {\bibinfo
  {volume} {93}},\ \bibinfo {pages} {104057} (\bibinfo {year}
  {2016})}\BibitemShut {NoStop}%
\bibitem [{\citenamefont {{Bowen}}\ \emph {et~al.}(2019)\citenamefont
  {{Bowen}}, \citenamefont {{Mewes}}, \citenamefont {{Noble}}, \citenamefont
  {{Avara}}, \citenamefont {{Campanelli}},\ and\ \citenamefont
  {{Krolik}}}]{bowen2019}%
  \BibitemOpen
  \bibfield  {author} {\bibinfo {author} {\bibfnamefont {D.~B.}\ \bibnamefont
  {{Bowen}}}, \bibinfo {author} {\bibfnamefont {V.}~\bibnamefont {{Mewes}}},
  \bibinfo {author} {\bibfnamefont {S.~C.}\ \bibnamefont {{Noble}}}, \bibinfo
  {author} {\bibfnamefont {M.}~\bibnamefont {{Avara}}}, \bibinfo {author}
  {\bibfnamefont {M.}~\bibnamefont {{Campanelli}}},\ and\ \bibinfo {author}
  {\bibfnamefont {J.~H.}\ \bibnamefont {{Krolik}}},\ }\href
  {https://doi.org/10.3847/1538-4357/ab2453} {\bibfield  {journal} {\bibinfo
  {journal} {{Astrophys. J.}}\ }\textbf {\bibinfo {volume} {879}},\ \bibinfo
  {eid} {76} (\bibinfo {year} {2019})},\ \Eprint
  {https://arxiv.org/abs/1904.12048} {arXiv:1904.12048 [astro-ph.HE]}
  \BibitemShut {NoStop}%
\bibitem [{\citenamefont {{Avara}}\ \emph {et~al.}(2024)\citenamefont
  {{Avara}}, \citenamefont {{Krolik}}, \citenamefont {{Campanelli}},
  \citenamefont {{Noble}}, \citenamefont {{Bowen}},\ and\ \citenamefont
  {{Ryu}}}]{avara2023accretion}%
  \BibitemOpen
  \bibfield  {author} {\bibinfo {author} {\bibfnamefont {M.~J.}\ \bibnamefont
  {{Avara}}}, \bibinfo {author} {\bibfnamefont {J.~H.}\ \bibnamefont
  {{Krolik}}}, \bibinfo {author} {\bibfnamefont {M.}~\bibnamefont
  {{Campanelli}}}, \bibinfo {author} {\bibfnamefont {S.~C.}\ \bibnamefont
  {{Noble}}}, \bibinfo {author} {\bibfnamefont {D.}~\bibnamefont {{Bowen}}},\
  and\ \bibinfo {author} {\bibfnamefont {T.}~\bibnamefont {{Ryu}}},\ }\href
  {https://doi.org/10.3847/1538-4357/ad5bda} {\bibfield  {journal} {\bibinfo
  {journal} {{Astrophys. J.}}\ }\textbf {\bibinfo {volume} {974}},\ \bibinfo
  {eid} {242} (\bibinfo {year} {2024})},\ \Eprint
  {https://arxiv.org/abs/2305.18538} {arXiv:2305.18538 [astro-ph.HE]}
  \BibitemShut {NoStop}%
\bibitem [{\citenamefont {Mignon-Risse}\ \emph {et~al.}(2023)\citenamefont
  {Mignon-Risse}, \citenamefont {Varniere},\ and\ \citenamefont
  {Casse}}]{mignon2023origin}%
  \BibitemOpen
  \bibfield  {author} {\bibinfo {author} {\bibfnamefont {R.}~\bibnamefont
  {Mignon-Risse}}, \bibinfo {author} {\bibfnamefont {P.}~\bibnamefont
  {Varniere}},\ and\ \bibinfo {author} {\bibfnamefont {F.}~\bibnamefont
  {Casse}},\ }\href@noop {} {\bibfield  {journal} {\bibinfo  {journal} {Mon.
  Not. R. Astron. Soc.}\ }\textbf {\bibinfo {volume} {520}},\ \bibinfo {pages}
  {1285} (\bibinfo {year} {2023})}\BibitemShut {NoStop}%
\bibitem [{\citenamefont {Farris}\ \emph {et~al.}(2011)\citenamefont {Farris},
  \citenamefont {Liu},\ and\ \citenamefont {Shapiro}}]{farris2011Binary}%
  \BibitemOpen
  \bibfield  {author} {\bibinfo {author} {\bibfnamefont {B.~D.}\ \bibnamefont
  {Farris}}, \bibinfo {author} {\bibfnamefont {Y.~T.}\ \bibnamefont {Liu}},\
  and\ \bibinfo {author} {\bibfnamefont {S.~L.}\ \bibnamefont {Shapiro}},\
  }\href {https://doi.org/10.1103/PhysRevD.84.024024} {\bibfield  {journal}
  {\bibinfo  {journal} {Phys. Rev. D}\ }\textbf {\bibinfo {volume} {84}},\
  \bibinfo {pages} {024024} (\bibinfo {year} {2011})},\ \Eprint
  {https://arxiv.org/abs/1105.2821} {arXiv:1105.2821} \BibitemShut {NoStop}%
\bibitem [{\citenamefont {{Farris}}\ \emph {et~al.}(2012)\citenamefont
  {{Farris}}, \citenamefont {{Gold}}, \citenamefont {{Paschalidis}},
  \citenamefont {{Etienne}},\ and\ \citenamefont {{Shapiro}}}]{Farris2012}%
  \BibitemOpen
  \bibfield  {author} {\bibinfo {author} {\bibfnamefont {B.~D.}\ \bibnamefont
  {{Farris}}}, \bibinfo {author} {\bibfnamefont {R.}~\bibnamefont {{Gold}}},
  \bibinfo {author} {\bibfnamefont {V.}~\bibnamefont {{Paschalidis}}}, \bibinfo
  {author} {\bibfnamefont {Z.~B.}\ \bibnamefont {{Etienne}}},\ and\ \bibinfo
  {author} {\bibfnamefont {S.~L.}\ \bibnamefont {{Shapiro}}},\ }\href
  {https://doi.org/10.1103/PhysRevLett.109.221102} {\bibfield  {journal}
  {\bibinfo  {journal} {\prl}\ }\textbf {\bibinfo {volume} {109}},\ \bibinfo
  {eid} {221102} (\bibinfo {year} {2012})},\ \Eprint
  {https://arxiv.org/abs/1207.3354} {arXiv:1207.3354 [astro-ph.HE]}
  \BibitemShut {NoStop}%
\bibitem [{\citenamefont {Combi}\ \emph {et~al.}(2021)\citenamefont {Combi},
  \citenamefont {Armengol}, \citenamefont {Campanelli}, \citenamefont
  {Ireland}, \citenamefont {Noble}, \citenamefont {Nakano},\ and\ \citenamefont
  {Bowen}}]{combi2021superposed}%
  \BibitemOpen
  \bibfield  {author} {\bibinfo {author} {\bibfnamefont {L.}~\bibnamefont
  {Combi}}, \bibinfo {author} {\bibfnamefont {F.~G.~L.}\ \bibnamefont
  {Armengol}}, \bibinfo {author} {\bibfnamefont {M.}~\bibnamefont
  {Campanelli}}, \bibinfo {author} {\bibfnamefont {B.}~\bibnamefont {Ireland}},
  \bibinfo {author} {\bibfnamefont {S.~C.}\ \bibnamefont {Noble}}, \bibinfo
  {author} {\bibfnamefont {H.}~\bibnamefont {Nakano}},\ and\ \bibinfo {author}
  {\bibfnamefont {D.}~\bibnamefont {Bowen}},\ }\href@noop {} {\bibfield
  {journal} {\bibinfo  {journal} {Phys. Rev. D}\ }\textbf {\bibinfo {volume}
  {104}},\ \bibinfo {pages} {044041} (\bibinfo {year} {2021})}\BibitemShut
  {NoStop}%
\bibitem [{\citenamefont {Armengol}\ \emph {et~al.}(2021)\citenamefont
  {Armengol}, \citenamefont {Combi}, \citenamefont {Campanelli}, \citenamefont
  {Noble}, \citenamefont {Krolik}, \citenamefont {Bowen}, \citenamefont
  {Avara}, \citenamefont {Mewes},\ and\ \citenamefont
  {Nakano}}]{armengol2021circumbinary}%
  \BibitemOpen
  \bibfield  {author} {\bibinfo {author} {\bibfnamefont {F.~G.~L.}\
  \bibnamefont {Armengol}}, \bibinfo {author} {\bibfnamefont {L.}~\bibnamefont
  {Combi}}, \bibinfo {author} {\bibfnamefont {M.}~\bibnamefont {Campanelli}},
  \bibinfo {author} {\bibfnamefont {S.~C.}\ \bibnamefont {Noble}}, \bibinfo
  {author} {\bibfnamefont {J.~H.}\ \bibnamefont {Krolik}}, \bibinfo {author}
  {\bibfnamefont {D.~B.}\ \bibnamefont {Bowen}}, \bibinfo {author}
  {\bibfnamefont {M.~J.}\ \bibnamefont {Avara}}, \bibinfo {author}
  {\bibfnamefont {V.}~\bibnamefont {Mewes}},\ and\ \bibinfo {author}
  {\bibfnamefont {H.}~\bibnamefont {Nakano}},\ }\href@noop {} {\bibfield
  {journal} {\bibinfo  {journal} {Astrophys.\ J.}\ }\textbf {\bibinfo {volume}
  {913}},\ \bibinfo {pages} {16} (\bibinfo {year} {2021})}\BibitemShut
  {NoStop}%
\bibitem [{\citenamefont {Guti{\'e}rrez}\ \emph {et~al.}(2022)\citenamefont
  {Guti{\'e}rrez}, \citenamefont {Combi}, \citenamefont {Noble}, \citenamefont
  {Campanelli}, \citenamefont {Krolik}, \citenamefont {Armengol},\ and\
  \citenamefont {Garc{\'\i}a}}]{gutierrez2022electromagnetic}%
  \BibitemOpen
  \bibfield  {author} {\bibinfo {author} {\bibfnamefont {E.~M.}\ \bibnamefont
  {Guti{\'e}rrez}}, \bibinfo {author} {\bibfnamefont {L.}~\bibnamefont
  {Combi}}, \bibinfo {author} {\bibfnamefont {S.~C.}\ \bibnamefont {Noble}},
  \bibinfo {author} {\bibfnamefont {M.}~\bibnamefont {Campanelli}}, \bibinfo
  {author} {\bibfnamefont {J.~H.}\ \bibnamefont {Krolik}}, \bibinfo {author}
  {\bibfnamefont {F.~L.}\ \bibnamefont {Armengol}},\ and\ \bibinfo {author}
  {\bibfnamefont {F.}~\bibnamefont {Garc{\'\i}a}},\ }\href@noop {} {\bibfield
  {journal} {\bibinfo  {journal} {Astrophys.\ J.}\ }\textbf {\bibinfo {volume}
  {928}},\ \bibinfo {pages} {137} (\bibinfo {year} {2022})}\BibitemShut
  {NoStop}%
\bibitem [{\citenamefont {Davelaar}\ and\ \citenamefont
  {Haiman}(2022{\natexlab{a}})}]{davelaar2022selfa}%
  \BibitemOpen
  \bibfield  {author} {\bibinfo {author} {\bibfnamefont {J.}~\bibnamefont
  {Davelaar}}\ and\ \bibinfo {author} {\bibfnamefont {Z.}~\bibnamefont
  {Haiman}},\ }\href@noop {} {\bibfield  {journal} {\bibinfo  {journal} {Phys.
  Rev. D}\ }\textbf {\bibinfo {volume} {105}},\ \bibinfo {pages} {103010}
  (\bibinfo {year} {2022}{\natexlab{a}})}\BibitemShut {NoStop}%
\bibitem [{\citenamefont {Davelaar}\ and\ \citenamefont
  {Haiman}(2022{\natexlab{b}})}]{davelaar2022selfb}%
  \BibitemOpen
  \bibfield  {author} {\bibinfo {author} {\bibfnamefont {J.}~\bibnamefont
  {Davelaar}}\ and\ \bibinfo {author} {\bibfnamefont {Z.}~\bibnamefont
  {Haiman}},\ }\href@noop {} {\bibfield  {journal} {\bibinfo  {journal} {Phys.
  Rev. Lett.}\ }\textbf {\bibinfo {volume} {128}},\ \bibinfo {pages} {191101}
  (\bibinfo {year} {2022}{\natexlab{b}})}\BibitemShut {NoStop}%
\bibitem [{\citenamefont {Giacomazzo}\ \emph {et~al.}(2012)\citenamefont
  {Giacomazzo}, \citenamefont {Baker}, \citenamefont {Miller}, \citenamefont
  {Reynolds},\ and\ \citenamefont {van Meter}}]{giacomazzo2012}%
  \BibitemOpen
  \bibfield  {author} {\bibinfo {author} {\bibfnamefont {B.}~\bibnamefont
  {Giacomazzo}}, \bibinfo {author} {\bibfnamefont {J.~G.}\ \bibnamefont
  {Baker}}, \bibinfo {author} {\bibfnamefont {M.~C.}\ \bibnamefont {Miller}},
  \bibinfo {author} {\bibfnamefont {C.~S.}\ \bibnamefont {Reynolds}},\ and\
  \bibinfo {author} {\bibfnamefont {J.~R.}\ \bibnamefont {van Meter}},\ }\href
  {https://doi.org/10.1088/2041-8205/752/1/l15} {\bibfield  {journal} {\bibinfo
   {journal} {Astrophys.\ J.}\ }\textbf {\bibinfo {volume} {752}},\ \bibinfo
  {pages} {L15} (\bibinfo {year} {2012})}\BibitemShut {NoStop}%
\bibitem [{\citenamefont {{Gold}}\ \emph {et~al.}(2014)\citenamefont {{Gold}},
  \citenamefont {{Paschalidis}}, \citenamefont {{Ruiz}}, \citenamefont
  {{Shapiro}}, \citenamefont {{Etienne}},\ and\ \citenamefont
  {{Pfeiffer}}}]{Gold2014b}%
  \BibitemOpen
  \bibfield  {author} {\bibinfo {author} {\bibfnamefont {R.}~\bibnamefont
  {{Gold}}}, \bibinfo {author} {\bibfnamefont {V.}~\bibnamefont
  {{Paschalidis}}}, \bibinfo {author} {\bibfnamefont {M.}~\bibnamefont
  {{Ruiz}}}, \bibinfo {author} {\bibfnamefont {S.~L.}\ \bibnamefont
  {{Shapiro}}}, \bibinfo {author} {\bibfnamefont {Z.~B.}\ \bibnamefont
  {{Etienne}}},\ and\ \bibinfo {author} {\bibfnamefont {H.~P.}\ \bibnamefont
  {{Pfeiffer}}},\ }\href {https://doi.org/10.1103/PhysRevD.90.104030}
  {\bibfield  {journal} {\bibinfo  {journal} {\prd}\ }\textbf {\bibinfo
  {volume} {90}},\ \bibinfo {eid} {104030} (\bibinfo {year} {2014})},\ \Eprint
  {https://arxiv.org/abs/1410.1543} {arXiv:1410.1543 [astro-ph.GA]}
  \BibitemShut {NoStop}%
\bibitem [{\citenamefont {Paschalidis}\ \emph {et~al.}(2021)\citenamefont
  {Paschalidis}, \citenamefont {Bright}, \citenamefont {Ruiz},\ and\
  \citenamefont {Gold}}]{paschalidis2021minidisk}%
  \BibitemOpen
  \bibfield  {author} {\bibinfo {author} {\bibfnamefont {V.}~\bibnamefont
  {Paschalidis}}, \bibinfo {author} {\bibfnamefont {J.}~\bibnamefont {Bright}},
  \bibinfo {author} {\bibfnamefont {M.}~\bibnamefont {Ruiz}},\ and\ \bibinfo
  {author} {\bibfnamefont {R.}~\bibnamefont {Gold}},\ }\href@noop {} {\bibfield
   {journal} {\bibinfo  {journal} {Astrophys.\ J.\ Lett.}\ }\textbf {\bibinfo
  {volume} {910}},\ \bibinfo {pages} {L26} (\bibinfo {year}
  {2021})}\BibitemShut {NoStop}%
\bibitem [{\citenamefont {Gold}\ \emph {et~al.}(2014)\citenamefont {Gold},
  \citenamefont {Paschalidis}, \citenamefont {Etienne}, \citenamefont
  {Shapiro},\ and\ \citenamefont {Pfeiffer}}]{Gold_2014a}%
  \BibitemOpen
  \bibfield  {author} {\bibinfo {author} {\bibfnamefont {R.}~\bibnamefont
  {Gold}}, \bibinfo {author} {\bibfnamefont {V.}~\bibnamefont {Paschalidis}},
  \bibinfo {author} {\bibfnamefont {Z.~B.}\ \bibnamefont {Etienne}}, \bibinfo
  {author} {\bibfnamefont {S.~L.}\ \bibnamefont {Shapiro}},\ and\ \bibinfo
  {author} {\bibfnamefont {H.~P.}\ \bibnamefont {Pfeiffer}},\ }\bibfield
  {journal} {\bibinfo  {journal} {Phys. Rev. D}\ }\textbf {\bibinfo {volume}
  {89}},\ \href {https://doi.org/10.1103/physrevd.89.064060}
  {10.1103/physrevd.89.064060} (\bibinfo {year} {2014})\BibitemShut {NoStop}%
\bibitem [{\citenamefont {Kelly}\ \emph {et~al.}(2017)\citenamefont {Kelly},
  \citenamefont {Baker}, \citenamefont {Etienne}, \citenamefont {Giacomazzo},\
  and\ \citenamefont {Schnittman}}]{Kelly2017}%
  \BibitemOpen
  \bibfield  {author} {\bibinfo {author} {\bibfnamefont {B.~J.}\ \bibnamefont
  {Kelly}}, \bibinfo {author} {\bibfnamefont {J.~G.}\ \bibnamefont {Baker}},
  \bibinfo {author} {\bibfnamefont {Z.~B.}\ \bibnamefont {Etienne}}, \bibinfo
  {author} {\bibfnamefont {B.}~\bibnamefont {Giacomazzo}},\ and\ \bibinfo
  {author} {\bibfnamefont {J.}~\bibnamefont {Schnittman}},\ }\href
  {https://doi.org/10.1103/PhysRevD.96.123003} {\bibfield  {journal} {\bibinfo
  {journal} {Phys. Rev. D}\ }\textbf {\bibinfo {volume} {96}},\ \bibinfo
  {pages} {123003} (\bibinfo {year} {2017})}\BibitemShut {NoStop}%
\bibitem [{\citenamefont {{Ressler}}\ \emph {et~al.}(2024)\citenamefont
  {{Ressler}}, \citenamefont {{Combi}}, \citenamefont {{Li}}, \citenamefont
  {{Ripperda}},\ and\ \citenamefont {{Yang}}}]{ressler2024}%
  \BibitemOpen
  \bibfield  {author} {\bibinfo {author} {\bibfnamefont {S.~M.}\ \bibnamefont
  {{Ressler}}}, \bibinfo {author} {\bibfnamefont {L.}~\bibnamefont {{Combi}}},
  \bibinfo {author} {\bibfnamefont {X.}~\bibnamefont {{Li}}}, \bibinfo {author}
  {\bibfnamefont {B.}~\bibnamefont {{Ripperda}}},\ and\ \bibinfo {author}
  {\bibfnamefont {H.}~\bibnamefont {{Yang}}},\ }\href
  {https://doi.org/10.3847/1538-4357/ad3ae2} {\bibfield  {journal} {\bibinfo
  {journal} {Astrophys. J.}\ }\textbf {\bibinfo {volume} {967}},\ \bibinfo
  {eid} {70} (\bibinfo {year} {2024})},\ \Eprint
  {https://arxiv.org/abs/2404.02193} {arXiv:2404.02193 [astro-ph.HE]}
  \BibitemShut {NoStop}%
\bibitem [{\citenamefont {{Ressler}}\ \emph
  {et~al.}(2025{\natexlab{a}})\citenamefont {{Ressler}}, \citenamefont
  {{Combi}}, \citenamefont {{Ripperda}},\ and\ \citenamefont
  {{Most}}}]{Ressler2025a}%
  \BibitemOpen
  \bibfield  {author} {\bibinfo {author} {\bibfnamefont {S.~M.}\ \bibnamefont
  {{Ressler}}}, \bibinfo {author} {\bibfnamefont {L.}~\bibnamefont {{Combi}}},
  \bibinfo {author} {\bibfnamefont {B.}~\bibnamefont {{Ripperda}}},\ and\
  \bibinfo {author} {\bibfnamefont {E.~R.}\ \bibnamefont {{Most}}},\ }\href
  {https://doi.org/10.3847/2041-8213/ad9eb5} {\bibfield  {journal} {\bibinfo
  {journal} {Astrophys. J. Lett.}\ }\textbf {\bibinfo {volume} {979}},\
  \bibinfo {eid} {L24} (\bibinfo {year} {2025}{\natexlab{a}})},\ \Eprint
  {https://arxiv.org/abs/2410.10944} {arXiv:2410.10944 [astro-ph.HE]}
  \BibitemShut {NoStop}%
\bibitem [{\citenamefont {{Ressler}}\ \emph
  {et~al.}(2025{\natexlab{b}})\citenamefont {{Ressler}}, \citenamefont
  {{Combi}}, \citenamefont {{Ripperda}},\ and\ \citenamefont
  {{Li}}}]{Ressler2025b}%
  \BibitemOpen
  \bibfield  {author} {\bibinfo {author} {\bibfnamefont {S.~M.}\ \bibnamefont
  {{Ressler}}}, \bibinfo {author} {\bibfnamefont {L.}~\bibnamefont {{Combi}}},
  \bibinfo {author} {\bibfnamefont {B.}~\bibnamefont {{Ripperda}}},\ and\
  \bibinfo {author} {\bibfnamefont {X.}~\bibnamefont {{Li}}},\ }\href
  {https://doi.org/10.3847/2041-8213/ae11ab} {\bibfield  {journal} {\bibinfo
  {journal} {Astrophys. J. Lett.}\ }\textbf {\bibinfo {volume} {993}},\
  \bibinfo {eid} {L22} (\bibinfo {year} {2025}{\natexlab{b}})},\ \Eprint
  {https://arxiv.org/abs/2509.18241} {arXiv:2509.18241 [astro-ph.HE]}
  \BibitemShut {NoStop}%
\bibitem [{\citenamefont {Poisson}(2004)}]{poisson}%
  \BibitemOpen
  \bibfield  {author} {\bibinfo {author} {\bibfnamefont {E.}~\bibnamefont
  {Poisson}},\ }\href@noop {} {\emph {\bibinfo {title} {A relativist's toolkit:
  the mathematics of black-hole mechanics}}}\ (\bibinfo  {publisher} {Cambridge
  university press},\ \bibinfo {year} {2004})\BibitemShut {NoStop}%
\bibitem [{\citenamefont {Ma}\ \emph {et~al.}(2021)\citenamefont {Ma},
  \citenamefont {Giesler}, \citenamefont {Scheel},\ and\ \citenamefont
  {Varma}}]{ma2021extending}%
  \BibitemOpen
  \bibfield  {author} {\bibinfo {author} {\bibfnamefont {S.}~\bibnamefont
  {Ma}}, \bibinfo {author} {\bibfnamefont {M.}~\bibnamefont {Giesler}},
  \bibinfo {author} {\bibfnamefont {M.~A.}\ \bibnamefont {Scheel}},\ and\
  \bibinfo {author} {\bibfnamefont {V.}~\bibnamefont {Varma}},\ }\href@noop {}
  {\bibfield  {journal} {\bibinfo  {journal} {Phys. Rev. D}\ }\textbf {\bibinfo
  {volume} {103}},\ \bibinfo {pages} {084029} (\bibinfo {year}
  {2021})}\BibitemShut {NoStop}%
\bibitem [{\citenamefont {Fermi}(1965)}]{fermi1962}%
  \BibitemOpen
  \bibfield  {author} {\bibinfo {author} {\bibfnamefont {E.}~\bibnamefont
  {Fermi}},\ }in\ \href@noop {} {\emph {\bibinfo {booktitle} {Enrico Fermi,
  Collected Papers (Note e Memorie)}}}\ (\bibinfo  {publisher} {Chicago
  University Press},\ \bibinfo {address} {Chicago},\ \bibinfo {year}
  {1962--1965})\BibitemShut {NoStop}%
\bibitem [{\citenamefont {Synge}(1960)}]{Synge1960}%
  \BibitemOpen
  \bibinfo {editor} {\bibfnamefont {J.~L.}\ \bibnamefont {Synge}},\ ed.,\
  \href@noop {} {\emph {\bibinfo {title} {Relativity: The General Theory}}}\
  (\bibinfo  {publisher} {Interscience Publishers},\ \bibinfo {address} {New
  York,},\ \bibinfo {year} {1960})\BibitemShut {NoStop}%
\bibitem [{\citenamefont {Manasse}\ and\ \citenamefont
  {Misner}(1963)}]{manasse1963fermi}%
  \BibitemOpen
  \bibfield  {author} {\bibinfo {author} {\bibfnamefont {F.}~\bibnamefont
  {Manasse}}\ and\ \bibinfo {author} {\bibfnamefont {C.~W.}\ \bibnamefont
  {Misner}},\ }\href@noop {} {\bibfield  {journal} {\bibinfo  {journal}
  {Journal of mathematical physics}\ }\textbf {\bibinfo {volume} {4}},\
  \bibinfo {pages} {735} (\bibinfo {year} {1963})}\BibitemShut {NoStop}%
\bibitem [{\citenamefont {Poisson}\ \emph {et~al.}(2011)\citenamefont
  {Poisson}, \citenamefont {Pound},\ and\ \citenamefont
  {Vega}}]{poisson2011motion}%
  \BibitemOpen
  \bibfield  {author} {\bibinfo {author} {\bibfnamefont {E.}~\bibnamefont
  {Poisson}}, \bibinfo {author} {\bibfnamefont {A.}~\bibnamefont {Pound}},\
  and\ \bibinfo {author} {\bibfnamefont {I.}~\bibnamefont {Vega}},\ }\href@noop
  {} {\bibfield  {journal} {\bibinfo  {journal} {Living Rev. Relativ.}\
  }\textbf {\bibinfo {volume} {14}},\ \bibinfo {pages} {7} (\bibinfo {year}
  {2011})}\BibitemShut {NoStop}%
\bibitem [{\citenamefont {Mashhoon}\ and\ \citenamefont
  {Muench}(2002)}]{mashhoon2002length}%
  \BibitemOpen
  \bibfield  {author} {\bibinfo {author} {\bibfnamefont {B.}~\bibnamefont
  {Mashhoon}}\ and\ \bibinfo {author} {\bibfnamefont {U.}~\bibnamefont
  {Muench}},\ }\href@noop {} {\bibfield  {journal} {\bibinfo  {journal}
  {Annalen der Physik}\ }\textbf {\bibinfo {volume} {11}},\ \bibinfo {pages}
  {532} (\bibinfo {year} {2002})}\BibitemShut {NoStop}%
\bibitem [{\citenamefont {Rindler}(1956)}]{rindler1956visual}%
  \BibitemOpen
  \bibfield  {author} {\bibinfo {author} {\bibfnamefont {W.}~\bibnamefont
  {Rindler}},\ }\href@noop {} {\bibfield  {journal} {\bibinfo  {journal} {Mon.
  Not. R. Astron. Soc.}\ }\textbf {\bibinfo {volume} {116}},\ \bibinfo {pages}
  {662} (\bibinfo {year} {1956})}\BibitemShut {NoStop}%
\bibitem [{\citenamefont {Kidder}(1995)}]{kidder1995coalescing}%
  \BibitemOpen
  \bibfield  {author} {\bibinfo {author} {\bibfnamefont {L.~E.}\ \bibnamefont
  {Kidder}},\ }\href@noop {} {\bibfield  {journal} {\bibinfo  {journal} {Phys.
  Rev. D}\ }\textbf {\bibinfo {volume} {52}},\ \bibinfo {pages} {821} (\bibinfo
  {year} {1995})}\BibitemShut {NoStop}%
\bibitem [{\citenamefont {Csizmadia}\ \emph {et~al.}(2012)\citenamefont
  {Csizmadia}, \citenamefont {Debreczeni}, \citenamefont {R{\'a}cz},\ and\
  \citenamefont {Vas{\'u}th}}]{csizmadia2012gravitational}%
  \BibitemOpen
  \bibfield  {author} {\bibinfo {author} {\bibfnamefont {P.}~\bibnamefont
  {Csizmadia}}, \bibinfo {author} {\bibfnamefont {G.}~\bibnamefont
  {Debreczeni}}, \bibinfo {author} {\bibfnamefont {I.}~\bibnamefont
  {R{\'a}cz}},\ and\ \bibinfo {author} {\bibfnamefont {M.}~\bibnamefont
  {Vas{\'u}th}},\ }\href@noop {} {\bibfield  {journal} {\bibinfo  {journal}
  {Class. Quantum Grav.}\ }\textbf {\bibinfo {volume} {29}},\ \bibinfo {pages}
  {245002} (\bibinfo {year} {2012})}\BibitemShut {NoStop}%
\bibitem [{Note1()}]{Note1}%
  \BibitemOpen
  \bibinfo {note} {\protect \url
  {https://github.com/shinsei90/CBwaves}}\BibitemShut {NoStop}%
\bibitem [{\citenamefont {Blanchet}(2014)}]{blanchet2014gravitational}%
  \BibitemOpen
  \bibfield  {author} {\bibinfo {author} {\bibfnamefont {L.}~\bibnamefont
  {Blanchet}},\ }\href@noop {} {\bibfield  {journal} {\bibinfo  {journal}
  {Living Rev. Relativ.}\ }\textbf {\bibinfo {volume} {17}},\ \bibinfo {pages}
  {1} (\bibinfo {year} {2014})}\BibitemShut {NoStop}%
\bibitem [{\citenamefont {Kacskovics}\ and\ \citenamefont
  {Vas{\'u}th}(2022)}]{kacskovics2022orbital}%
  \BibitemOpen
  \bibfield  {author} {\bibinfo {author} {\bibfnamefont {B.}~\bibnamefont
  {Kacskovics}}\ and\ \bibinfo {author} {\bibfnamefont {M.}~\bibnamefont
  {Vas{\'u}th}},\ }\href@noop {} {\bibfield  {journal} {\bibinfo  {journal}
  {Class. Quantum Grav.}\ }\textbf {\bibinfo {volume} {39}},\ \bibinfo {pages}
  {095007} (\bibinfo {year} {2022})}\BibitemShut {NoStop}%
\bibitem [{\citenamefont {Buonanno}\ \emph {et~al.}(2008)\citenamefont
  {Buonanno}, \citenamefont {Kidder},\ and\ \citenamefont
  {Lehner}}]{buonanno2008estimating}%
  \BibitemOpen
  \bibfield  {author} {\bibinfo {author} {\bibfnamefont {A.}~\bibnamefont
  {Buonanno}}, \bibinfo {author} {\bibfnamefont {L.~E.}\ \bibnamefont
  {Kidder}},\ and\ \bibinfo {author} {\bibfnamefont {L.}~\bibnamefont
  {Lehner}},\ }\href@noop {} {\bibfield  {journal} {\bibinfo  {journal} {Phys.
  Rev. D}\ }\textbf {\bibinfo {volume} {77}},\ \bibinfo {pages} {026004}
  (\bibinfo {year} {2008})}\BibitemShut {NoStop}%
\bibitem [{\citenamefont {Lehner}\ and\ \citenamefont
  {Pretorius}(2014)}]{lehner2014numerical}%
  \BibitemOpen
  \bibfield  {author} {\bibinfo {author} {\bibfnamefont {L.}~\bibnamefont
  {Lehner}}\ and\ \bibinfo {author} {\bibfnamefont {F.}~\bibnamefont
  {Pretorius}},\ }\href@noop {} {\bibfield  {journal} {\bibinfo  {journal}
  {Annu. Rev. Astron. Astrophys.}\ }\textbf {\bibinfo {volume} {52}},\ \bibinfo
  {pages} {661} (\bibinfo {year} {2014})}\BibitemShut {NoStop}%
\bibitem [{\citenamefont {Gerosa}\ and\ \citenamefont
  {Kesden}(2016)}]{gerosa2016precession}%
  \BibitemOpen
  \bibfield  {author} {\bibinfo {author} {\bibfnamefont {D.}~\bibnamefont
  {Gerosa}}\ and\ \bibinfo {author} {\bibfnamefont {M.}~\bibnamefont
  {Kesden}},\ }\href@noop {} {\bibfield  {journal} {\bibinfo  {journal} {Phys.
  Rev. D}\ }\textbf {\bibinfo {volume} {93}},\ \bibinfo {pages} {124066}
  (\bibinfo {year} {2016})}\BibitemShut {NoStop}%
\bibitem [{\citenamefont {Barausse}\ \emph {et~al.}(2012)\citenamefont
  {Barausse}, \citenamefont {Morozova},\ and\ \citenamefont
  {Rezzolla}}]{barausse2012mass}%
  \BibitemOpen
  \bibfield  {author} {\bibinfo {author} {\bibfnamefont {E.}~\bibnamefont
  {Barausse}}, \bibinfo {author} {\bibfnamefont {V.}~\bibnamefont {Morozova}},\
  and\ \bibinfo {author} {\bibfnamefont {L.}~\bibnamefont {Rezzolla}},\
  }\href@noop {} {\bibfield  {journal} {\bibinfo  {journal} {Astrophys.\ J.}\
  }\textbf {\bibinfo {volume} {758}},\ \bibinfo {pages} {63} (\bibinfo {year}
  {2012})}\BibitemShut {NoStop}%
\bibitem [{\citenamefont {Hofmann}\ \emph {et~al.}(2016)\citenamefont
  {Hofmann}, \citenamefont {Barausse},\ and\ \citenamefont
  {Rezzolla}}]{hofmann2016final}%
  \BibitemOpen
  \bibfield  {author} {\bibinfo {author} {\bibfnamefont {F.}~\bibnamefont
  {Hofmann}}, \bibinfo {author} {\bibfnamefont {E.}~\bibnamefont {Barausse}},\
  and\ \bibinfo {author} {\bibfnamefont {L.}~\bibnamefont {Rezzolla}},\
  }\href@noop {} {\bibfield  {journal} {\bibinfo  {journal} {Astrophys.\ J.\
  Lett.}\ }\textbf {\bibinfo {volume} {825}},\ \bibinfo {pages} {L19} (\bibinfo
  {year} {2016})}\BibitemShut {NoStop}%
\bibitem [{\citenamefont {Kesden}\ \emph {et~al.}(2010)\citenamefont {Kesden},
  \citenamefont {Sperhake},\ and\ \citenamefont {Berti}}]{kesden2010final}%
  \BibitemOpen
  \bibfield  {author} {\bibinfo {author} {\bibfnamefont {M.}~\bibnamefont
  {Kesden}}, \bibinfo {author} {\bibfnamefont {U.}~\bibnamefont {Sperhake}},\
  and\ \bibinfo {author} {\bibfnamefont {E.}~\bibnamefont {Berti}},\
  }\href@noop {} {\bibfield  {journal} {\bibinfo  {journal} {Phys. Rev. D}\
  }\textbf {\bibinfo {volume} {81}},\ \bibinfo {pages} {084054} (\bibinfo
  {year} {2010})}\BibitemShut {NoStop}%
\bibitem [{\citenamefont {Nakamura}\ \emph {et~al.}(1987)\citenamefont
  {Nakamura}, \citenamefont {Oohara},\ and\ \citenamefont
  {Kojima}}]{nakamura1987General}%
  \BibitemOpen
  \bibfield  {author} {\bibinfo {author} {\bibfnamefont {T.}~\bibnamefont
  {Nakamura}}, \bibinfo {author} {\bibfnamefont {K.}~\bibnamefont {Oohara}},\
  and\ \bibinfo {author} {\bibfnamefont {Y.}~\bibnamefont {Kojima}},\ }\href
  {https://doi.org/10/c8hpjc} {\bibfield  {journal} {\bibinfo  {journal} {Prog.
  Theor. Phys. Suppl.}\ }\textbf {\bibinfo {volume} {90}},\ \bibinfo {pages}
  {1} (\bibinfo {year} {1987})}\BibitemShut {NoStop}%
\bibitem [{\citenamefont {Shibata}\ and\ \citenamefont
  {Nakamura}(1995)}]{shibata1995Evolution}%
  \BibitemOpen
  \bibfield  {author} {\bibinfo {author} {\bibfnamefont {M.}~\bibnamefont
  {Shibata}}\ and\ \bibinfo {author} {\bibfnamefont {T.}~\bibnamefont
  {Nakamura}},\ }\href {https://doi.org/10/c7dsgx} {\bibfield  {journal}
  {\bibinfo  {journal} {Phys. Rev. D}\ }\textbf {\bibinfo {volume} {52}},\
  \bibinfo {pages} {5428} (\bibinfo {year} {1995})}\BibitemShut {NoStop}%
\bibitem [{\citenamefont {Baumgarte}\ and\ \citenamefont
  {Shapiro}(1999)}]{baumgarte1999Numerical}%
  \BibitemOpen
  \bibfield  {author} {\bibinfo {author} {\bibfnamefont {T.~W.}\ \bibnamefont
  {Baumgarte}}\ and\ \bibinfo {author} {\bibfnamefont {S.~L.}\ \bibnamefont
  {Shapiro}},\ }\href {https://doi.org/10/dvp73g} {\bibfield  {journal}
  {\bibinfo  {journal} {Phys. Rev. D}\ }\textbf {\bibinfo {volume} {59}},\
  \bibinfo {pages} {024007} (\bibinfo {year} {1999})}\BibitemShut {NoStop}%
\bibitem [{\citenamefont {Combi}\ and\ \citenamefont
  {Ressler}(2024)}]{softcombi}%
  \BibitemOpen
  \bibfield  {author} {\bibinfo {author} {\bibfnamefont {L.}~\bibnamefont
  {Combi}}\ and\ \bibinfo {author} {\bibfnamefont {S.}~\bibnamefont
  {Ressler}},\ }\href {https://doi.org/10.5281/zenodo.10841021} {\bibinfo
  {title} {{Semi-analytical metric approximation of a binary black hole
  spacetime}}} (\bibinfo {year} {2024})\BibitemShut {NoStop}%
\bibitem [{\citenamefont {Lichnerowicz}(1944)}]{lichnerowicz1944espaces}%
  \BibitemOpen
  \bibfield  {author} {\bibinfo {author} {\bibfnamefont {A.}~\bibnamefont
  {Lichnerowicz}},\ }\href@noop {} {\bibfield  {journal} {\bibinfo  {journal}
  {Bulletin de la Soci{\'e}t{\'e} Math{\'e}matique de France}\ }\textbf
  {\bibinfo {volume} {72}},\ \bibinfo {pages} {146} (\bibinfo {year}
  {1944})}\BibitemShut {NoStop}%
\bibitem [{\citenamefont {Arnowitt}\ \emph {et~al.}(2008)\citenamefont
  {Arnowitt}, \citenamefont {Deser},\ and\ \citenamefont
  {Misner}}]{arnowitt2008Dynamics}%
  \BibitemOpen
  \bibfield  {author} {\bibinfo {author} {\bibfnamefont {R.~L.}\ \bibnamefont
  {Arnowitt}}, \bibinfo {author} {\bibfnamefont {S.}~\bibnamefont {Deser}},\
  and\ \bibinfo {author} {\bibfnamefont {C.~W.}\ \bibnamefont {Misner}},\
  }\href {https://doi.org/10/fpmkcf} {\bibfield  {journal} {\bibinfo  {journal}
  {Gen. Rel. Grav.}\ }\textbf {\bibinfo {volume} {40}},\ \bibinfo {pages}
  {1997} (\bibinfo {year} {2008})}\BibitemShut {NoStop}%
\bibitem [{\citenamefont {Baumgarte}\ and\ \citenamefont
  {Shapiro}(2010)}]{baumgarte2010Numerical}%
  \BibitemOpen
  \bibfield  {author} {\bibinfo {author} {\bibfnamefont {T.}~\bibnamefont
  {Baumgarte}}\ and\ \bibinfo {author} {\bibfnamefont {S.}~\bibnamefont
  {Shapiro}},\ }\href@noop {} {\emph {\bibinfo {title} {Numerical
  {{Relativity}}}}}\ (\bibinfo  {publisher} {{Cambridge University Press}},\
  \bibinfo {address} {{Cambridge}},\ \bibinfo {year} {2010})\BibitemShut
  {NoStop}%
\bibitem [{\citenamefont {Bona}\ \emph {et~al.}(1995)\citenamefont {Bona},
  \citenamefont {Masso}, \citenamefont {Seidel},\ and\ \citenamefont
  {Stela}}]{bona1995New}%
  \BibitemOpen
  \bibfield  {author} {\bibinfo {author} {\bibfnamefont {C.}~\bibnamefont
  {Bona}}, \bibinfo {author} {\bibfnamefont {J.}~\bibnamefont {Masso}},
  \bibinfo {author} {\bibfnamefont {E.}~\bibnamefont {Seidel}},\ and\ \bibinfo
  {author} {\bibfnamefont {J.}~\bibnamefont {Stela}},\ }\href
  {https://doi.org/10/c7m2mk} {\bibfield  {journal} {\bibinfo  {journal} {Phys.
  Rev. Lett.}\ }\textbf {\bibinfo {volume} {75}},\ \bibinfo {pages} {600}
  (\bibinfo {year} {1995})}\BibitemShut {NoStop}%
\bibitem [{\citenamefont {Alcubierre}\ \emph {et~al.}(2003)\citenamefont
  {Alcubierre}, \citenamefont {Bruegmann}, \citenamefont {Diener},
  \citenamefont {Koppitz}, \citenamefont {Pollney}, \citenamefont {Seidel},\
  and\ \citenamefont {Takahashi}}]{alcubierre2003Gauge}%
  \BibitemOpen
  \bibfield  {author} {\bibinfo {author} {\bibfnamefont {M.}~\bibnamefont
  {Alcubierre}}, \bibinfo {author} {\bibfnamefont {B.}~\bibnamefont
  {Bruegmann}}, \bibinfo {author} {\bibfnamefont {P.}~\bibnamefont {Diener}},
  \bibinfo {author} {\bibfnamefont {M.}~\bibnamefont {Koppitz}}, \bibinfo
  {author} {\bibfnamefont {D.}~\bibnamefont {Pollney}}, \bibinfo {author}
  {\bibfnamefont {E.}~\bibnamefont {Seidel}},\ and\ \bibinfo {author}
  {\bibfnamefont {R.}~\bibnamefont {Takahashi}},\ }\href
  {https://doi.org/10/ctxz84} {\bibfield  {journal} {\bibinfo  {journal} {Phys.
  Rev. D}\ }\textbf {\bibinfo {volume} {67}},\ \bibinfo {pages} {084023}
  (\bibinfo {year} {2003})}\BibitemShut {NoStop}%
\bibitem [{\citenamefont {Goodale}\ \emph {et~al.}(2003)\citenamefont
  {Goodale}, \citenamefont {Allen}, \citenamefont {Lanfermann}, \citenamefont
  {Mass{\'o}}, \citenamefont {Radke}, \citenamefont {Seidel},\ and\
  \citenamefont {Shalf}}]{goodale2003Cactus}%
  \BibitemOpen
  \bibfield  {author} {\bibinfo {author} {\bibfnamefont {T.}~\bibnamefont
  {Goodale}}, \bibinfo {author} {\bibfnamefont {G.}~\bibnamefont {Allen}},
  \bibinfo {author} {\bibfnamefont {G.}~\bibnamefont {Lanfermann}}, \bibinfo
  {author} {\bibfnamefont {J.}~\bibnamefont {Mass{\'o}}}, \bibinfo {author}
  {\bibfnamefont {T.}~\bibnamefont {Radke}}, \bibinfo {author} {\bibfnamefont
  {E.}~\bibnamefont {Seidel}},\ and\ \bibinfo {author} {\bibfnamefont
  {J.}~\bibnamefont {Shalf}},\ }in\ \href {https://doi.org/10/c3bzjm} {\emph
  {\bibinfo {booktitle} {High {{Performance Computing}} for {{Computational
  Science}} \textemdash{} {{VECPAR}} 2002}}},\ \bibinfo {editor} {edited by\
  \bibinfo {editor} {\bibfnamefont {J.~M. L.~M.}\ \bibnamefont {Palma}},
  \bibinfo {editor} {\bibfnamefont {A.~A.}\ \bibnamefont {Sousa}}, \bibinfo
  {editor} {\bibfnamefont {J.}~\bibnamefont {Dongarra}},\ and\ \bibinfo
  {editor} {\bibfnamefont {V.}~\bibnamefont {Hern{\'a}ndez}}}\ (\bibinfo
  {publisher} {{Springer Berlin Heidelberg}},\ \bibinfo {address} {{Berlin,
  Heidelberg}},\ \bibinfo {year} {2003})\ pp.\ \bibinfo {pages}
  {197--227}\BibitemShut {NoStop}%
\bibitem [{\citenamefont {Loffler}\ \emph {et~al.}(2012)\citenamefont {Loffler}
  \emph {et~al.}}]{loffler2012Einstein}%
  \BibitemOpen
  \bibfield  {author} {\bibinfo {author} {\bibfnamefont {F.}~\bibnamefont
  {Loffler}} \emph {et~al.},\ }\href {https://doi.org/10/ggcgbr} {\bibfield
  {journal} {\bibinfo  {journal} {Class. Quantum Grav.}\ }\textbf {\bibinfo
  {volume} {29}},\ \bibinfo {pages} {115001} (\bibinfo {year}
  {2012})}\BibitemShut {NoStop}%
\bibitem [{\citenamefont {Babiuc-Hamilton}\ \emph {et~al.}(2019)\citenamefont
  {Babiuc-Hamilton}, \citenamefont {Brandt}, \citenamefont {Diener},
  \citenamefont {Elley}, \citenamefont {Etienne}, \citenamefont {Ficarra},
  \citenamefont {Haas}, \citenamefont {Witek}, \citenamefont {Alcubierre},
  \citenamefont {Alic} \emph {et~al.}}]{babiuc2019einstein}%
  \BibitemOpen
  \bibfield  {author} {\bibinfo {author} {\bibfnamefont {M.}~\bibnamefont
  {Babiuc-Hamilton}}, \bibinfo {author} {\bibfnamefont {S.~R.}\ \bibnamefont
  {Brandt}}, \bibinfo {author} {\bibfnamefont {P.}~\bibnamefont {Diener}},
  \bibinfo {author} {\bibfnamefont {M.}~\bibnamefont {Elley}}, \bibinfo
  {author} {\bibfnamefont {Z.}~\bibnamefont {Etienne}}, \bibinfo {author}
  {\bibfnamefont {G.}~\bibnamefont {Ficarra}}, \bibinfo {author} {\bibfnamefont
  {R.}~\bibnamefont {Haas}}, \bibinfo {author} {\bibfnamefont {H.}~\bibnamefont
  {Witek}}, \bibinfo {author} {\bibfnamefont {M.}~\bibnamefont {Alcubierre}},
  \bibinfo {author} {\bibfnamefont {D.}~\bibnamefont {Alic}}, \emph {et~al.},\
  }\href@noop {} {\bibfield  {journal} {\bibinfo  {journal} {Zenodo}\ }
  (\bibinfo {year} {2019})}\BibitemShut {NoStop}%
\bibitem [{\citenamefont {Schnetter}\ \emph {et~al.}(2004)\citenamefont
  {Schnetter}, \citenamefont {Hawley},\ and\ \citenamefont
  {Hawke}}]{schnetter2004Evolutions}%
  \BibitemOpen
  \bibfield  {author} {\bibinfo {author} {\bibfnamefont {E.}~\bibnamefont
  {Schnetter}}, \bibinfo {author} {\bibfnamefont {S.~H.}\ \bibnamefont
  {Hawley}},\ and\ \bibinfo {author} {\bibfnamefont {I.}~\bibnamefont
  {Hawke}},\ }\href {https://doi.org/10/frnf47} {\bibfield  {journal} {\bibinfo
   {journal} {Class. Quantum Grav.}\ }\textbf {\bibinfo {volume} {21}},\
  \bibinfo {pages} {1465} (\bibinfo {year} {2004})}\BibitemShut {NoStop}%
\bibitem [{\citenamefont {Thornburg}(2004)}]{thornburg2004Fast}%
  \BibitemOpen
  \bibfield  {author} {\bibinfo {author} {\bibfnamefont {J.}~\bibnamefont
  {Thornburg}},\ }\href {https://doi.org/10/cmnzst} {\bibfield  {journal}
  {\bibinfo  {journal} {Cl. Quantum Grav}\ }\textbf {\bibinfo {volume} {21}},\
  \bibinfo {pages} {743} (\bibinfo {year} {2004})}\BibitemShut {NoStop}%
\bibitem [{\citenamefont {M{\"o}sta}\ \emph {et~al.}(2013)\citenamefont
  {M{\"o}sta}, \citenamefont {Mundim}, \citenamefont {Faber}, \citenamefont
  {Haas}, \citenamefont {Noble}, \citenamefont {Bode}, \citenamefont
  {L{\"o}ffler}, \citenamefont {Ott}, \citenamefont {Reisswig},\ and\
  \citenamefont {Schnetter}}]{mosta2013GRHydro}%
  \BibitemOpen
  \bibfield  {author} {\bibinfo {author} {\bibfnamefont {P.}~\bibnamefont
  {M{\"o}sta}}, \bibinfo {author} {\bibfnamefont {B.~C.}\ \bibnamefont
  {Mundim}}, \bibinfo {author} {\bibfnamefont {J.~A.}\ \bibnamefont {Faber}},
  \bibinfo {author} {\bibfnamefont {R.}~\bibnamefont {Haas}}, \bibinfo {author}
  {\bibfnamefont {S.~C.}\ \bibnamefont {Noble}}, \bibinfo {author}
  {\bibfnamefont {T.}~\bibnamefont {Bode}}, \bibinfo {author} {\bibfnamefont
  {F.}~\bibnamefont {L{\"o}ffler}}, \bibinfo {author} {\bibfnamefont {C.~D.}\
  \bibnamefont {Ott}}, \bibinfo {author} {\bibfnamefont {C.}~\bibnamefont
  {Reisswig}},\ and\ \bibinfo {author} {\bibfnamefont {E.}~\bibnamefont
  {Schnetter}},\ }\href {https://doi.org/10/ggcgbm} {\bibfield  {journal}
  {\bibinfo  {journal} {Class. Quantum Grav.}\ }\textbf {\bibinfo {volume}
  {31}},\ \bibinfo {pages} {015005} (\bibinfo {year} {2013})}\BibitemShut
  {NoStop}%
\bibitem [{\citenamefont {Combi}\ and\ \citenamefont
  {Siegel}(2023)}]{combi2023grmhd}%
  \BibitemOpen
  \bibfield  {author} {\bibinfo {author} {\bibfnamefont {L.}~\bibnamefont
  {Combi}}\ and\ \bibinfo {author} {\bibfnamefont {D.~M.}\ \bibnamefont
  {Siegel}},\ }\href@noop {} {\bibfield  {journal} {\bibinfo  {journal}
  {Astrophys.\ J.}\ }\textbf {\bibinfo {volume} {944}},\ \bibinfo {pages} {28}
  (\bibinfo {year} {2023})}\BibitemShut {NoStop}%
\bibitem [{\citenamefont {Baiotti}\ \emph {et~al.}(2005)\citenamefont
  {Baiotti}, \citenamefont {Hawke}, \citenamefont {Montero}, \citenamefont
  {Loffler}, \citenamefont {Rezzolla}, \citenamefont {Stergioulas},
  \citenamefont {Font},\ and\ \citenamefont
  {Seidel}}]{baiotti2005Threedimensional}%
  \BibitemOpen
  \bibfield  {author} {\bibinfo {author} {\bibfnamefont {L.}~\bibnamefont
  {Baiotti}}, \bibinfo {author} {\bibfnamefont {I.}~\bibnamefont {Hawke}},
  \bibinfo {author} {\bibfnamefont {P.~J.}\ \bibnamefont {Montero}}, \bibinfo
  {author} {\bibfnamefont {F.}~\bibnamefont {Loffler}}, \bibinfo {author}
  {\bibfnamefont {L.}~\bibnamefont {Rezzolla}}, \bibinfo {author}
  {\bibfnamefont {N.}~\bibnamefont {Stergioulas}}, \bibinfo {author}
  {\bibfnamefont {J.~A.}\ \bibnamefont {Font}},\ and\ \bibinfo {author}
  {\bibfnamefont {E.}~\bibnamefont {Seidel}},\ }\href
  {https://doi.org/10/cq6b4b} {\bibfield  {journal} {\bibinfo  {journal} {Phys.
  Rev. D}\ }\textbf {\bibinfo {volume} {71}},\ \bibinfo {pages} {024035}
  (\bibinfo {year} {2005})}\BibitemShut {NoStop}%
\bibitem [{\citenamefont {{Brown}}\ \emph {et~al.}(2009)\citenamefont
  {{Brown}}, \citenamefont {{Diener}}, \citenamefont {{Sarbach}}, \citenamefont
  {{Schnetter}},\ and\ \citenamefont {{Tiglio}}}]{Brown:2008sb}%
  \BibitemOpen
  \bibfield  {author} {\bibinfo {author} {\bibfnamefont {D.}~\bibnamefont
  {{Brown}}}, \bibinfo {author} {\bibfnamefont {P.}~\bibnamefont {{Diener}}},
  \bibinfo {author} {\bibfnamefont {O.}~\bibnamefont {{Sarbach}}}, \bibinfo
  {author} {\bibfnamefont {E.}~\bibnamefont {{Schnetter}}},\ and\ \bibinfo
  {author} {\bibfnamefont {M.}~\bibnamefont {{Tiglio}}},\ }\href
  {https://doi.org/10.1103/PhysRevD.79.044023} {\bibfield  {journal} {\bibinfo
  {journal} {\prd}\ }\textbf {\bibinfo {volume} {79}},\ \bibinfo {eid} {044023}
  (\bibinfo {year} {2009})},\ \Eprint {https://arxiv.org/abs/0809.3533}
  {arXiv:0809.3533 [gr-qc]} \BibitemShut {NoStop}%
\bibitem [{\citenamefont {{Ansorg}}\ \emph {et~al.}(2004)\citenamefont
  {{Ansorg}}, \citenamefont {{Br{\"u}gmann}},\ and\ \citenamefont
  {{Tichy}}}]{Ansorg:2004ds}%
  \BibitemOpen
  \bibfield  {author} {\bibinfo {author} {\bibfnamefont {M.}~\bibnamefont
  {{Ansorg}}}, \bibinfo {author} {\bibfnamefont {B.}~\bibnamefont
  {{Br{\"u}gmann}}},\ and\ \bibinfo {author} {\bibfnamefont {W.}~\bibnamefont
  {{Tichy}}},\ }\href {https://doi.org/10.1103/PhysRevD.70.064011} {\bibfield
  {journal} {\bibinfo  {journal} {\prd}\ }\textbf {\bibinfo {volume} {70}},\
  \bibinfo {eid} {064011} (\bibinfo {year} {2004})},\ \Eprint
  {https://arxiv.org/abs/gr-qc/0404056} {arXiv:gr-qc/0404056 [gr-qc]}
  \BibitemShut {NoStop}%
\bibitem [{\citenamefont {Brandt}\ and\ \citenamefont
  {Br{\"u}gmann}(1997)}]{brandt1997simple}%
  \BibitemOpen
  \bibfield  {author} {\bibinfo {author} {\bibfnamefont {S.}~\bibnamefont
  {Brandt}}\ and\ \bibinfo {author} {\bibfnamefont {B.}~\bibnamefont
  {Br{\"u}gmann}},\ }\href@noop {} {\bibfield  {journal} {\bibinfo  {journal}
  {Phys. Rev. Lett.}\ }\textbf {\bibinfo {volume} {78}},\ \bibinfo {pages}
  {3606} (\bibinfo {year} {1997})}\BibitemShut {NoStop}%
\bibitem [{\citenamefont {Gourgoulhon}(2007)}]{gourgoulhon2007construction}%
  \BibitemOpen
  \bibfield  {author} {\bibinfo {author} {\bibfnamefont {E.}~\bibnamefont
  {Gourgoulhon}},\ }in\ \href@noop {} {\emph {\bibinfo {booktitle} {Journal of
  Physics: Conference Series}}},\ Vol.~\bibinfo {volume} {91}\ (\bibinfo
  {organization} {IOP Publishing},\ \bibinfo {year} {2007})\ p.\ \bibinfo
  {pages} {012001}\BibitemShut {NoStop}%
\bibitem [{\citenamefont {Font}\ \emph {et~al.}(2002)\citenamefont {Font},
  \citenamefont {Goodale}, \citenamefont {Iyer}, \citenamefont {Miller},
  \citenamefont {Rezzolla}, \citenamefont {Seidel}, \citenamefont
  {Stergioulas}, \citenamefont {Suen},\ and\ \citenamefont
  {Tobias}}]{font2002Threedimensional}%
  \BibitemOpen
  \bibfield  {author} {\bibinfo {author} {\bibfnamefont {J.~A.}\ \bibnamefont
  {Font}}, \bibinfo {author} {\bibfnamefont {T.}~\bibnamefont {Goodale}},
  \bibinfo {author} {\bibfnamefont {S.}~\bibnamefont {Iyer}}, \bibinfo {author}
  {\bibfnamefont {M.~A.}\ \bibnamefont {Miller}}, \bibinfo {author}
  {\bibfnamefont {L.}~\bibnamefont {Rezzolla}}, \bibinfo {author}
  {\bibfnamefont {E.}~\bibnamefont {Seidel}}, \bibinfo {author} {\bibfnamefont
  {N.}~\bibnamefont {Stergioulas}}, \bibinfo {author} {\bibfnamefont {W.-M.}\
  \bibnamefont {Suen}},\ and\ \bibinfo {author} {\bibfnamefont
  {M.}~\bibnamefont {Tobias}},\ }\href {https://doi.org/10/fd6pjj} {\bibfield
  {journal} {\bibinfo  {journal} {Phys. Rev. D}\ }\textbf {\bibinfo {volume}
  {65}},\ \bibinfo {pages} {084024} (\bibinfo {year} {2002})}\BibitemShut
  {NoStop}%
\bibitem [{\citenamefont {Harten}\ \emph {et~al.}(1987)\citenamefont {Harten},
  \citenamefont {Engquist}, \citenamefont {Osher},\ and\ \citenamefont
  {Chakravarthy}}]{harten1987Uniformly}%
  \BibitemOpen
  \bibfield  {author} {\bibinfo {author} {\bibfnamefont {A.}~\bibnamefont
  {Harten}}, \bibinfo {author} {\bibfnamefont {B.}~\bibnamefont {Engquist}},
  \bibinfo {author} {\bibfnamefont {S.}~\bibnamefont {Osher}},\ and\ \bibinfo
  {author} {\bibfnamefont {S.~R.}\ \bibnamefont {Chakravarthy}},\ }\href
  {https://doi.org/10/bv8p2k} {\bibfield  {journal} {\bibinfo  {journal} {J.
  Comput. Phys.}\ }\textbf {\bibinfo {volume} {71}},\ \bibinfo {pages} {231}
  (\bibinfo {year} {1987})}\BibitemShut {NoStop}%
\bibitem [{\citenamefont {Tchekhovskoy}\ \emph {et~al.}(2007)\citenamefont
  {Tchekhovskoy}, \citenamefont {McKinney},\ and\ \citenamefont
  {Narayan}}]{tchekhovskoy2007wham}%
  \BibitemOpen
  \bibfield  {author} {\bibinfo {author} {\bibfnamefont {A.}~\bibnamefont
  {Tchekhovskoy}}, \bibinfo {author} {\bibfnamefont {J.~C.}\ \bibnamefont
  {McKinney}},\ and\ \bibinfo {author} {\bibfnamefont {R.}~\bibnamefont
  {Narayan}},\ }\href {https://doi.org/10/fsrrjb} {\bibfield  {journal}
  {\bibinfo  {journal} {Mon. Not. R. Astron. Soc.}\ }\textbf {\bibinfo {volume}
  {379}},\ \bibinfo {pages} {469} (\bibinfo {year} {2007})}\BibitemShut
  {NoStop}%
\bibitem [{\citenamefont {Siegel}\ \emph {et~al.}(2018)\citenamefont {Siegel},
  \citenamefont {M{\"o}sta}, \citenamefont {Desai},\ and\ \citenamefont
  {Wu}}]{siegel2018Recovery}%
  \BibitemOpen
  \bibfield  {author} {\bibinfo {author} {\bibfnamefont {D.~M.}\ \bibnamefont
  {Siegel}}, \bibinfo {author} {\bibfnamefont {P.}~\bibnamefont {M{\"o}sta}},
  \bibinfo {author} {\bibfnamefont {D.}~\bibnamefont {Desai}},\ and\ \bibinfo
  {author} {\bibfnamefont {S.}~\bibnamefont {Wu}},\ }\href
  {https://doi.org/10/ghsv4x} {\bibfield  {journal} {\bibinfo  {journal}
  {Astrophys. J.}\ }\textbf {\bibinfo {volume} {859}},\ \bibinfo {pages} {71}
  (\bibinfo {year} {2018})}\BibitemShut {NoStop}%
\bibitem [{\citenamefont {Farris}\ \emph {et~al.}(2012)\citenamefont {Farris},
  \citenamefont {Gold}, \citenamefont {Paschalidis}, \citenamefont {Etienne},\
  and\ \citenamefont {Shapiro}}]{farris2012binary}%
  \BibitemOpen
  \bibfield  {author} {\bibinfo {author} {\bibfnamefont {B.~D.}\ \bibnamefont
  {Farris}}, \bibinfo {author} {\bibfnamefont {R.}~\bibnamefont {Gold}},
  \bibinfo {author} {\bibfnamefont {V.}~\bibnamefont {Paschalidis}}, \bibinfo
  {author} {\bibfnamefont {Z.~B.}\ \bibnamefont {Etienne}},\ and\ \bibinfo
  {author} {\bibfnamefont {S.~L.}\ \bibnamefont {Shapiro}},\ }\href@noop {}
  {\bibfield  {journal} {\bibinfo  {journal} {Phys. Rev. Lett.}\ }\textbf
  {\bibinfo {volume} {109}},\ \bibinfo {pages} {221102} (\bibinfo {year}
  {2012})}\BibitemShut {NoStop}%
\bibitem [{\citenamefont {Del~Zanna}\ \emph {et~al.}(2007)\citenamefont
  {Del~Zanna}, \citenamefont {Zanotti}, \citenamefont {Bucciantini},\ and\
  \citenamefont {Londrillo}}]{del2007echo}%
  \BibitemOpen
  \bibfield  {author} {\bibinfo {author} {\bibfnamefont {L.}~\bibnamefont
  {Del~Zanna}}, \bibinfo {author} {\bibfnamefont {O.}~\bibnamefont {Zanotti}},
  \bibinfo {author} {\bibfnamefont {N.}~\bibnamefont {Bucciantini}},\ and\
  \bibinfo {author} {\bibfnamefont {P.}~\bibnamefont {Londrillo}},\ }\href@noop
  {} {\bibfield  {journal} {\bibinfo  {journal} {Astronomy \& Astrophysics}\
  }\textbf {\bibinfo {volume} {473}},\ \bibinfo {pages} {11} (\bibinfo {year}
  {2007})}\BibitemShut {NoStop}%
\bibitem [{\citenamefont {White}\ \emph {et~al.}(2016)\citenamefont {White},
  \citenamefont {Stone},\ and\ \citenamefont {Gammie}}]{white2016extension}%
  \BibitemOpen
  \bibfield  {author} {\bibinfo {author} {\bibfnamefont {C.~J.}\ \bibnamefont
  {White}}, \bibinfo {author} {\bibfnamefont {J.~M.}\ \bibnamefont {Stone}},\
  and\ \bibinfo {author} {\bibfnamefont {C.~F.}\ \bibnamefont {Gammie}},\
  }\href@noop {} {\bibfield  {journal} {\bibinfo  {journal} {Astrophys.\ J.\
  Suppl.\ Ser.}\ }\textbf {\bibinfo {volume} {225}},\ \bibinfo {pages} {22}
  (\bibinfo {year} {2016})}\BibitemShut {NoStop}%
\bibitem [{\citenamefont {Sadiq}\ \emph {et~al.}(2018)\citenamefont {Sadiq},
  \citenamefont {Zlochower},\ and\ \citenamefont {Nakano}}]{sadiq2018}%
  \BibitemOpen
  \bibfield  {author} {\bibinfo {author} {\bibfnamefont {J.}~\bibnamefont
  {Sadiq}}, \bibinfo {author} {\bibfnamefont {Y.}~\bibnamefont {Zlochower}},\
  and\ \bibinfo {author} {\bibfnamefont {H.}~\bibnamefont {Nakano}},\
  }\href@noop {} {\bibfield  {journal} {\bibinfo  {journal} {Phys. Rev. D}\
  }\textbf {\bibinfo {volume} {97}},\ \bibinfo {pages} {084007} (\bibinfo
  {year} {2018})}\BibitemShut {NoStop}%
\bibitem [{\citenamefont {Baumgarte}\ and\ \citenamefont
  {Naculich}(2007)}]{baumgarte2007analytical}%
  \BibitemOpen
  \bibfield  {author} {\bibinfo {author} {\bibfnamefont {T.~W.}\ \bibnamefont
  {Baumgarte}}\ and\ \bibinfo {author} {\bibfnamefont {S.~G.}\ \bibnamefont
  {Naculich}},\ }\href@noop {} {\bibfield  {journal} {\bibinfo  {journal}
  {Phys. Rev. D}\ }\textbf {\bibinfo {volume} {75}},\ \bibinfo {pages} {067502}
  (\bibinfo {year} {2007})}\BibitemShut {NoStop}%
\bibitem [{\citenamefont {Alic}\ \emph {et~al.}(2013)\citenamefont {Alic},
  \citenamefont {Kastaun},\ and\ \citenamefont
  {Rezzolla}}]{alic2013Constraint}%
  \BibitemOpen
  \bibfield  {author} {\bibinfo {author} {\bibfnamefont {D.}~\bibnamefont
  {Alic}}, \bibinfo {author} {\bibfnamefont {W.}~\bibnamefont {Kastaun}},\ and\
  \bibinfo {author} {\bibfnamefont {L.}~\bibnamefont {Rezzolla}},\ }\href
  {https://doi.org/10/gpb5n5} {\bibfield  {journal} {\bibinfo  {journal} {Phys.
  Rev. D}\ }\textbf {\bibinfo {volume} {88}},\ \bibinfo {pages} {064049}
  (\bibinfo {year} {2013})}\BibitemShut {NoStop}%
\bibitem [{\citenamefont {Hawking}\ and\ \citenamefont
  {Ellis}(2023)}]{hawking2023large}%
  \BibitemOpen
  \bibfield  {author} {\bibinfo {author} {\bibfnamefont {S.~W.}\ \bibnamefont
  {Hawking}}\ and\ \bibinfo {author} {\bibfnamefont {G.~F.}\ \bibnamefont
  {Ellis}},\ }\href@noop {} {\emph {\bibinfo {title} {The large scale structure
  of space-time}}}\ (\bibinfo  {publisher} {Cambridge university press},\
  \bibinfo {year} {2023})\BibitemShut {NoStop}%
\bibitem [{\citenamefont {Pook-Kolb}\ \emph {et~al.}(2021)\citenamefont
  {Pook-Kolb}, \citenamefont {Hennigar},\ and\ \citenamefont
  {Booth}}]{pook2021happens}%
  \BibitemOpen
  \bibfield  {author} {\bibinfo {author} {\bibfnamefont {D.}~\bibnamefont
  {Pook-Kolb}}, \bibinfo {author} {\bibfnamefont {R.~A.}\ \bibnamefont
  {Hennigar}},\ and\ \bibinfo {author} {\bibfnamefont {I.}~\bibnamefont
  {Booth}},\ }\href@noop {} {\bibfield  {journal} {\bibinfo  {journal} {Phys.
  Rev. Lett.}\ }\textbf {\bibinfo {volume} {127}},\ \bibinfo {pages} {181101}
  (\bibinfo {year} {2021})}\BibitemShut {NoStop}%
\bibitem [{\citenamefont {{Dreyer}}\ \emph {et~al.}(2003)\citenamefont
  {{Dreyer}}, \citenamefont {{Krishnan}}, \citenamefont {{Shoemaker}},\ and\
  \citenamefont {{Schnetter}}}]{Dreyer:2002mx}%
  \BibitemOpen
  \bibfield  {author} {\bibinfo {author} {\bibfnamefont {O.}~\bibnamefont
  {{Dreyer}}}, \bibinfo {author} {\bibfnamefont {B.}~\bibnamefont
  {{Krishnan}}}, \bibinfo {author} {\bibfnamefont {D.}~\bibnamefont
  {{Shoemaker}}},\ and\ \bibinfo {author} {\bibfnamefont {E.}~\bibnamefont
  {{Schnetter}}},\ }\href {https://doi.org/10.1103/PhysRevD.67.024018}
  {\bibfield  {journal} {\bibinfo  {journal} {\prd}\ }\textbf {\bibinfo
  {volume} {67}},\ \bibinfo {eid} {024018} (\bibinfo {year} {2003})},\ \Eprint
  {https://arxiv.org/abs/gr-qc/0206008} {arXiv:gr-qc/0206008 [gr-qc]}
  \BibitemShut {NoStop}%
\bibitem [{\citenamefont {Schnetter}\ \emph {et~al.}(2006)\citenamefont
  {Schnetter}, \citenamefont {Krishnan},\ and\ \citenamefont
  {Beyer}}]{schnetter2006introduction}%
  \BibitemOpen
  \bibfield  {author} {\bibinfo {author} {\bibfnamefont {E.}~\bibnamefont
  {Schnetter}}, \bibinfo {author} {\bibfnamefont {B.}~\bibnamefont
  {Krishnan}},\ and\ \bibinfo {author} {\bibfnamefont {F.}~\bibnamefont
  {Beyer}},\ }\href@noop {} {\bibfield  {journal} {\bibinfo  {journal} {Phys.
  Rev. D}\ }\textbf {\bibinfo {volume} {74}},\ \bibinfo {pages} {024028}
  (\bibinfo {year} {2006})}\BibitemShut {NoStop}%
\bibitem [{\citenamefont {Farris}\ \emph {et~al.}(2010)\citenamefont {Farris},
  \citenamefont {Liu},\ and\ \citenamefont {Shapiro}}]{farris2010binary}%
  \BibitemOpen
  \bibfield  {author} {\bibinfo {author} {\bibfnamefont {B.~D.}\ \bibnamefont
  {Farris}}, \bibinfo {author} {\bibfnamefont {Y.~T.}\ \bibnamefont {Liu}},\
  and\ \bibinfo {author} {\bibfnamefont {S.~L.}\ \bibnamefont {Shapiro}},\
  }\href@noop {} {\bibfield  {journal} {\bibinfo  {journal} {Phys. Rev. D}\
  }\textbf {\bibinfo {volume} {81}},\ \bibinfo {pages} {084008} (\bibinfo
  {year} {2010})}\BibitemShut {NoStop}%
\bibitem [{\citenamefont {Cattorini}\ \emph {et~al.}(2021)\citenamefont
  {Cattorini}, \citenamefont {Giacomazzo}, \citenamefont {Haardt},\ and\
  \citenamefont {Colpi}}]{cattorini2021fully}%
  \BibitemOpen
  \bibfield  {author} {\bibinfo {author} {\bibfnamefont {F.}~\bibnamefont
  {Cattorini}}, \bibinfo {author} {\bibfnamefont {B.}~\bibnamefont
  {Giacomazzo}}, \bibinfo {author} {\bibfnamefont {F.}~\bibnamefont {Haardt}},\
  and\ \bibinfo {author} {\bibfnamefont {M.}~\bibnamefont {Colpi}},\
  }\href@noop {} {\bibfield  {journal} {\bibinfo  {journal} {Phys. Rev. D}\
  }\textbf {\bibinfo {volume} {103}},\ \bibinfo {pages} {103022} (\bibinfo
  {year} {2021})}\BibitemShut {NoStop}%
\bibitem [{\citenamefont {Edgar}(2004)}]{edgar2004review}%
  \BibitemOpen
  \bibfield  {author} {\bibinfo {author} {\bibfnamefont {R.}~\bibnamefont
  {Edgar}},\ }\href@noop {} {\bibfield  {journal} {\bibinfo  {journal} {New
  Astronomy Reviews}\ }\textbf {\bibinfo {volume} {48}},\ \bibinfo {pages}
  {843} (\bibinfo {year} {2004})}\BibitemShut {NoStop}%
\bibitem [{\citenamefont {Reisswig}\ \emph {et~al.}(2013)\citenamefont
  {Reisswig}, \citenamefont {Haas}, \citenamefont {Ott}, \citenamefont
  {Abdikamalov}, \citenamefont {M{\"o}sta}, \citenamefont {Pollney},\ and\
  \citenamefont {Schnetter}}]{reisswig2013ThreeDimensional}%
  \BibitemOpen
  \bibfield  {author} {\bibinfo {author} {\bibfnamefont {C.}~\bibnamefont
  {Reisswig}}, \bibinfo {author} {\bibfnamefont {R.}~\bibnamefont {Haas}},
  \bibinfo {author} {\bibfnamefont {C.~D.}\ \bibnamefont {Ott}}, \bibinfo
  {author} {\bibfnamefont {E.}~\bibnamefont {Abdikamalov}}, \bibinfo {author}
  {\bibfnamefont {P.}~\bibnamefont {M{\"o}sta}}, \bibinfo {author}
  {\bibfnamefont {D.}~\bibnamefont {Pollney}},\ and\ \bibinfo {author}
  {\bibfnamefont {E.}~\bibnamefont {Schnetter}},\ }\href
  {https://doi.org/10/gpb5tx} {\bibfield  {journal} {\bibinfo  {journal} {Phys.
  Rev. D}\ }\textbf {\bibinfo {volume} {87}},\ \bibinfo {pages} {064023}
  (\bibinfo {year} {2013})}\BibitemShut {NoStop}%
\bibitem [{\citenamefont {Ng}\ \emph {et~al.}(2024)\citenamefont {Ng},
  \citenamefont {Jiang}, \citenamefont {Musolino}, \citenamefont {Ecker},
  \citenamefont {Tootle},\ and\ \citenamefont {Rezzolla}}]{ng2024hybrid}%
  \BibitemOpen
  \bibfield  {author} {\bibinfo {author} {\bibfnamefont {H.~H.-Y.}\
  \bibnamefont {Ng}}, \bibinfo {author} {\bibfnamefont {J.-L.}\ \bibnamefont
  {Jiang}}, \bibinfo {author} {\bibfnamefont {C.}~\bibnamefont {Musolino}},
  \bibinfo {author} {\bibfnamefont {C.}~\bibnamefont {Ecker}}, \bibinfo
  {author} {\bibfnamefont {S.~D.}\ \bibnamefont {Tootle}},\ and\ \bibinfo
  {author} {\bibfnamefont {L.}~\bibnamefont {Rezzolla}},\ }\href@noop {}
  {\bibfield  {journal} {\bibinfo  {journal} {Phys. Rev. D}\ }\textbf {\bibinfo
  {volume} {109}},\ \bibinfo {pages} {064061} (\bibinfo {year}
  {2024})}\BibitemShut {NoStop}%
\bibitem [{Note2()}]{Note2}%
  \BibitemOpen
  \bibinfo {note} {\protect \url
  {https://gitlab.com/combi.luciano/analyticalbbh}}\BibitemShut {NoStop}%
\bibitem [{\citenamefont {{Ryan}}\ \emph {et~al.}(2025)\citenamefont {{Ryan}},
  \citenamefont {{van Eerten}}, \citenamefont {{Piro}}, \citenamefont
  {{Troja}}, \citenamefont {{O'Connor}},\ and\ \citenamefont
  {{Ricci}}}]{afterglowpy}%
  \BibitemOpen
  \bibfield  {author} {\bibinfo {author} {\bibfnamefont {G.}~\bibnamefont
  {{Ryan}}}, \bibinfo {author} {\bibfnamefont {H.}~\bibnamefont {{van
  Eerten}}}, \bibinfo {author} {\bibfnamefont {L.}~\bibnamefont {{Piro}}},
  \bibinfo {author} {\bibfnamefont {E.}~\bibnamefont {{Troja}}}, \bibinfo
  {author} {\bibfnamefont {B.}~\bibnamefont {{O'Connor}}},\ and\ \bibinfo
  {author} {\bibfnamefont {R.}~\bibnamefont {{Ricci}}},\ }\href@noop {}
  {\bibinfo {title} {{afterglowpy: Compute and fit GRB afterglows}}},\ \bibinfo
  {howpublished} {Astrophysics Source Code Library, record ascl:2505.014}
  (\bibinfo {year} {2025}),\ \Eprint {https://arxiv.org/abs/2505.014}
  {ascl:2505.014} \BibitemShut {NoStop}%
\bibitem [{\citenamefont {Dasgupta}\ \emph {et~al.}(2018)\citenamefont
  {Dasgupta}, \citenamefont {Mirizzi},\ and\ \citenamefont
  {Sen}}]{dasgupta_simple_2018}%
  \BibitemOpen
  \bibfield  {author} {\bibinfo {author} {\bibfnamefont {B.}~\bibnamefont
  {Dasgupta}}, \bibinfo {author} {\bibfnamefont {A.}~\bibnamefont {Mirizzi}},\
  and\ \bibinfo {author} {\bibfnamefont {M.}~\bibnamefont {Sen}},\ }\href
  {https://doi.org/10.1103/PhysRevD.98.103001} {\bibfield  {journal} {\bibinfo
  {journal} {\prd}\ }\textbf {\bibinfo {volume} {98}},\ \bibinfo {pages}
  {103001} (\bibinfo {year} {2018})}\BibitemShut {NoStop}%
\bibitem [{\citenamefont {Abbar}(2020)}]{abbar_searching_2020}%
  \BibitemOpen
  \bibfield  {author} {\bibinfo {author} {\bibfnamefont {S.}~\bibnamefont
  {Abbar}},\ }\href {https://doi.org/10.1088/1475-7516/2020/05/027} {\bibfield
  {journal} {\bibinfo  {journal} {\jcapp}\ }\textbf {\bibinfo {volume}
  {2020}},\ \bibinfo {pages} {027} (\bibinfo {year} {2020})}\BibitemShut
  {NoStop}%
\end{thebibliography}

\end{document}